\pdfoutput=1
\newcommand*{\ATLASLATEXPATH}{latex/}

\documentclass[UKenglish,texlive=2016,cernpreprint]{\ATLASLATEXPATH atlasdoc}

\usepackage{multirow}
\usepackage{pdflscape}
\usepackage{afterpage}
\usepackage{rotating}

\usepackage{\ATLASLATEXPATH atlaspackage}
\usepackage{\ATLASLATEXPATH atlasbiblatex}

\usepackage{\ATLASLATEXPATH atlascontribute}

\usepackage{\ATLASLATEXPATH atlasphysics}

\graphicspath{{logos/}{figures/}}

\usepackage{Moriond2017_HZZllll_Couplings-defs}

\AtlasTitle{Measurement of the Higgs boson coupling properties in the $H\rightarrow ZZ^{*} \rightarrow 4\ell$ decay channel at $\sqrt{s}$ = 13 TeV with the ATLAS detector}

\author{The ATLAS Collaboration}

\PreprintIdNumber{CERN-EP-2017-206}

\AtlasJournalRef{JHEP 03 (2018) 095}
\AtlasDOI{10.1007/JHEP03(2018)095}

\AtlasAbstract{The coupling properties of the Higgs boson are studied in the four-lepton ($e$, $\mu$) decay channel using 36.1 fb$^{-1}$ of $pp$ collision data from the LHC at a centre-of-mass energy of 13 TeV collected by the ATLAS detector. Cross sections are measured for the main production modes in several exclusive regions of the Higgs boson production phase space and are interpreted in terms of coupling modifiers. The inclusive cross section times branching ratio for $H \rightarrow ZZ^*$ decay and for a Higgs boson absolute rapidity below 2.5 is measured to be $1.73^{+0.24}_{-0.23}$(stat.)$^{+0.10}_{-0.08}$(exp.)$\pm 0.04$(th.)~pb 
  compared to the Standard Model prediction of $1.34\pm0.09$~pb. In addition, the tensor structure of the Higgs boson couplings is studied using an effective Lagrangian approach for the description of interactions beyond the Standard Model. Constraints are placed on the non-Standard-Model CP-even and CP-odd couplings to $Z$ bosons and on the CP-odd coupling to gluons.}

\hypersetup{pdftitle={ATLAS document},pdfauthor={The ATLAS Collaboration}}

\begin{document}

\maketitle




\section{Introduction}
\label{sec:intro}

The observation of the Higgs boson by the ATLAS and CMS experiments~\cite{HIGG-2012-27, CMS-HIG-12-028} with the LHC \runa~data at centre-of-mass energies of $\sqrt{ \mathrm{s} }=$~7~TeV and 8~TeV has been a major step towards the understanding of the mechanism of electroweak (EW) symmetry breaking~\cite{Englert:1964et,Higgs:1964pj,Guralnik:1964eu}. Further measurements of the spin, parity and couplings of the new particle have shown no significant deviation from the predictions for the Standard Model (SM) Higgs boson~\cite{HIGG-2013-17, CMS-HIG-14-018, HIGG-2014-06, CMS-HIG-14-009, HIGG-2015-07}. The increased centre-of-mass energy and higher integrated luminosity of the LHC \runb~data
allows the study of the Higgs boson properties in greater detail and an improved search for deviations from the SM predictions.

In this paper, the measurement of the Higgs boson coupling properties is performed in the four-lepton decay channel, \htollllbrief{}, where $\ell \equiv e\text{ or }\mu$, using \Lum of \runb~$pp$ collision data collected by the ATLAS experiment at a centre-of-mass energy of 13~TeV. This channel provides a clear signature and high signal-to-background ratio. The largest background is the continuum $(Z^{(*)}/\gamma^*)(Z^{(*)}/\gamma^*)$
production, referred to as \zzstar\ hereafter. For the studied  four-lepton invariant mass range of 118~GeV~$<m_{4\ell}<$~129~GeV,
there are also small but non-negligible background contributions from $Z+\mathrm{jets}$ and $t\bar{t}$ production with two prompt leptons.

The Higgs boson spin, parity and coupling properties have been studied in this channel with~\runa~data by the ATLAS and CMS experiments~\cite{HIGG-2013-21, HIGG-2013-17, CMS-HIG-14-035, CMS-HIG-13-002, CMS-HIG-14-018} and recently with~\runb~data by the CMS experiment~\cite{CMS-HIG-16-041, CMS-HIG-17-011}. In this paper, the Higgs boson couplings to SM particles are studied using two analysis approaches. In the first approach, the Higgs boson production cross sections are analyzed based on 
different production modes in several exclusive regions of the production phase space, testing whether it is compatible with the SM predictions. An interpretation in terms of coupling modifiers within the $\kappa$ framework~\cite{Heinemeyer:2013tqa,YR4}  is given, assuming a SM tensor structure ($J^P=0^+$) for all couplings. In the second approach, the tensor structure of the Higgs boson couplings
is studied, probing for admixtures of CP-even and CP-odd interactions in theories beyond the SM (BSM) in addition to the corresponding SM interactions. Both analyses are performed assuming that the studied resonance is a single particle state with spin-0 and a mass of 125.09~GeV based on experimental results obtained with the LHC \runa~data~\cite{HIGG-2014-14}. It is assumed that the total width of the resonance is small compared the experimental resolution and the interference effects between the signal and SM backgrounds are neglected due to the small contribution.

The paper is organized as follows. A brief introduction of the ATLAS detector is given in Section~\ref{sec:detector}. The analysis strategy describing the two analysis approaches is outlined in Section~\ref{sec:strategy}. In Section~\ref{sec:samples}
the data as well as the simulated signal 
and background samples are described.
The selection and categorization of the Higgs boson candidate events, as well as the discriminating observables used in the measurement, are described in Section~\ref{sec:selection}, while the signal and background modelling is detailed in Sections~\ref{sec:signal_bkg} and~\ref{sec:background}, respectively. The experimental and theoretical systematic uncertainties (Section~\ref{sec:systematics}) are taken into account for the statistical interpretation of the data, with the results presented in Section~\ref{sec:results}. Concluding remarks are given in Section~\ref{sec:summary}.

\section{ATLAS detector}
\label{sec:detector}

The ATLAS detector~\cite{PERF-2007-01} is a multi-purpose particle detector with a forward-backward symmetric cylindrical geometry.\footnote{ATLAS uses a right-handed coordinate system with its origin at the nominal interaction point (IP) in the centre of the detector and the $z$-axis along the beam pipe. The $x$-axis points from the IP to the centre of the LHC ring, and the $y$-axis points upward. Cylindrical coordinates $(r, \phi)$ are used in the transverse plane, with $\phi$ being the azimuthal angle around the beam pipe. The pseudorapidity is defined in terms of the polar angle $\theta$ as $\eta = - \ln \tan (\theta / 2)$.} It consists of an inner tracking detector (ID) in a 2~T axial magnetic field covering the pseudorapidity range $|\eta| < 2.5$. A new innermost silicon pixel layer~\cite{ATLAS-TDR-19} (IBL) was added to the ID after the \runa~data-taking. The ID is surrounded by the electromagnetic and hadronic calorimeters up to $|\eta|=$~4.9 
and by the  muon spectrometer (MS)
extending up to $|\eta| = 2.7$. The magnetic field for the MS is provided by a set of toroids with a field integral ranging between 2~Tm and 6~Tm 
across most of the detector.
The trigger and data-acquisition system is based on two levels of online event selection: a hardware-based first-level trigger and a software-based high-level trigger employing algorithms similar to those for the offline particle reconstruction.

\section{Analysis strategy}
\label{sec:strategy}

The Higgs boson couplings to 
heavy SM vector bosons ($W$ and $Z$) and gluons are studied by measuring the cross sections for different production modes and by probing BSM contributions in tensor couplings. In both approaches,
the reconstructed Higgs boson candidate events are classified into different categories.
The categories are defined to be sensitive to different Higgs boson production modes, which in turn also provides sensitivity to the BSM contributions. The event yields in each category serve as the final discriminant for both the cross section and the tensor structure studies. 
There are nine reconstructed event categories defined for the cross-section measurement, one of which is additionally split into two separate ones for the tensor structure studies to improve their sensitivity. For the cross-section measurement, there are also additional discriminating observables introduced in reconstructed event categories with a sufficiently high number of events. 
These observables are defined using dedicated boosted decision trees (\BDTs)~\cite{Hocker:2007ht}.

\subsection{Classification of the Higgs boson production modes}
\label{subsec:strategy_SM}

\begin{figure}[htb!]
   \hskip-0.4cm\includegraphics[width=1.04\linewidth]{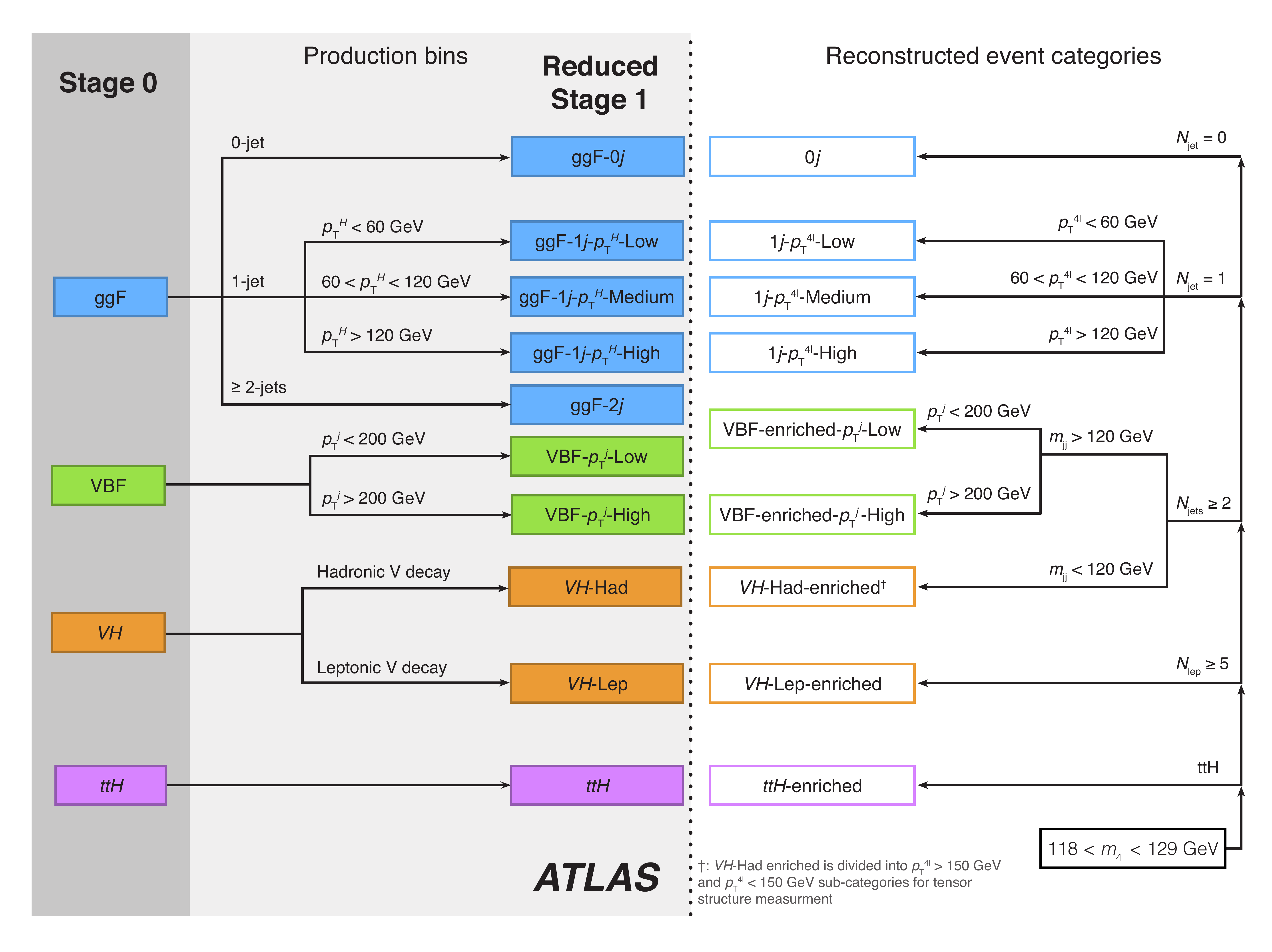}
   \caption{The phase-space regions (production bins) for the measurement of the Higgs boson production cross sections which are defined at the particle level for Stage 0 and 1, and the corresponding reconstructed event categories. Description of production bins is given in Section~\ref{sec:strategy}, while reconstructed event categories are described in Section~\ref{sec:selection}.}
      \label{fig:stxs_bins}
  \end{figure}

The Higgs boson production cross section times the branching ratio of the decay into $Z$ boson pairs, \sBRZZ,  is measured in several dedicated mutually exclusive regions of the phase space based on the production process. For simplicity, these regions are called ``production bins''.
Theoretical uncertainties have a reduced impact on \sBRZZ results and enter primarily for the interpretation of results in terms of Higgs boson couplings. The definitions of the production bins shown in Figure~\ref{fig:stxs_bins}~(shaded area) are based on particle-level events produced by dedicated event generators closely following the framework of simplified template cross sections~\cite{YR4}. The bins are chosen in such a way that the measurement precision is maximized and at the same time possible BSM contributions can be isolated. All production bins are defined for Higgs bosons with rapidity $|y_H|<$~2.5 and no requirements placed on the particle-level leptons. Two sets of production bins are considered since a more inclusive phase-space region usually reduces the statistical uncertainty of the measurement but at the cost of a larger theoretical uncertainty. 

For the first set (Stage 0)~\cite{YR4}, production bins are simply defined according to the Higgs boson production vertex: gluon--gluon fusion (\STXSggF), vector boson fusion (\STXSVBF) and associated production with top quark pairs (\STXSttH) or vector bosons (\STXSVH), where $V$ is a $W$ or a $Z$ boson. 
The \bbH Higgs boson production bin is not included because there is  insufficient sensitivity to measure this process with the current integrated luminosity. This production mode has an acceptance similar to gluon--gluon fusion, and their contributions are therefore considered together in the analysis.
The sum of their contributions is referred to in the following as gluon--gluon fusion.

For the second set (reduced Stage 1), a more exclusive set of production bins is defined.
This set is obtained by the merging of those production bins of the original Stage-1 set from Ref.~\cite{YR4} which cannot be measured separately in the $H\to ZZ^* \to 4\ell$ channel with the current data sample.
The gluon--gluon fusion process is split into events with zero, one or at least two particle-level jets.  The particle-level jets are built from all stable particles (all particles with c$\mathrm{\tau} >$ 1 mm) including neutrinos, photons and leptons from hadron decays or produced in the shower. All decay products from the Higgs boson, as well as the leptons and neutrinos from decays of the signal $V$ bosons are removed, while decay products from hadronically decaying signal $V$ bosons are included in the inputs to the particle-level jet building. The anti-$k_t$ jet reconstruction algorithm~\cite{Cacciari:2008gp}, implemented in the FastJet package~\cite{Cacciari:2011ma}, with a radius parameter
$R=0.4$ is used and jets are required to have $\pt>$~30~GeV.  The 1-jet bin is further split into three bins with the Higgs boson transverse momentum $\pt^H$ below 60~GeV, between 60~GeV and 120~GeV, and above 120~GeV. The reduced Stage-1 gluon--gluon fusion bins are correspondingly denoted by \STXSggFZeroJ, \STXSggFOneJL, \STXSggFOneJM, \STXSggFOneJH and \STXSggFTwoJ. The \STXSVBF production bin is split into two bins with the transverse momentum of the leading jet, $\pt^{j1}$, below and above 200~GeV (\STXSVBFLow and \STXSVBFHigh, respectively). The former bin is expected to be dominated by SM events, while the latter is sensitive to potential BSM contributions. For \STXSVH production, separate bins with hadronically (\STXSVHHad) and leptonically (\STXSVHLep) decaying vector bosons are considered. The leptonic $V$ boson decays include the decays into $\tau$ leptons and into neutrino pairs. The \STXSttH production bin remains the same as for Stage 0. 

Figure~\ref{fig:stxs_bins} also summarizes the corresponding categories of reconstructed events in which the cross-section measurements are performed and which are described in more detail in Section~\ref{sec:selection}. There is a dedicated reconstructed event category for each production bin except for \STXSggFTwoJ. This process contributes strongly to all reconstructed event categories containing events with at least two jets, and can therefore be measured in these categories, with the highest sensitivity expected in \CatVBFLow category.

\subsection{Tensor structure of Higgs boson couplings}
\label{subsec:strategy_BSM}

In order to study the tensor structure of the Higgs boson couplings to SM gauge bosons, interactions of the Higgs boson with these SM particles are described in terms of the effective Lagrangian of the Higgs characterization model~\cite{HC}, 
\begin{align}\label{eq:L_EFT}
  \mathcal{L}^{V}_{0} = & \bigg\{ 
  \kappa_{\textrm{SM}} \left[\frac{1}{2}g_{HZZ} Z_{\mu}Z^{\mu} + g_{HWW} W^{+}_{\mu}W^{-\mu}\right] \nonumber \\
  &-\frac{1}{4} \left[
    \kappa_{Hgg}g_{Hgg}G^{a}_{\mu\nu}G^{a,\mu\nu} + \tan{\alpha}\kappa_{Agg}g_{Agg}G^{a}_{\mu\nu}\tilde{G}^{a,\mu\nu}\right]\nonumber \\
  &-\frac{1}{4} \frac{1}{\Lambda}\left[
    \kappa_{HZZ}Z_{\mu\nu}Z^{\mu\nu} + \tan{\alpha}\kappa_{AZZ}Z_{\mu\nu}\tilde{Z}^{\mu\nu}\right] \nonumber \\
  &-\frac{1}{2} \frac{1}{\Lambda}\left[
    \kappa_{HWW}W^{+}_{\mu\nu}W^{-\mu\nu} + \tan{\alpha}\kappa_{AWW}W^{+}_{\mu\nu}\tilde{W}^{-\mu\nu}\right]  \bigg\} \mathcal{X}_{0}  .	  
\end{align} 	

The additional terms in the Lagrangian involving couplings to fermions are not considered since the present analysis is not sensitive to these couplings. The model is based on an effective field theory description which assumes there are no new BSM particles below the energy scale $\Lambda$. The cut-off scale $\Lambda$ is set to 1~TeV, supported by the current experimental results showing no evidence of new physics below this scale. The notation of Eq.~\eqref{eq:L_EFT} follows the notation of Eq.~(2.4) in Ref.~\cite{HC}
with $\mathcal{X}_{0}$ defining a new bosonic state of spin 0
and with the difference that the dimensionless coupling parameters $\kappa$ are redefined by dividing them by $\cos{\alpha}$, where $\alpha$ is the mixing angle between the $0^+$ and $0^-$ CP states implying CP-violation for $\alpha \neq 0$ and $\alpha \neq \pi$.
In this way the prediction for the SM Higgs boson is given by \kSM~=~1 and \kHgg~=~1 with the values of the BSM couplings set to zero. In this analysis, only the effective Lagrangian terms with coupling parameters \kHVV, \kAVV and \kAgg are considered as possible BSM admixtures to the corresponding SM interactions. These terms describe the CP-even (scalar) and CP-odd (pseudo-scalar) BSM interaction with vector bosons and the CP-odd BSM interaction with gluons, respectively. The BSM couplings are assumed to be the same for $W$ and
$Z$ bosons (i.e.  \kHWW~=~\kHZZ$\equiv$~\kHVV and   \kAWW~=~\kAZZ$\equiv$~\kAVV). 
The value of $\alpha$ is arbitrarily set to $\pi/4$ such that the CP-odd couplings can be more simply denoted by \kAVV~$\tan{\alpha} \Rightarrow$~\kAVV and \kAgg~$\tan{\alpha}~\Rightarrow$~\kAgg.

In the previous \runa~analysis~\cite{HIGG-2013-21}, the Higgs-related BSM interactions with heavy vector bosons were studied only in Higgs boson decays. In this analysis, the impact of BSM contributions on both the decay rates and the production cross sections in different production modes is taken into account. The \kHVV and \kAVV parameters contribute the most to \VH and \VBF Higgs boson production in the four-lepton decay mode since the coupling is present in both the production and decay vertices. 
The \kAgg parameter  mostly affects the \ggF production.

\section{Signal and background simulation}
\label{sec:samples}

The production of the SM Higgs boson via \ggF, \VBF and \VH (including  $gg\rightarrow ZH$) production mechanisms was modelled with the\progname{POWHEG-BOX} v2 Monte Carlo (MC) event generator~\cite{powheg5,Luisoni:2013kna}, interfaced to\progname{EvtGen}~v1.2.0~\cite{Lange:2001uf} for properties of the bottom and charm hadron decays, using the PDF4LHC next-to-leading-order (NLO) set of parton distribution functions (PDF)~\cite{Butterworth:2015oua}.
The gluon--gluon fusion Higgs boson production is accurate to next-to-next-to-leading order (NNLO) in the strong coupling, 
using the\progname{POWHEG} method for merging the NLO Higgs + jet cross section with the parton
shower, and the\progname{MiNLO} method~\cite{Hamilton:2013fea} to simultaneously achieve NLO accuracy for inclusive
Higgs boson production. A reweighting procedure, employing the Higgs boson rapidity, was applied using the\progname{HNNLO} program~\cite{Catani:2007vq,Grazzini:2008tf}.
The matrix elements of the \VBF and \VH production mechanisms were calculated up to NLO in QCD.
For \VH production, the\progname{MiNLO} method was used to merge 0- and 1-jet events ~\cite{Hamilton:2012rf}.
The $gg\rightarrow ZH$ contribution was modelled at leading order (LO) in QCD.
The production of a Higgs boson in association with a top (bottom) quark pair was
simulated at NLO with\progname{MadGraph5\_aMC@NLO} v2.2.3 (v2.3.3)~\cite{Alwall:2014hca,Wiesemann:2014ioa}
, using the CT10nlo PDF set~\cite{Lai:2010vv} for \ttH production and the NNPDF23 PDF set~\cite{Ball:2014uwa} for \bbH production. For the \ggF, \VBF, \VH and \bbH production mechanisms, the\progname{PYTHIA 8}~\cite{Sjostrand:2007gs} generator was used for the \htollllbrief{} decay as well as for the parton shower model 
using a set of tuned parameters called the AZNLO tune~\cite{STDM-2012-23}. For the \ttH production mechanism, the\progname{Herwig++}~\cite{Bahr:2008pv} event generator was used with the UEEE5 tune~\cite{Seymour:2013qka}. All signal samples were simulated for the Higgs boson with a mass $m_H=$~125.00~GeV. Wherever relevant, the signal mass distribution is shifted to the reference value of 125.09~GeV.

The Higgs boson production cross sections
and decay branching ratios, as well as their
uncertainties, were taken from Refs.~\cite{Dittmaier:2011ti,Dittmaier:2012vm,Heinemeyer:2013tqa,Anastasiou:2015ema,Anastasiou:2016cez,Actis:2008ug,Anastasiou:2008tj,Grazzini:2013mca,Stewart:2011cf, Ciccolini:2007jr,Ciccolini:2007ec,Bolzoni:2010xr, Brein:2003wg,Denner:2011id,Altenkamp:2012sx,Beenakker:2002nc,Dawson:2003zu, Yu:2014cka,Frixione:2015zaa,Dawson:2003kb,Dittmaier:2003ej,Harlander:2003ai,Djouadi:1997yw,Djouadi:2006bz,Bredenstein:2006rh,Bredenstein:2006ha,Boselli:2015aha,Butterworth:2015oua,Dulat:2015mca,Harland-Lang:2014zoa,Ball:2014uwa}. The \ggF production was calculated with next-to-next-to-next-to-leading order (N$^{3}$LO) accuracy in QCD and has NLO electroweak (EW) corrections applied. For \VBF production, full NLO QCD and EW calculations were used with approximate NNLO QCD corrections.
The \VH production was calculated at NNLO in QCD and NLO EW corrections are applied. The \ttH and \bbH processes were calculated to NLO accuracy in QCD. The branching ratio for the \htollllbrief{} decay with $m_H=$~125.09~GeV was predicted to be 0.0125\%~\cite{Djouadi:1997yw} in the SM using\progname{PROPHECY4F}~\cite{Bredenstein:2006rh,Bredenstein:2006ha}, which includes the complete NLO~QCD and EW corrections, and the interference effects between identical final-state fermions.
Table~\ref{tab:MCSignal} summarizes the production cross sections and branching ratios for the
\htollllbrief{} decay for $m_H=125.09$~\gev .

\begin{table*}[htbp]
  \centering
  \caption{The predicted SM Higgs boson production cross sections ($\sigma$) for \ggF,
    \VBF and associated production with a $W$ or $Z$ boson or with a $t\bar{t}$ or $b\bar{b}$ 
     pair in $pp$ collisions for $m_H=125.09$~\gev\ at    $\sqrt{\mathrm{s}}=13~\tev$~\cite{Dittmaier:2011ti,Dittmaier:2012vm,Heinemeyer:2013tqa,Anastasiou:2015ema,Anastasiou:2016cez,Actis:2008ug,Anastasiou:2008tj,Grazzini:2013mca,Stewart:2011cf, Ciccolini:2007jr,Ciccolini:2007ec,Bolzoni:2010xr, Brein:2003wg,Denner:2011id,Altenkamp:2012sx,Beenakker:2002nc,Dawson:2003zu, Yu:2014cka,Frixione:2015zaa,Dawson:2003kb,Dittmaier:2003ej,Harlander:2003ai,Djouadi:1997yw,Djouadi:2006bz,Bredenstein:2006rh,Bredenstein:2006ha,Boselli:2015aha,Butterworth:2015oua,Dulat:2015mca,Harland-Lang:2014zoa,Ball:2014uwa}.  The quoted uncertainties
    correspond to the total theoretical systematic uncertainties calculated by adding in quadrature
    the QCD scale and PDF$+\alpha_\mathrm{s}$ uncertainties.  The decay branching ratio (\BR) with the associated
    uncertainty for \htoZZ\ and \htollllbrief\ with $\ell = e, \mu$, is also given.
    \label{tab:MCSignal}}
  \vspace{0.1cm}
  {\renewcommand{\arraystretch}{1.1}
  
  \begin{tabular}{ll|r}
    \hline \hline
 
 \multicolumn{2}{l|}{Production process}&  \multicolumn{1}{c}{$\sigma$~[pb]}\\
 \hline

 \ggF& $\left(gg\to H \right)$&                   
 $48.5 \pm 2.4\phantom{00}$    \\
 
 \VBF& $\left(qq'\to Hqq' \right)$&               
 $3.78  \pm 0.08\phantom{0}$    \\
 
 \WH&  $\left(q\bar{q'}\to WH \right)$  &          
 $1.369 \pm 0.028$ \\
 
 \ZH&  $\left(q\bar{q}/gg\to ZH \right)$ &         
 $0.88 \pm 0.04\phantom{0}$    \\
 
 \ttH&  $\left(q\bar{q}/gg\to t\bar{t}H \right)$ & 
 $0.51 \pm 0.05\phantom{0}$ \\
 
 \bbH&  $\left(q\bar{q}/gg\to b\bar{b}H \right)$ & 
 $0.49 \pm 0.12\phantom{0}$ \\

\hline

 \multicolumn{2}{l|}{Decay process}&  \multicolumn{1}{c}{\BR~[$\cdot$ 10$^{-4}$]}\\
 \hline
 \multicolumn{2}{l|}{\htoZZ}       &   $264  \pm 6\phantom{.000}$ \\
 \multicolumn{2}{l|}{\htollllbrief}&  $\phantom{00}1.250 \pm 0.027$ \\

     \hline\hline
  \end{tabular}
  }
\end{table*}

Additional \ggF, \VBF and \VH signal samples with different values of the BSM couplings 
\kAgg, \kHVV and \kAVV
were generated with\progname{MadGraph5\_aMC@NLO} and are used for the signal modelling
as a function of the BSM couplings as explained in Section~\ref{sec:signal_bkg}.
The \ggF simulation includes samples at NLO QCD accuracy for zero, one and two additional partons merged with the\progname{FxFx} merging scheme~\cite{Alwall:2014hca,Frederix:2012ps},
while the \VBF and \VH simulations are accurate to LO in $\alpha_\mathrm{s}$. Equivalent \VBF and \VH processes were also generated at NLO QCD accuracy and used to estimate the relative uncertainties of higher-order QCD effects as a function of the BSM coupling parameters.

The \zzstar\ continuum background from  quark--antiquark annihilation was modelled using\progname{Sherpa} 2.2.2~\cite{Gleisberg:2008ta,Gleisberg:2008fv,Cascioli:2011va},
which provides a matrix element calculation accurate to NLO in $\alpha_\mathrm{s}$  for 0-, and 1-jet  final states and LO accuracy for 2- and 3-jet final states. 
The merging was performed with the\progname{Sherpa} parton shower~\cite{Schumann:2007mg} using the ME+PS@NLO prescription~\cite{Hoeche:2012yf}.
The NLO EW corrections were applied as a function of the invariant mass of the $ZZ^*$ system $m_{\zzstar}$ ~\cite{Biedermann:2016yvs,Biedermann:2016lvg}.

The gluon-induced \zzstar\ production was modelled by\progname{gg2VV}~\cite{Kauer:2015dma} at LO in QCD.
The higher-order QCD effects for the $gg\rightarrow ZZ^{*}$ continuum production have been calculated
for massless quark loops \cite{Caola:2015psa,Caola:2015rqy,Campbell:2016ivq} in the heavy top-quark
approximation ~\cite{Melnikov:2015laa}, including the $gg\rightarrow H^{*} \rightarrow ZZ$ processes ~\cite{Bonvini:2013jha,Li:2015jva}. The simulated LO samples are scaled by the $K$-factor of 1.7$\pm$1.0, defined as the ratio of the higher-order and the leading-order cross section predictions.

The \emph{WZ} background was modelled using\progname{POWHEG-BOX} v2 interfaced to\progname{PYTHIA 8}
and\progname{EvtGen}~v1.2.0  
for properties of the bottom and charm hadron decays. The
triboson backgrounds \emph{ZZZ}, \emph{WZZ}, and \emph{WWZ} with four or more prompt leptons were modelled
using\progname{Sherpa} 2.1.1.
The simulation of $t\bar{t}+Z$ events with both top quarks decaying semi-leptonically and the Z boson decaying leptonically was performed with\progname{MadGraph} interfaced to\progname{PYTHIA 8}
and the total cross section was
normalized to the prediction which includes the two dominant terms at both the LO and
the NLO in a mixed perturbative expansion in the QCD  and EW couplings ~\cite{Frixione:2015zaa}.

The modelling of events containing $Z$ bosons with associated jets
was performed using the\progname{Sherpa}~2.2.2 generator. Matrix elements were calculated for up to two partons at NLO and four partons at LO using
\progname{Comix}~\cite{Gleisberg:2008fv} and\progname{OpenLoops}~\cite{Cascioli:2011va},
and merged with the\progname{Sherpa} parton shower~\cite{Schumann:2007mg} using
the\progname{ME+PS@NLO} prescription~\cite{Hoeche:2012yf}.  The NNPDF3.0 NNLO PDF set was used in conjunction
with dedicated parton shower parameters tuning developed by the\progname{Sherpa} authors. 
Simulated samples were normalized to the data-driven estimate described in Section~\ref{sec:background}. As a cross-check, this estimate was compared to the theory prediction obtained  with
\progname{FEWZ}~\cite{Melnikov:2006kv,Anastasiou:2003ds} at NNLO in $\alpha_s$.

The $t\bar{t}$ background was modelled using\progname{POWHEG-BOX} v2 interfaced to\progname{PYTHIA 6}~\cite{pythia}
for parton showering, hadronisation, and the underlying event and to\progname{EvtGen}~v1.2.0 for properties of the bottom and charm hadron decays.

Generated events were processed through the ATLAS detector simulation~\cite{SOFT-2010-01} within the
\progname{Geant4} framework~\cite{GEANT4} and reconstructed the same way as the data.
Additional $pp$ interactions in the same and nearby
bunch crossings (pile-up) are included in the simulation.  The pile-up events were generated using~\progname{PYTHIA~8} with the A2 set of tuned parameters~\cite{ATLAS:2012uec} and the  MSTW2008LO PDF set~\cite{Martin:2009iq}. The simulation samples were weighted to
reproduce the observed distribution of the mean number of interactions per bunch crossing in the
data.

\section{Event selection}
\label{sec:selection}

\subsection{Event reconstruction}
\label{subsec:object_reconstruction}

The selection and categorization of the Higgs boson candidate events rely on the
reconstruction and identification of electrons, muons and jets, closely following the analyses reported in Refs.~\cite{HIGG-2013-21, ATLAS-CONF-2016-079}.

Collision vertices are reconstructed from ID tracks with transverse momentum $\pt>$~400~MeV. The vertex with the highest $\sum{\pt^2}$ of reconstructed tracks is selected as the primary vertex. Events are required to have at least one collision vertex with at least two associated tracks. 

Electron candidates are reconstructed from ID tracks that are matched to energy clusters in the electromagnetic calorimeter~\cite{ATL-PHYS-PUB-2015-041}. A Gaussian-sum filter algorithm~\cite{ATLAS-CONF-2014-032} is used to compensate for radiative energy losses in the ID. Electron identification is based on a likelihood discriminant combining the measured track properties, electromagnetic shower shapes and quality of the track--cluster matching. 
The ``loose'' likelihood criteria applied in combination with track hit requirements provide an electron efficiency of 95\%
~\cite{ATL-PHYS-PUB-2015-041}. Electrons are required to have $\et>$~7~GeV and $|\eta|<$~2.47, with their energy calibrated as described in Ref.~\cite{PERF-2013-05}.

Muon candidate reconstruction~\cite{PERF-2015-10} within $|\eta|<$~2.5 is primarily performed by a global fit of fully reconstructed tracks in the ID and the MS.
In the central detector region ($|\eta|<$~0.1),
which has a limited MS geometrical coverage, muons are also identified by matching a fully reconstructed ID track to either an MS track segment (segment-tagged muons) or a calorimetric energy deposit consistent with a minimum-ionizing particle (calorimeter-tagged muons). For these two  cases, the muon momentum is determined by  the ID track alone. In the forward MS region (2.5~$<|\eta|<~$~2.7) outside the ID coverage,  MS tracks with hits in 
the three MS layers are accepted and combined with forward ID tracklets, if they exist (stand-alone muons).  Calorimeter-tagged muons are required to have $\pt>$~15~GeV. For all other muon candidates, the minimum transverse momentum is 5~GeV instead of the 6~GeV threshold in the\linebreak \runa~publication~\cite{HIGG-2013-21}, increasing the signal acceptance in the four-muon final state by about 7\%. At most one calorimeter-tagged or stand-alone muon is allowed per event.

Jets are reconstructed from noise-suppressed topological clusters~\cite{PERF-2014-07} in the calorimeter using the anti-$k_t$ algorithm with a radius parameter $R$~=~0.4.
The jet four-momentum is corrected for the calorimeter's non-compensating response, signal losses due to noise threshold effects, energy lost in non-instrumented regions, and contributions from pile-up~\cite{ATLAS-CONF-2014-018}. Jets are required to have $\pt>$~30~GeV and $|\eta|<$~4.5. Jets from pile-up are rejected using a jet-vertex-tagger discriminant~\cite{PERF-2014-03} based on the fraction of the jet's tracks that come from the primary vertex. Jets with $|\eta|<$~2.5 containing  $b$-hadrons are identified using the MV2c20 $b$-tagging algorithm~\cite{PERF-2012-04, ATL-PHYS-PUB-2015-022} at an operating point with  70\% $b$-tagging efficiency.

Ambiguities are resolved
if electron, muon or jet candidates are reconstructed from the same detector information.
If a reconstructed electron and muon share the same ID track, the muon is rejected if it is calorimeter-tagged; otherwise the electron is rejected. Reconstructed jets geometrically overlapping in a cone of radius R = 0.2 with electrons or muons are also removed. 

\subsection{Selection of the Higgs boson candidates}
\label{subsec:event_selection}

Events are triggered by a combination of unprescaled single-lepton, dilepton and trilepton triggers with $\pt$ and $\et$ thresholds increasing slightly during the data-taking periods due to an increasing peak luminosity.
The lowest-threshold triggers are complemented by triggers with higher thresholds but looser lepton selection criteria. The global trigger efficiency for signal events passing the final selection is 98\%.

At least two same-flavour and opposite-charge lepton pairs are required in the final state, 
resulting in one or more possible lepton quadruplets in each event.
The three highest-$\pt$ leptons in each quadruplet must have transverse momenta above 20~GeV, 15~GeV and 10~GeV, respectively. The lepton pair with the invariant mass $m_{12}$ ($m_{34}$) closest (second closest) to the $Z$ boson mass in each quadruplet is referred to as the leading (subleading) lepton pair. 
Based on the lepton flavour, each quadruplet is classified into one of the following decay channels: 4$\mu$, 2$e$2$\mu$, 2$\mu$2$e$ and 4$e$, with the first two leptons always representing  the leading lepton pair.
In each subchannel, only the quadruplet containing the leading lepton pair with an invariant mass closest to the $Z$ boson mass is accepted. 

The leading lepton pair must satisfy 50~GeV~$<m_{12}<$~106~GeV. The subleading lepton pair is required to have a mass $m_{\mathrm{min}}<m_{34}<$~115~GeV, where $m_{\mathrm{min}}$ is 12~GeV for the four-lepton invariant mass $m_{4\ell}$ below 140~GeV, rising linearly to 50~GeV at $m_{4\ell}=$~190~GeV and then remaining at 50~GeV for all higher $m_{4\ell}$ values. In the 4$e$ and 4$\mu$ channels, the two alternative opposite-charge lepton pairings within a quadruplet must have a dilepton mass above 5~GeV to suppress the $J/\psi$ background.
The two lepton pairs within the quadruplet must have an angular separation of $\Delta R = \sqrt{(\Delta y)^2 +(\Delta \phi)^2}>$~0.1 (0.2) for same-flavour (different-flavour) lepton pairs.
Each electron (muon) must have a transverse impact parameter significance $|d_0|/\sigma(d_0)$ below 5 (3) to suppress the background from heavy-flavour hadrons. Reducible background from the $Z$+jets and $t\bar{t}$ processes is further suppressed by imposing track-based and calorimeter-based isolation criteria on each lepton. The scalar sum of the $\pt$ of the tracks lying within a cone of 
$\Delta R =$~0.3 (0.2) around the muon (electron) is required to be smaller than 15\% of the lepton $\pt$ ($\et$). Similarly, the sum of the calorimeter $\et$ deposits in a cone of $\Delta R =$~0.2 around the muon (electron) is required to be less than 
30\% (20\%) of the lepton $\et$. The calorimeter-based isolation requirement is applied after correcting for the pile-up and underlying-event contributions as well as removing the energy deposits from the remaining three leptons.  If there is more than one decay channel per event with a quadruplet satisfying the above selection criteria, the quadruplet from the channel with highest efficiency is chosen as the Higgs boson candidate. The signal selection efficiencies in the fiducial region with $|y_H|$<2.5 are 33\%, 25\%,19\% and 17\%, in the 4$\mu$, 2$e$2$\mu$, 2$\mu$2$e$ and 4$e$ channels, respectively.

In case of \STXSVHLep or \STXSttH production, there may be additional leptons present in the event,
together with the selected quadruplet.  There is therefore a possibility that some of the quadruplet
leptons do not originate from a Higgs boson decay, but rather from the $V$ boson or top quark
decays. To improve the lepton pairing in such cases, a matrix-element-based pairing method is used
for all events containing at least one additional lepton with $\pt>$12~GeV and which satisfies the same 
identification and isolation criteria as the four quadruplet leptons. For all possible quadruplet
combinations which pass the above selection, a matrix element for the Higgs boson decay is computed
at LO using the \progname{MadGraph5\_aMC@NLO}~\cite{Alwall:2014hca} generator. The quadruplet with the largest
matrix element value is selected as the final Higgs boson candidate.

In order to improve the four-lepton mass reconstruction, the reconstructed final-state radiation (FSR) photons in $Z$ boson decays are accounted for using the same strategy as in the \runa~data analysis~\cite{HIGG-2013-02,HIGG-2013-21}. After the FSR correction, the lepton four-momenta of the leading lepton pair are recomputed by means of a $Z$-mass-constrained kinematic fit. The fit uses a Breit--Wigner $Z$ line shape, and a single Gaussian function per lepton to model the momentum response function  for the expected resolution of each lepton. The $Z$ boson mass constraint improves the resolution of the four-lepton invariant mass $m_{4\ell}$ by about 15\%. The expected mass resolution for the Higgs boson with a mass $m_H = $~125.09~GeV is 1.6~GeV, 1.7~GeV, 2.1~GeV and 2.4~GeV in the 4$\mu$, 2$e$2$\mu$, 2$\mu$2$e$ and 4$e$ channels, respectively. Finally, to compensate for an increased  reducible background due to lowering the muon $\pt$ threshold to 5~GeV, the four quadruplet leptons are required to originate from a common vertex point. A requirement corresponding to a signal efficiency of 99.5\% is imposed in all decay channels on the $\chi^2$ value from the fit of the four lepton tracks to their common vertex.

The Higgs boson candidates within a mass window of 118~GeV~$<m_{4\ell}<$~129~GeV are selected to study the properties of the Higgs boson.

\subsection{Categorization of reconstructed Higgs boson event candidates}
\label{subsec:event_categorization}

In order to gain sensitivity to different Higgs boson production modes, reconstructed events are classified into several exclusive categories based on the presence of jets and additional leptons in the final state as outlined in 
Figure~\ref{fig:stxs_bins}. 
The classification of events is performed in the following order. First, events are classified as enriched in the \ttH  process (\CatttH) by requiring at least one \textit{b}-tagged jet in the event. In addition, there must be at least four additional jets or one additional lepton with $\pt>$~12~GeV together with at least two jets. The additional lepton is required to satisfy the same isolation, impact parameter and angular separation requirements as the leptons in the quadruplet. Events with additional leptons but not satisfying the above jet requirements compose the next category enriched in \VH production with leptonic vector boson decays (\CatVHLep). 

The remaining events are classified according to their jet multiplicity into events with no jets, exactly one jet or at least two jets. Among events with at least two jets there are significant contributions from the \VBF and \VH production modes in addition to \ggF. These events are divided into two categories according to the invariant mass $m_{jj}$ of the two leading jets. The requirement of $m_{jj}\le$~120~GeV enhances the \VH production mode with hadronically decaying vector bosons (\CatVHHad). For $m_{jj}>$~120~GeV, the \VBF Higgs boson signal is enhanced, and these events are further classified according to the transverse momentum of the leading jet into events with $\pt^{j1}$ below (\CatVBFLow) and above 200~GeV \linebreak (\CatVBFHigh). Events with zero or one jet in the final state are expected to be dominated by the \ggF process. Following the particle-level definition of production bins from Section~\ref{subsec:strategy_SM}, the 1-jet~category is further split into three categories with the four-lepton transverse momentum $\pt^{4\ell}$ smaller than 60~GeV (\CatOneJL), between 60 and 120~GeV (\CatOneJM), and larger than 120~GeV (\CatOneJH). The largest number of \ggF events and the highest \ggF purity are expected in the zero-jet category (\CatZeroJ).

For the tensor structure measurement, the BSM interactions are expected to be more prominent at higher Higgs boson and jet transverse momenta. 
Thus, in addition to the splitting of events with a \VBF-like topology according to $\pt^{j1}$, the \CatVHHad category is further divided into two categories with four-lepton transverse momentum $\pt^{4\ell}$ below and above 150~GeV: \CatVHHadLow and \CatVHHadHigh, respectively.

\begin{figure}
\begin{center}	\includegraphics[width=0.99\columnwidth]{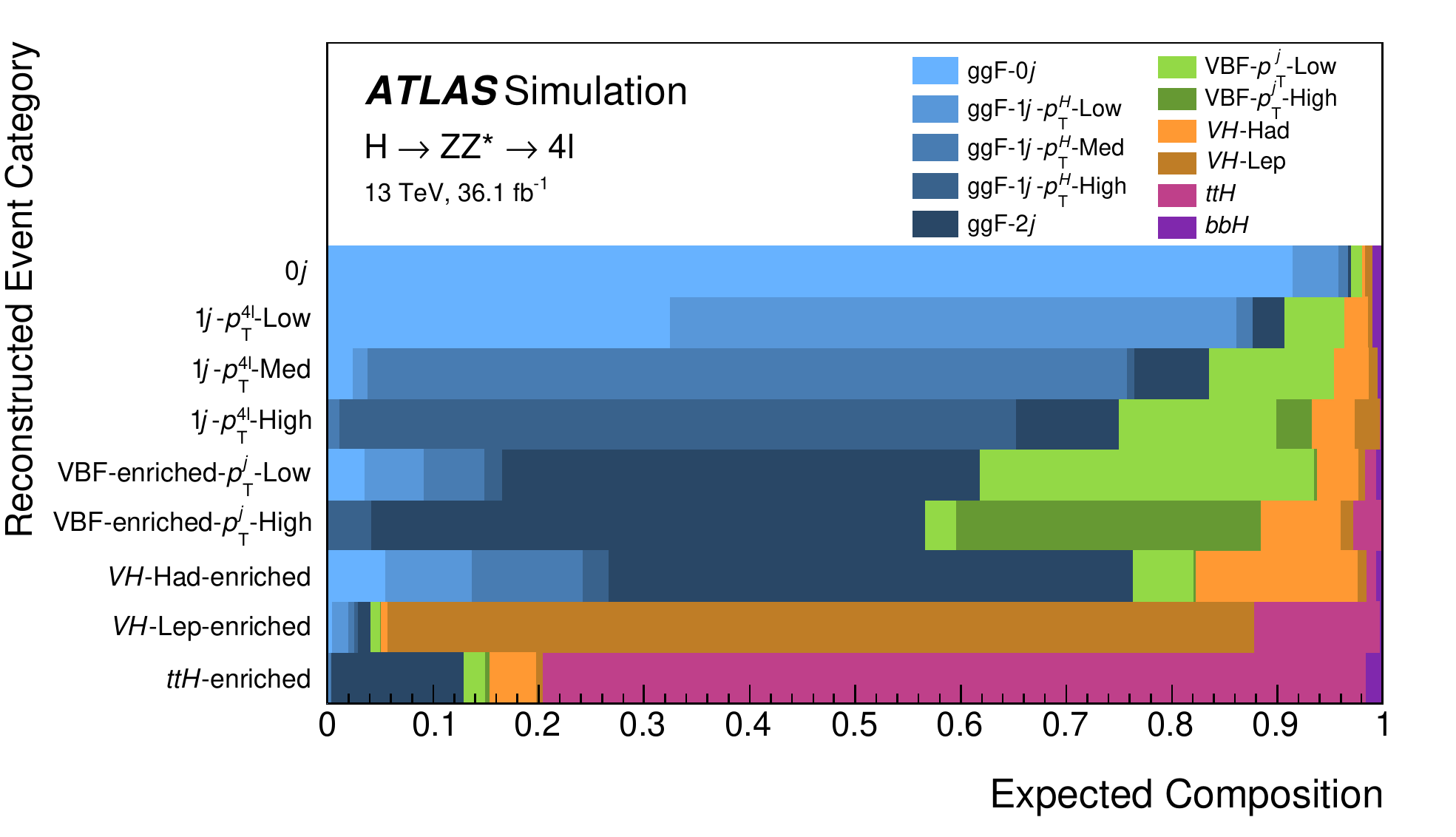}
\caption{Signal composition in terms of the reduced Stage-1 production bins in each reconstructed event category. The \ggF and \bbH contributions are shown separately but both contribute to the same (\ggF) production bin.}
\label{fig:sig_compositionSTXS}
\end{center}
\end{figure}

The expected number of signal events is shown in Table~\ref{tab:event_yields_signal} for each Stage-0 production bin and separately for each reconstructed event category. The \ggF and \bbH contributions are shown separately in order to compare their relative contributions, but both are included in the same (\ggF) production bin. The highest \bbH event yield is expected in the \CatZeroJ category since the jets tend to be more forward than in the \ttH process, thus escaping the acceptance of the \ttH selection criteria. The included systematic uncertainties are detailed in Section~\ref{sec:systematics}. The signal composition in terms of the reduced Stage-1 production bins is shown in Figure~\ref{fig:sig_compositionSTXS}. The separation of contributions from different production bins, such as the sizeable contribution of the \STXSggFTwoJ component in reconstructed categories with two or more jets, is further improved by means of boosted decision tree observables, as described in the following.
\begin{table*}[htbb!]
  \centering 
  \caption{The expected number of SM Higgs boson events with a mass $m_H=$~125.09~\GeV\ in the mass range $118~<~m_{4\ell}~<~129$~\GeV\  for an integrated luminosity of \Lum and $\sqrt{\mathrm{s}}=$~13~TeV in each reconstructed event category, shown separately for each Stage-0 production bin. The \ggF and \bbH contributions are shown separately but both contribute to the same (\ggF) production bin. Statistical and systematic uncertainties are added in quadrature.}
  \label{tab:event_yields_signal}
   
    \vspace{0.1cm}
  \setlength{\tabcolsep}{1cm}
  {\renewcommand{\arraystretch}{1.1}

    \resizebox{\linewidth}{!}{
  \begin{tabular}{* {1} {@{\hspace{2pt}}l@{\hspace{2pt}}}* {5} {@{\hspace{2pt}}c@{\hspace{2pt}}}}
      \hline\hline
    \noalign{\vspace{0.05cm}}
    Reconstructed & 
    \multicolumn{5}{c}{SM Higgs boson production mode}\\
     event category    & \ggF  & \VBF & \VH & \ttH & \bbH \\ \hline               

      \CatZeroJ  & $  25.9 \pm   2.5~~ $ & $  0.29 \pm  0.09 $ & $ 0.253 \pm 0.025 $ & $ 0.00025 \pm 0.00019 $ & $  0.29 \pm  0.14 $ \\
      \CatOneJL  & $   8.0 \pm   1.1 $ & $ 0.514 \pm 0.034 $ & $ 0.230 \pm 0.018 $ & $ 0.0007 \pm 0.0005 $ & $  0.09 \pm  0.05 $ \\
      \CatOneJM  & $   4.5 \pm   0.7 $ & $  0.64 \pm  0.09 $ & $ 0.227 \pm 0.019 $ & $ 0.0010 \pm 0.0005 $ & $ 0.026 \pm 0.013 $ \\
      \CatOneJH  & $  1.10 \pm  0.24 $ & $  0.27 \pm  0.04 $ & $ 0.095 \pm 0.007 $ & $ 0.00080 \pm 0.00024 $ & $ 0.0036 \pm 0.0018 $ \\
     \CatVBFLow  & $   3.9 \pm   0.8 $ & $  2.03 \pm  0.19 $ & $ 0.285 \pm 0.024 $ & $ 0.065 \pm 0.009 $ & $ 0.045 \pm 0.023 $ \\
    \CatVBFHigh  & $  0.33 \pm  0.09 $ & $ 0.185 \pm 0.024 $ & $ 0.050 \pm 0.004 $ & $ 0.0159 \pm 0.0027 $ & $ 0.00058 \pm 0.00029 $ \\
   \CatVHHadLow  & $   2.3 \pm   0.5 $ & $ 0.169 \pm 0.014 $ & $ 0.418 \pm 0.023 $ & $ 0.022 \pm 0.004 $ & $ 0.025 \pm 0.013 $ \\
  \CatVHHadHigh  & $  0.42 \pm  0.09 $ & $ 0.048 \pm 0.008 $ & $ 0.162 \pm 0.005 $ & $ 0.0090 \pm 0.0015 $ &
  < 0.0001 %
  \\
      \CatVHLep  & $ 0.0129 \pm 0.0018 $ & $ 0.00310 \pm 0.00021 $ & $ 0.263 \pm 0.018 $ & $ 0.038 \pm 0.005 $ & $ 0.0009 \pm 0.0005 $ \\
        \CatttH  & $ 0.050 \pm 0.016 $ & $ 0.010 \pm 0.006 $ & $ 0.0196 \pm 0.0031 $ & $ 0.301 \pm 0.032 $ & $ 0.0064 \pm 0.0035 $ \\
        \hline
          Total  & $    47 \pm     4~~ $ & $  4.16 \pm  0.23 $ & $  2.00 \pm  0.11 $ & $  0.45 \pm  0.05 $ & $  0.49 \pm  0.24 $ \\

                   \hline\hline

        \end{tabular}
}
}

\end{table*}

\subsection{Additional discriminating observables}
\label{subsec:discriminants}

In order to further increase the sensitivity of the cross-section measurements in the production bins (Section~\ref{subsec:strategy_SM}), \BDT discriminants are introduced in reconstructed event categories with a sufficiently high number of events. The \BDTs are trained on simulated samples to distinguish a particular Higgs boson production process from either the background or the other production processes, based on several discriminating observables as summarized in Table~\ref{tab:BDT_discriminants}.
It is assumed for the training that all input distributions are governed by the SM predictions.  
\begin{table}[!htbp]
\caption{The \BDT discriminants and their corresponding input variables used for the measurement of cross sections per production bin. The jets are denoted by ``$j$''. See the text for variable definitions.}
\label{tab:BDT_discriminants}
\begin{center}

  \vspace{0.1cm}
  {\renewcommand{\arraystretch}{1.1}

\begin{tabular}{lll} 
\hline\hline 
Reconstructed event category&  \BDT discriminant& Input variables\\
\hline

\CatZeroJ&    \BDTZeroJ&     $p_{\mathrm{T}}^{4\ell}$, $\eta_{4\ell}$, $D_{ZZ^*}$\\
\CatOneJL&    \BDTOneJL&     $p_{\mathrm{T}}^j$, $\eta_j$, $\Delta R (j, 4\ell)$\\
\CatOneJM&    \BDTOneJM&     $p_{\mathrm{T}}^j$, $\eta_j$, $\Delta R (j, 4\ell)$\\
\CatOneJH&            -&     -\\

\CatVBFLow&  \BDTVBFLow&    $m_{jj}$, $\Delta\eta_{jj}$, $\pt^{j1}$, $\pt^{j2}$, 
			    $\eta_{4\ell}^{*}$, $\Delta R_{jZ}^{\mathrm{min}}$, $(\pt^{4\ell jj})_{ \mathrm{constrained}}$ \\
\CatVBFHigh&          -&     -\\

\CatVHHad&   \BDTVHHad&     $m_{jj}$, $\Delta\eta_{jj}$, $\pt^{j1}$, $\pt^{j2}$, 
			    $\eta_{4\ell}^{*}$, $\Delta R_{jZ}^{\mathrm{min}}$, $\eta_{j1}$\\
			    
\multicolumn{1}{l}{\CatVHLep}&  -&  -\\

\multicolumn{1}{l}{\CatttH}&  -&  -\\

\hline\hline

\end{tabular}
}
\end{center}
\end{table}

A \BDT discriminant in the \CatZeroJ category is built to separate the Higgs boson signal from the non-resonant $ZZ^*$ background, relying on the four-lepton transverse momentum and rapidity as well as on the kinematic discriminant $D_{ZZ^*}$~\cite{HIGG-2013-21}, defined as the difference between the logarithms of the signal and background matrix elements squared. In the two $1$-jet categories with $\pt^{4\ell}$ below 120~GeV, a \BDT discriminant combining information about the jet transverse momentum ($p_{\mathrm{T}}^j$), rapidity ($\eta_j$) and angular separation between the jet and the four-lepton system ($\Delta R (j, 4\ell)$) is introduced to distinguish between \ggF and \VBF Higgs boson production. In the \CatVBFLow (\CatVHHad) category, the separation of the \VBF (\VH) from \ggF (\ggF and \VBF) production mechanism is achieved by means of the following input variables: $m_{jj}$, pseudorapidity separation ($\Delta\eta_{jj}$) and transverse momenta of the two leading jets ($\pt^{j1}$ and $\pt^{j2}$),  the difference between the pseudorapidity of the four-lepton system and the average pseudorapidity of the two leading jets ($\eta_{4\ell}^{*}$), as well as
the minimum angular separation between the leading lepton pair and the two leading jets ($\Delta R_{jZ}^{\mathrm{min}}$). In addition, the pseudorapidity of the leading jet ($\eta_{j1}$) is used as an input in the \CatVHHad category, while the constrained transverse momentum of the Higgs--dijet system, defined as $(\pt^{4\ell jj})_{ \mathrm{constrained}} = \pt^{4\ell jj}$~(50~GeV) for $\pt^{4\ell jj} >$~50~GeV ($\pt^{4\ell jj} <$~50~GeV) is employed for the \VBF-enriched category. The transverse momentum $\pt^{4\ell jj}$ of the Higgs--dijet system below 50~GeV is replaced by the minimum value of $\pt^{4\ell jj} =$~50~GeV  in order to reduce the QCD scale variation uncertainty.

The \BDT discriminants improve the expected cross-section measurement statistical uncertainties by 15\%, 35\% and 25\% for the \ggF, \VBF and \VH Stage-0 production bins, respectively.

\section{Signal modelling}
\label{sec:signal_bkg}

The observables used for the measurements of the cross sections in the production bins introduced in Section~\ref{sec:strategy} are \BDT
discriminants for five of the selected reconstructed event categories, described in Section~\ref{subsec:discriminants}, together with event yields for the remaining four event categories. For the SM Higgs boson signal, the shapes of the BDT distributions and the fractions of events in each category are predicted using simulation. 

No BDT discriminants are used for the measurement of the tensor structure of the Higgs boson couplings. This measurement is based on event yields in the ten event categories introduced in Section~\ref{sec:selection}.
A dedicated signal model is introduced to describe the impact of BSM contributions. The model is based on a morphing technique~\cite{ATL-PHYS-PUB-2015-047} which
provides a parameterization to evaluate the signal response as a function of the BSM coupling parameters.
The expected number of signal events $n_{\mathrm{S}}(\vec{\kappa}_{\mathrm{target}})$ at a given target point in the BSM parameter space, defined by a set of BSM coupling values $\vec{\kappa}_\mathrm{target}\equiv\{\kappa_\mathrm{SM},\kappa_{\mathrm{BSM}\_1},..,\kappa_{\mathrm{BSM}\_n}\}$, is obtained by
\begin{equation*}
n_{\mathrm{S}}(\vec{\kappa}_\mathrm{target}) = \sum_{i}{w_{i}(\vec{\kappa}_\mathrm{target},\vec{\kappa}_{i}) \cdot n_{\mathrm{S}}{(\vec{\kappa}_{i}})} .
\end{equation*}
This corresponds to a linear weighted ($w_{i}(\vec{\kappa}_\mathrm{target},\vec{\kappa}_{i})$) combination of a minimal set of base inputs $n_{\mathrm{S}}(\vec{\kappa}_i)$, with coupling values $\vec{\kappa}_{i} = \{\kappa^i_{\mathrm{SM}}, \kappa^i_{\mathrm{BSM}\_1}, .., \kappa^i_{\mathrm{BSM}\_n}\}$
for each input $i$ selected in such a way as to span  the full coupling parameter space. The functional form of the weight $w_i$ and the value assigned to each input is defined by the coupling structure of the BSM signal matrix element as described in Ref.~\cite{ATL-PHYS-PUB-2015-047}.
The inputs for the \ggF, \VBF and \VH production processes are obtained from the simulation samples described in Section~\ref{sec:samples}. The values $\kappa_i$ for each input sample are chosen to cover most parts of the interesting BSM parameter space and to therefore ensure a reasonably small statistical uncertainty for any target point in the BSM parameter space within the range of coupling values under study. These samples are then used to predict the expected variations of event yields in each reconstructed event category relative to the SM prediction. The limited number of events in the simulated BSM samples is estimated to impact the measurement results by less than 5\%. In combination with all other systematic uncertainties, the impact on the final result is negligible in the couplings range under study. Therefore, this uncertainty is not taken into account in the results presented in Section~\ref{sec:results}. Since the BSM input samples were generated with the SM value $\Gamma_\mathrm{SM}$ for the total decay width of the Higgs boson, an additional correction corresponding to 
ratio of the total width with BSM to the SM width
is applied to the $\sigma\cdot\mbr(H\rightarrow ZZ)$ value for the samples with non-vanishing BSM coupling parameters. The correction is of the order 
of $-11\%$ for  \kAgg~=~$\pm$~0.8,  $-2\%$ for \kAVV~=~$\pm$~8 and about $+14\%$ ($-17\%$) for  \kHVV~=~$-8$~(+8).

The  \ttH and \bbH BSM processes are not simulated.
Since the Higgs boson coupling to top or bottom quarks in the effective coupling to gluons is included in \kHgg and \kAgg, and there is little sensitivity to \ttH production in the \htollllbrief{} channel, it is assumed that the production vertex of the \ttH and \bbH processes is not affected by the BSM parameters. The impact of the BSM parameters on the Higgs boson decay is accounted for by scaling the corresponding decay branching ratio.
The BSM parameters also affect $\mbr(H\rightarrow Z\gamma)$ and $\mbr(H\rightarrow \gamma\gamma)$ but the impact on the signal model predictions is found to be negligible and is not considered in the analyses.

\section{Background contributions}
\label{sec:background}

The main source of background in the \htollllbrief\ decay channel is non-resonant
\zzstar\ production with the same final state as the signal. This process, as well as a minor contribution from $t\bar{t}V$ and triboson production, is modelled using simulation normalized to the highest-order SM prediction available.
Additional reducible background sources are the $Z+$jets, $t\bar{t}$ and \emph{WZ} processes whose contributions in the signal region (SR) are estimated using dedicated signal-depleted control regions (CRs) in data, separately for events with different flavours of the subleading lepton pair (i.e.~\llmumu~or \llee, where $\ell\ell$ denotes the leading and $\mu\mu$ or $ee$ the subleading lepton pair).   
No requirement is imposed on the four-lepton invariant mass in the control data. The backgrounds are first estimated for the inclusive event selection, i.e.~prior to event categorization, and then divided into separate contributions in each reconstructed event category. 

\subsection{Background estimation for the inclusive selection}
\label{sec:background_inclusive}

The reducible \llmumu~background with at least one jet containing a muon from secondary decays of pions/kaons or heavy-flavour hadrons originates from $Z+$jets, $t\bar{t}$ and \emph{WZ} production. 
The $Z+$jets background comprises a heavy-flavour ($Z$+HF) component containing jets with \textit{b}- or \textit{c}-quark content and a light-flavour ($Z$+LF) component from pion or kaon decays.
These components of the $Z+$jets background and the $t\bar{t}$ contribution are extracted using orthogonal CRs
formed by relaxing the ${\chi}^2$ requirement on the vertex fit, and by inverting or relaxing isolation and/or impact parameter requirements on the subleading muon pair.
In these regions an unbinned maximum-likelihood fit to $m_{12}$ is performed. 
The numbers of $t\bar{t}$, $Z$+HF and $Z$+LF events estimated in these CRs are each extrapolated 
to the SR using a simulation-based transfer factor which depends on the efficiency of the isolation and impact parameter selection criteria. The contribution from \emph{WZ} production is estimated using simulation.

The reducible \llee\ background originating mainly from the $Z$+jets, $t\bar{t}$ and \emph{WZ} production is classified into processes with misidentified 
jets faking an electron ($f$), electrons from photon conversions
($\gamma$) and  electrons from semileptonic decays of heavy quarks ($q$). The contribution of the $q$ component is obtained from simulation, while the $f$ and the $\gamma$ components are obtained from 
the $3\ell+X$ CR containing $2\mu 2e$ and $4e$ final states. 
In this CR, three leptons pass the full analysis selection, while the most  probable candidate for a fake electron, the lowest-$E_{\textrm{T}}$ electron (denoted by $X$) in the subleading electron pair, has only the track hit requirement of the electron identification applied.  In order to suppress the $ZZ^*$ contribution, only electrons with same-sign charge are considered for the subleading electron pair in this CR.  A template fit to the number of track hits ($n_\mathrm{InnerPix}$) in the innermost or next-to-innermost\footnote{A hit in the next-to-innermost pixel layer is used when the electron falls in a region that is either not instrumented with an IBL module or the IBL module is not operating.} 
pixel layer for the associated track is used to separate the  $\gamma$ and $f$ background components. 
The templates for the $\gamma$ and $f$ background contributions
are obtained from  simulated $Z+X$ events with an on-shell $Z$ boson decay candidate accompanied by an electron $X$ selected using the same criteria as in the $3\ell+X$ CR. The simulated $Z+X$ events are also used to obtain the efficiencies needed to extrapolate the $f$ and $\gamma$ background contributions from the CR to the SR, after correcting the simulation to match the data in dedicated control samples of $Z+X$ events.

\subsection{Background estimation per reconstructed event category}
\label{sec:background_per_cat}

The background event yields and \BDT output distributions 
are determined separately for each event category. %
The reducible \llee\  background normalization is obtained by applying the data-driven approach described above for the inclusive sample in each separate category.
The fraction of the reducible \llmumu\ background per category with respect to the inclusive yield
is obtained from simulation, separately for the $Z$+jets and $t\bar{t}$ background.  The
\llmumu\ simulation was checked against data in CRs with relaxed selection criteria and is
found to predict the fraction of reducible background events in each category well within the
statistical uncertainty.  

Since the data-driven background estimates provide the event yields for
the full $m_{4\ell}$ range,
the effect of the $m_{4\ell}$ mass window requirement has to be taken into account. For this purpose, the $m_{4\ell}$ distributions of reducible
backgrounds in each category are smoothed with the kernel density estimation
method~\cite{Cranmer:2000du} and then integrated to obtain the fraction of events within the mass
window. The yields of the backgrounds in each category are shown in Table~\ref{tab:redbkg_yields},
together with the associated systematic uncertainties. 
\begin{table}[!t]
\caption{Estimates of reducible background yields in each reconstructed event category in the signal region for
  \Lum at $\sqrt{\mathrm{s}}=$~13~TeV, together with the associated correlated and uncorrelated systematic
  uncertainties.  The total error in each category is composed of the combined statistical and
  systematic uncertainty of the inclusive background estimate, as well as an additional statistical
  uncertainty in the fraction of the reducible background in each category.  The uncertainty due to
  the inclusive background estimate is considered as correlated (penultimate column), while the
  statistical uncertainty due to the event categorization (last column) is uncorrelated across the reconstructed event categories.}
  \label{tab:redbkg_yields}
  
\begin{center}
  \vspace{0.1cm}
  {\renewcommand{\arraystretch}{1.1}

  \resizebox{\linewidth}{!}{
  \begin{tabular}{l c c c r r}\hline\hline
Reconstructed    &   
\multicolumn{3}{c}{Reducible background}& 
\multicolumn{2}{c}{Uncertainty}\\

event category    &   
$\ell\ell$+$\mu\mu$  &  
$\ell\ell$+$ee$ & 
Total& 
Corr.& Uncorr.\\ 

\hline   
      \CatZeroJ	 & 
      $ 0.96 \pm 0.21 $ & 
      $ 1.25 \pm 0.23 $ & 
      $ 2.21 \pm 0.33 $ & $\pm$13\% & $\pm$7\%\\
      
      \CatOneJL	 & 
      $ 0.21 \pm 0.05 $ & 
      $ 0.30 \pm 0.06 $ & 
      $ 0.52 \pm 0.08 $ &  $\pm$13\% & $\pm$10\%\\
      
      \CatOneJM	&  
      $ 0.19 \pm 0.12 $ & 
      $ 0.16 \pm 0.04 $ & 
      $ 0.35 \pm 0.13 $ &  $\pm$13\% & $\pm$40\%\\
      
      \CatOneJH	 & 
      $ 0.0049 \pm 0.0025 $ & 
      $ 0.036 \pm 0.008 $ & 
      $ 0.041 \pm 0.009 $ &  $\pm$13\% & $\pm$18\%\\
      
\CatVBFLow &	 
$ 0.14 \pm 0.04 $ & 
$ 0.128 \pm 0.025 $ & 
$ 0.27 \pm 0.05 $ &  $\pm$13\% & $\pm$15\%\\

\CatVBFHigh &	 
$ 0.019 \pm 0.010 $ & 
$ 0.018 \pm 0.004 $ & 
$ 0.037 \pm 0.009 $ &  $\pm$13\% & $\pm$28\%\\

      \CatVHHadLow  &
      $ 0.057 \pm 0.015 $ &
      $ 0.067 \pm 0.015 $ &
      $ 0.124 \pm 0.021 $ &  $\pm$13\% & $\pm$14\%\\

      \CatVHHadHigh &
      $ 0.0035 \pm 0.0023 $ &
      $ 0.011 \pm 0.004 $ &
      $ 0.015 \pm 0.004 $ &  $\pm$13\% & $\pm$34\%\\

\CatVHLep &	 
$ 0.003 \pm 0.004 $ & 
$ 0.0005 \pm 0.0008 $ & 
$ 0.0031 \pm 0.0031 $ &  $\pm$13\% & $\pm$100\%\\

\CatttH  &	 
$ 0.009 \pm 0.004 $ & 
$ 0.022 \pm 0.005 $ & 
$ 0.031 \pm 0.007 $ &  $\pm$13\% & $\pm$22\%\\

\hline\hline
\end{tabular}
}}
  \end{center}
  \vspace{-0.4cm}
\end{table}
Three sources of uncertainty are considered. First, the systematic uncertainty of the inclusive background estimate from the determination of the selection efficiencies related to the lepton identification, isolation and impact parameter significance. This uncertainty is evaluated by comparing data with an on-shell $Z$ boson decay candidate accompanied by an electron or a muon to the simulation. Second, the inclusive background estimate has also a relatively small (4\%) statistical uncertainty from the control data. The total uncertainty of the inclusive reducible background estimate from both of these sources is considered as correlated across the experimental categories. Third, there is an additional uncorrelated uncertainty in the fraction of the reducible background in each experimental category due to the statistical precision of the simulated samples.

The shapes of the \BDT discriminant distributions for the reducible background are determined from simulation by combining the simulated $t\bar{t}$ and $Z$+jets distributions according to the relative fractions measured in data.
To increase the statistical precision of the simulated samples, the isolation requirements and $m_{4\ell}$ range are relaxed. 
The mass window requirement is relaxed in the \CatZeroJ category to $115~<~m_{4\ell}~<~130$~\GeV\ and to $110~<~m_{4\ell}~<~200$~\GeV\ for all other categories. Instead of both leptons, at least one lepton in the subleading pair is required to meet the isolation criteria. These looser selection criteria have no impact on the shape of the \BDT distributions. 
The statistical precision of the simulated samples and the uncertainty in the relative fractions of $Z$+jets and $t\bar{t}$ contributions are taken into account as systematic shape variations.

\section{Systematic uncertainties}
\label{sec:systematics}

The systematic uncertainties in this analysis are grouped into experimental and theoretical 
uncertainties. The first category includes uncertainties in the modelling
of lepton and jet reconstruction,
identification efficiencies, energy resolution and scale,  and in the total integrated
luminosity. Uncertainties from the procedure used to derive the data-driven background estimates are also
included in this category. The second category includes uncertainties in the theoretical modelling of the signal and
the background processes.

The uncertainties can affect the signal acceptance, efficiency and discriminant distributions as well as the background
estimates. 
The dominant sources of uncertainty and their effect are described in the following subsections. The impact of these uncertainties on the cross-section measurements in different production bins is summarized in Table~\ref{tab:RankingSummarXS}.

\begin{table*}[!htbp]
  \centering
  \caption{Impact of the dominant systematic uncertainties (in percent) on the measured inclusive 
  and the Stage-0 production mode cross sections \sBRZZ. Signal theory uncertainties include only acceptance effects and no uncertainty in predicted cross sections.}
  \label{tab:RankingSummarXS}
 
    \vspace{0.1cm}
  {\renewcommand{\arraystretch}{1.1}

 \resizebox{\linewidth}{!}{
  \begin{tabular}{c |rrrrr |rrrr}
    \hline\hline
    & 
    \multicolumn{5}{c|}{Experimental uncertainties [\%]}&
    \multicolumn{4}{c}{Theory uncertainties [\%]}\\

    Production & \multicolumn{1}{c}{Lumi} & 
    \multicolumn{1}{c}{$e$, $\mu$,}  &  
    \multicolumn{1}{c}{Jets, flavour}&             
    \multicolumn{1}{c}{Higgs}&   
     \multicolumn{1}{c|}{Reducible}       & 
     \multicolumn{1}{c}{\zzstar}  &      
     \multicolumn{3}{c}{Signal theory}                    \\
     \addlinespace[-0.5ex]

    bin&  
    &       
    \multicolumn{1}{c}{pile-up}       &  
    \multicolumn{1}{c}{tagging} & 
     \multicolumn{1}{c}{mass}            & 
    \multicolumn{1}{c|}{backgr.}                    & 
    \multicolumn{1}{c}{backgr.} &     
     \multicolumn{1}{c}{PDF} & 
     \multicolumn{1}{c}{QCD scale} & 
     \multicolumn{1}{c}{Shower}              \\

    \hline
    \multicolumn{10}{l}{Inclusive cross section}\\
    \hline
      & $ 4.1 $ & 
      $ 3.1 $ & 
      $ 0.7 $ & 
      $ 0.8 $ & 
      $ 0.9 $ & 
      $ 1.9 $ & 
      $ 0.3 $ & 
      $ 0.8 $ & 
      $ 1.2 $\\
       
    \hline
    \multicolumn{10}{l}{Stage-0 production bin cross sections}\\
    \hline
    \STXSggF                 & 
    $ 4.3 $ & 
    $ 3.4 $ & 
    $ 1.1 $ & 
    $ 1.2 $ & 
    $ 1.1 $ & 
    $ 1.8 $ & 
    $ 0.5 $ & 
    $ 1.8 $ & 
    $ 1.4 $\\

    \STXSVBF                  & 
    $ 2.6 $ & 
    $ 2.7 $ & 
    $ 10 $ & 
    $ 1.3 $ & 
    $ 0.9 $ & 
    $ 2.2 $ & 
    $ 1.6 $ & 
    $ 11 $ & 
    $ 5.3 $\\
    
    \STXSVH                 & 
    $ 3.0 $ & 
    $ 2.7 $ & 
    $ 11 $ & 
    $ 1.6 $ & 
    $ 1.7 $ & 
    $ 5.9 $ & 
    $ 2.1 $ & 
    $ 12 $ & 
    $ 3.7 $\\
    
    \STXSttH                  & 
    $ 3.6 $ & 
    $ 2.9 $ & 
    $ 19 $ & 
    $ <0.1 $ & 
    $ 2.4 $ & 
    $ 1.9 $ & 
    $ 3.3 $ & 
    $ 7.9 $ & 
    $ 2.1 $\\
    \hline\hline
  \end{tabular}
  }}
\end{table*}

\subsection{Experimental uncertainties}
\label{subsec:systematics_exp}

The  uncertainty in the combined 2015+2016 integrated luminosity is 3.2\%. It is derived, following a methodology similar to the one described in Ref.~\cite{DAPR-2013-01}, from a preliminary calibration of the luminosity scale using $x$--$y$ beam-separation scans performed in August 2015 and May 2016. 

The uncertainty in the predicted yields due to pile-up modelling  is about 2\%  and is derived by varying the
average number of pile-up events in the simulation to cover  the 
uncertainty in the ratio of the predicted to measured inelastic cross sections~\cite{STDM-2015-05}.

 The electron (muon) reconstruction and identification efficiencies, and the energy (momentum) scale and resolution are
derived from data using large samples of $J/\psi\rightarrow \ell\ell$ and $Z\rightarrow \ell\ell$ decays~\cite{PERF-2015-10,ATLAS-CONF-2014-032,PERF-2013-05}.
Typical uncertainties in the predicted yield due to the identification efficiencies are in the range $0.5$--$1.0\%$ for muons and
$1.0$--$1.3 \%$ for electrons. The uncertainty in the expected yields  coming from  the muon and electron isolation efficiencies are also taken into account, with the typical size being 2\%.
The uncertainties in the electron and muon energy scale and resolution are small and have a
negligible impact on the measurements presented in Section~\ref{sec:results}.

The uncertainties in the jet energy scale and resolution are in the range of 3--7\% and 2--4\%, respectively~\cite{PERF-2011-03, ATL-PHYS-PUB-2015-015}.
Given the analysis categories, the impact of these uncertainties are more relevant for the  \VH, \VBF and \ttH production modes cross-section measurements (10--20\%)
and for all the reduced Stage-1  cross-section measurements, including the \ggF process split into the different \textit{n-jet} exclusive production bins (5--20\%),
while they are negligible for the inclusive and the \ggF (Stage-0) cross-section measurements. 

The uncertainties  associated with the efficiency of the $b$-tagging algorithm, which are derived from  $t\bar{t}$ events, are at the level of a few percent over most of the jet \pt\ range~\cite{PERF-2012-04}.
This uncertainty is only relevant in the \CatttH category, with its expected impact being approximately 5\% in the \ttH cross-section measurement.

The impact of the  precision  of the Higgs boson mass measurement, $m_{H} = 125.09 \pm 0.24$ GeV~\cite{HIGG-2014-14}, on the signal acceptance due to the mass window requirement defining the signal region is negligible.
A small dependency of the \BDTZeroJ shape on $m_{H}$ is observed for the signal (below 2\% in the highest BDT bins)  and is included in the signal model. This uncertainty affects the measurement of \ggF production, as well as the measurements in other production bins with large \ggF contamination.

The uncertainties from the data-driven measurement of reducible background contributions are detailed in Section~\ref{sec:background}. Their impact on the cross-section measurements is also summarized in Table~\ref{tab:RankingSummarXS}.

\subsection{Theoretical uncertainties}
\label{subsec:systematics_theory_couplings}

The theoretical modelling of the signal and background processes is affected by uncertainties from QCD scale variations, modelling of parton showers and multiple-particle interactions, and PDF uncertainties.

The impact of the theory systematic uncertainties on the signal depends on the kind of measurement
that is performed.  For the signal strength measurements and the tensor structure analysis, each
source of theory uncertainty affects both the fiducial acceptance and the predicted SM cross
section.  For the cross-section measurements, only effects on the acceptance need to be considered.

One of the dominant sources of theoretical uncertainty is the prediction of the \ggF
process in the different $n$-jet categories.
The \ggF process is the major background in the 2-jet categories that are used to measure the cross section of the \VBF and \VH production modes.  To estimate the QCD scale variation
and migration effects on the $n$-jet \ggF cross sections, the approach described in Ref.~\cite{YR4}
is used, which exploits the latest predictions for the inclusive jet cross sections and the
exclusive jet bin fractions. In particular, the uncertainty from the choice of the factorization and
renormalization scales, the choice of resummation scales, and the migrations between the 0-jet and 1-jet phase-space bins or between the 1-jet and $\geq 2$-jet bins are considered. 
The impact of QCD scale variations on the Higgs boson \pT\ distribution is taken into account as an 
additional uncertainty. The uncertainty in the Higgs boson \pT\ at higher order originating from the assumption of 
infinite top and bottom quark masses in
the heavy-quark loop is also taken into account by comparing the \pT\ distribution predictions to finite-mass calculations.
An additional uncertainty in the acceptance of the \ggF process in \VBF topologies due to missing higher orders in QCD 
in the calculation
is estimated by variations of the resummation and factorization scales using fixed-order calculations with MCFM~\cite{MCFM}.
For the other production modes, the QCD scale uncertainties are obtained by varying the scale by
factors of two. The configuration with the largest impact is chosen to define the uncertainty in each experimental
category as the relative difference between the prediction in this and the nominal configuration. QCD scale uncertainties are treated as uncorrelated among the different
production modes.

The uncertainties in the acceptances due to the modelling of parton showers and multiple-parton
interactions are estimated with AZNLO tune eigenvector variations and by comparing the acceptance
using the parton showering algorithm from \progname{PYTHIA 8} with
\progname{Herwig7} for the \ggF, \VBF and \VH processes, while \progname{Herwig++} is compared with \progname{PYTHIA 8} for the \ttH process.  The uncertainty
due to each AZNLO tune variation is taken as correlated among the different production modes while
the difference between  the parton showering algorithms is treated as an uncorrelated
uncertainty. Uncertainties due to the modelling of the \ggF production in association with $b$-quarks affect the measurement in the \ttH production bin only negligibly compared to the statistical precision.  They are therefore not taken into account for the final result.

The impact of the PDF uncertainties is estimated with the
eigenvector variations of the \progname{PDF4LHC\_nlo\_30} Hessian PDF set. The modification of the predictions
for each eigenvector variation is added as a separate source of uncertainty in the model.  The same procedure is
applied for the \ggF, \VBF and \VH processes, enabling correlations to be taken into account in the fit model.

The same procedure is used to estimate the impact of the sources of theoretical uncertainty described above on the shape of \BDT discriminants. In addition, for \VBF Higgs production, the changes in the $\Delta\eta_{jj}$ distribution 
as predicted at NNLO compared to NLO in QCD~\cite{Cacciari:2015jma} are considered and shown to have a negligible impact on the \BDT distributions. For \ggF production, a further cross-check is performed by
comparing the \BDTOneJL, \BDTOneJM, \BDTVBFLow and \BDTVHHad
shapes in the corresponding
categories as predicted by\progname{Powheg NNLOPS} and\progname{MadGraph5\_aMC@NLO} (with the FxFx
merging scheme). 
The \BDT shapes from the two generators agree within the statistical uncertainties and, therefore, no additional shape uncertainty is included.

For BSM interactions parameterized via the effective Lagrangian terms, the theoretical uncertainties
in the PDF set and the missing higher-order QCD and EW corrections are generally assumed to
factorize with respect to the new physics. However, it has been shown~\cite{Maltoni:2013sma} that the
$K$-factors corresponding to the NLO to LO cross-section ratio, as well as several kinematic
quantities that affect the categorization of reconstructed events, such as  the jet transverse momenta,
receive higher-order corrections that can differ from those computed for the SM process and depend
on the value of the BSM couplings. Therefore, an uncertainty is assigned to the $K$-factor obtained from the SM samples.
For this purpose, the $K$-factor for a given \VBF and \VH BSM process is evaluated as the ratio of NLO to LO event
yields in simulated BSM samples, separately
for each reconstructed event category.  The uncertainty in the SM $K$-factor is then defined as the relative difference of the $K$-factors computed for the BSM and SM processes. The obtained uncertainties range from 10\% to 40\% depending on
the category and are considered as being correlated across all categories. This is one of the dominant sources of
uncertainty for the tensor structure measurements. No such uncertainty is considered for the \ggF BSM
samples as these are simulated at NLO.

The dominant theoretical uncertainty in the expected \zzstar\ background yield in the signal mass
window is obtained by varying the factorization and renormalization QCD scales by factors of two. 
The configuration with the largest impact is chosen to define the uncertainty in each experimental
category as the relative difference between the prediction in this and the nominal configuration. This uncertainty is about 4\% for the inclusive event yield and is as large as 30\% for
the categories where additional jets are required. The impact of the QCD scale
uncertainty on the \BDT discriminant shapes is approximately 1--2\%.  The PDF
uncertainty on the \zzstar\ event yield in each category and on the \BDT distributions, obtained
using the MC replicas of the\progname{NNPDF3.0} PDF set, was found to be approximately 1--2\%.
The impact of the parton shower modelling uncertainty on the \zzstar\ event yield is estimated to be approximately 1--5\%, with
the largest value reached in the categories where the presence of one or more jets is required.  In
addition, the event yield and \BDT discriminant shapes in each event category are compared to the data
in a sideband around the signal region ($m_{4l}<$~115~\gev\ or 130~\gev$<m_{4l}<$~170~\gev ). Good agreement
between the \progname{Sherpa} predictions and the data is found.

\section{Results}
\label{sec:results}

\begin{figure}[!htbp]
  \centering 
  \includegraphics[width=0.55\linewidth]{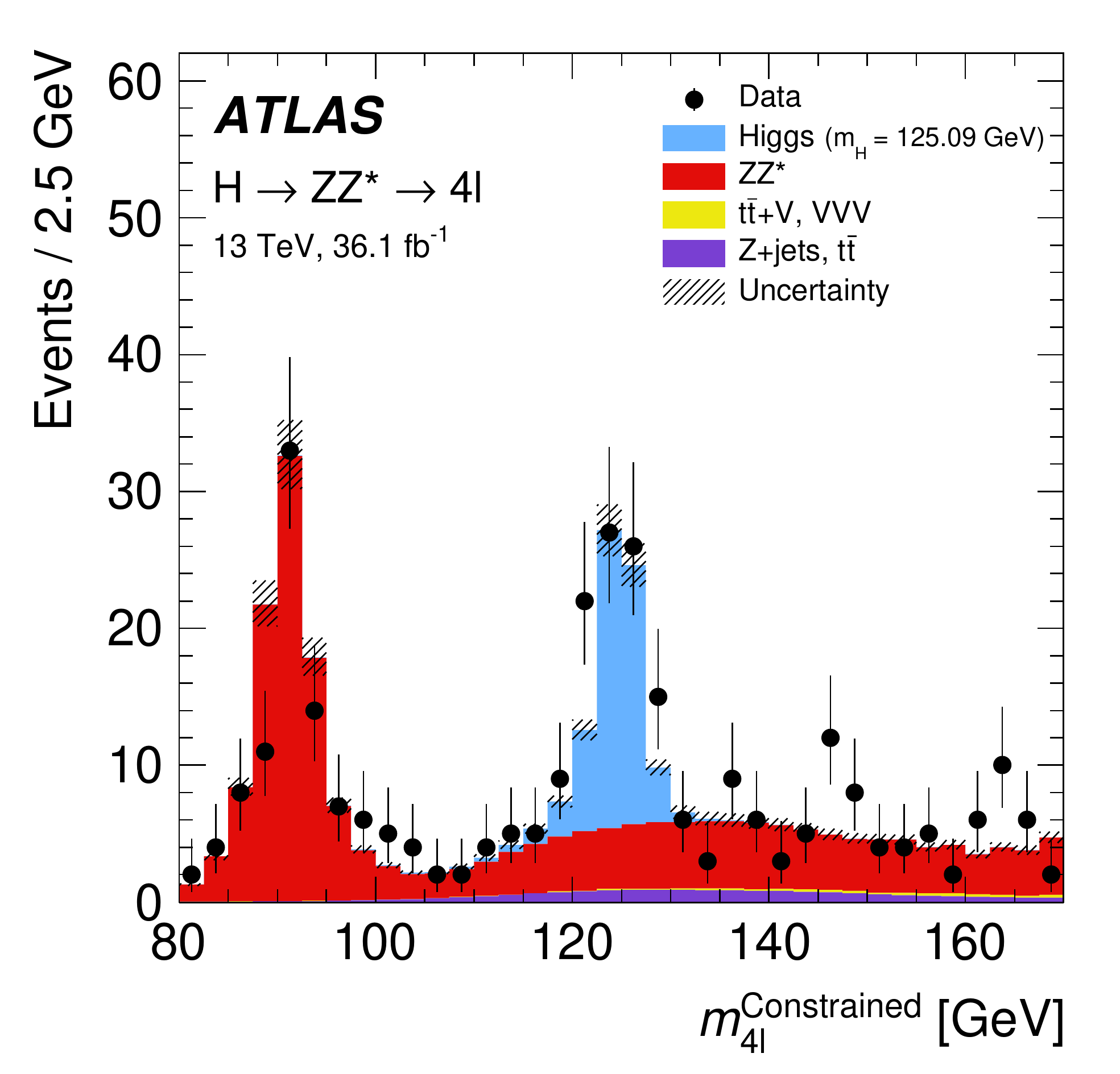}
  \caption{The expected and observed four-lepton invariant mass distribution for the selected Higgs boson candidates with a constrained $Z$ boson mass, shown for an integrated luminosity of \Lum and at $\sqrt{\mathrm{s}}=$~13~TeV assuming the SM Higgs boson signal with a mass $m_{H}$ = 125.09~\gev .
}
    \label{fig:m4l}
\end{figure}

\begin{figure}[!htbp]
\begin{center}
\subfloat[\label{fig:njet}]{\includegraphics[width=0.4\textwidth]{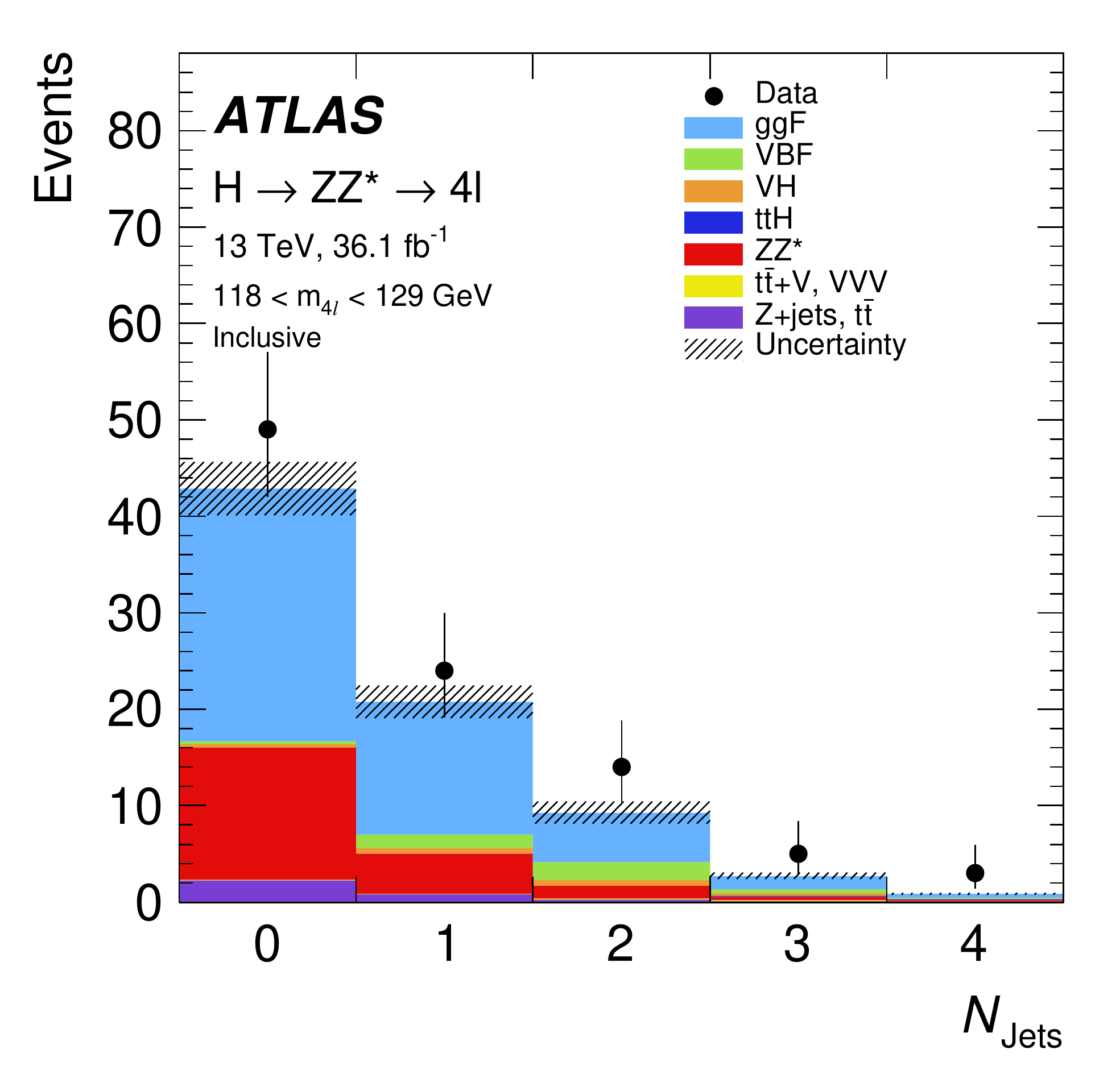}}
\subfloat[\label{fig:pt4l}]{\includegraphics[width=0.4\textwidth]{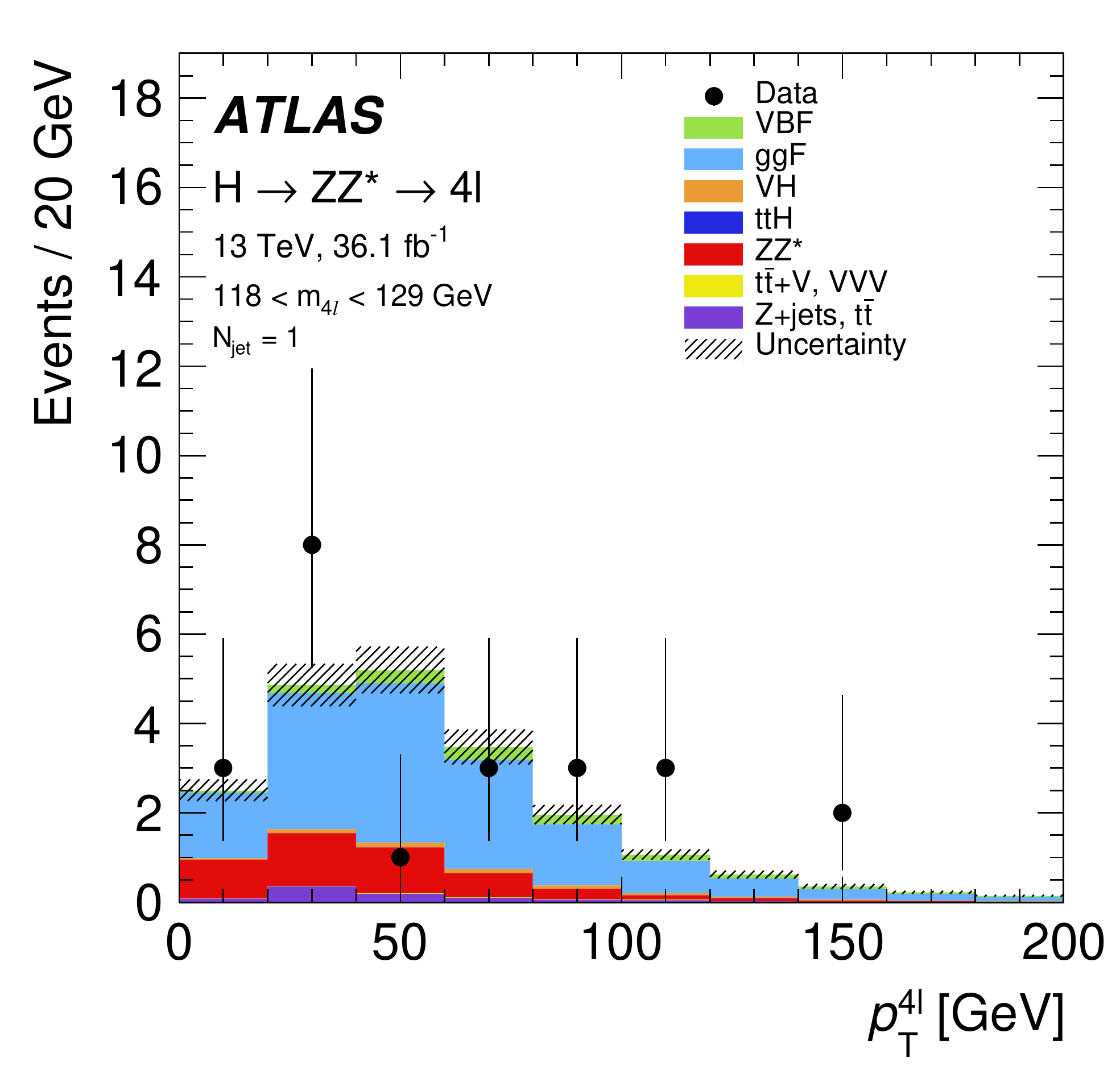}}\\
\subfloat[\label{fig:mjj}]{\includegraphics[width=0.4\textwidth]{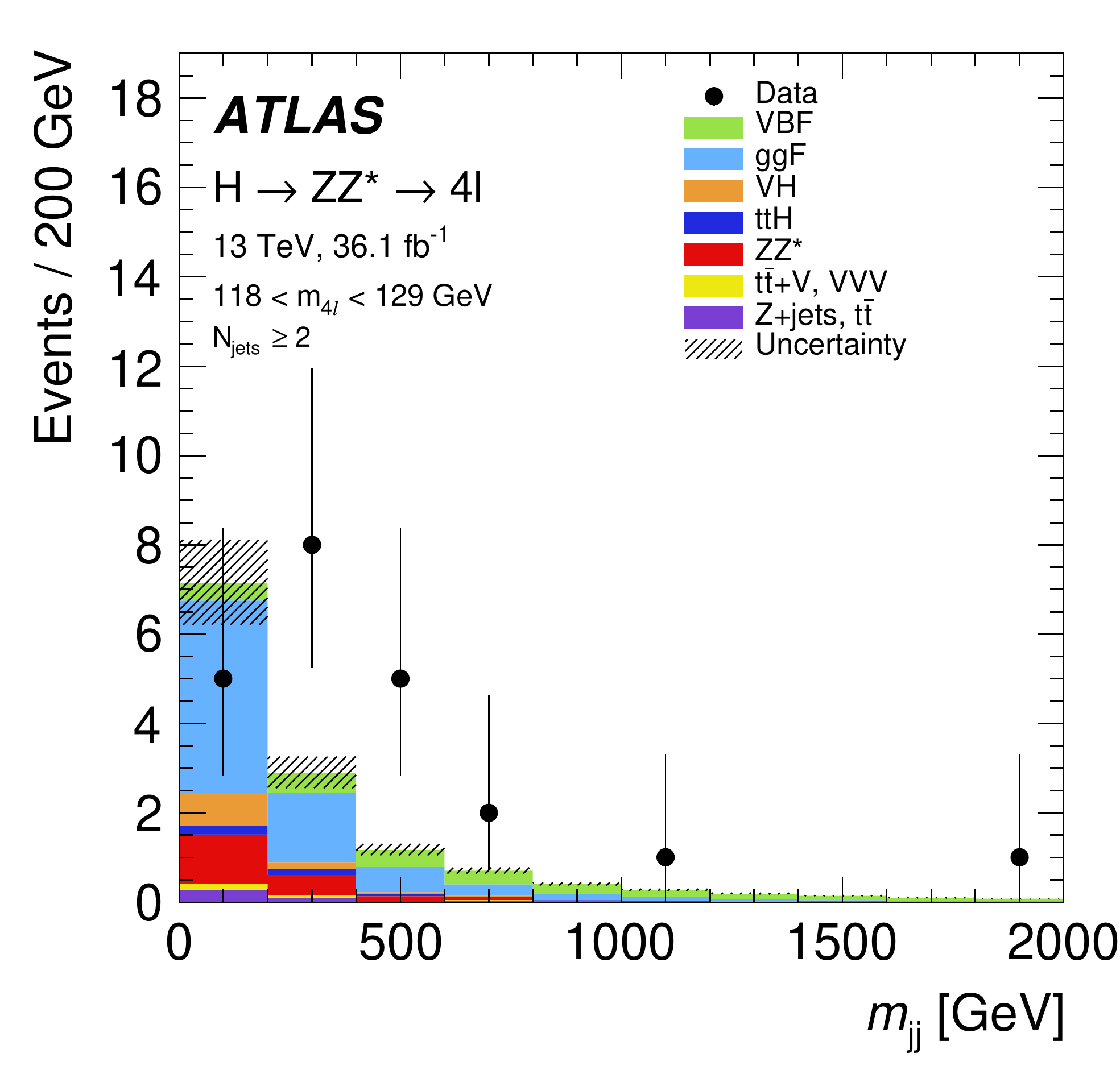}}
\subfloat[\label{fig:ptj1}]{\includegraphics[width=0.4\textwidth]{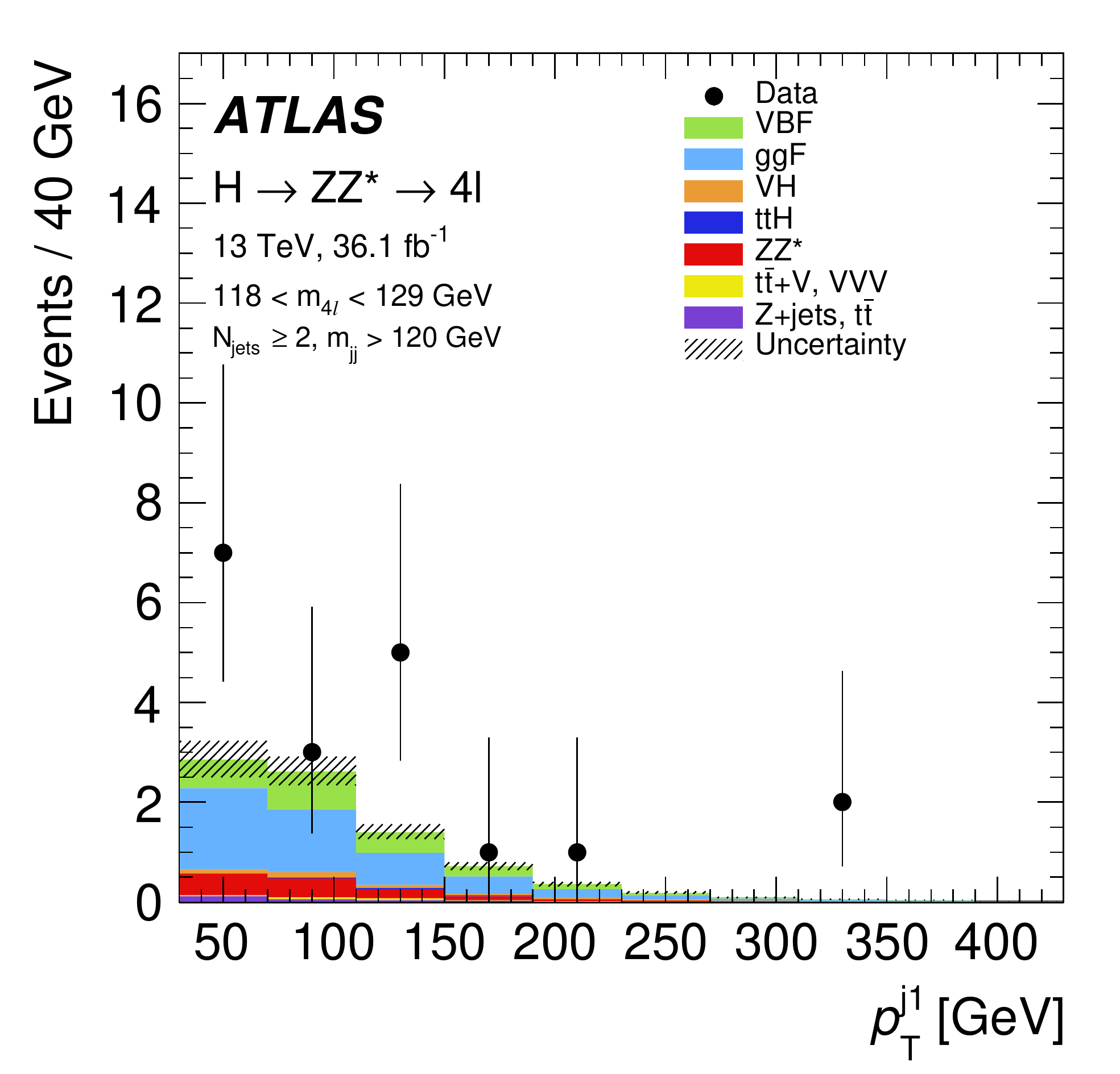}}\\
\subfloat[\label{fig:pt4lVh}]{\includegraphics[width=0.4\textwidth]{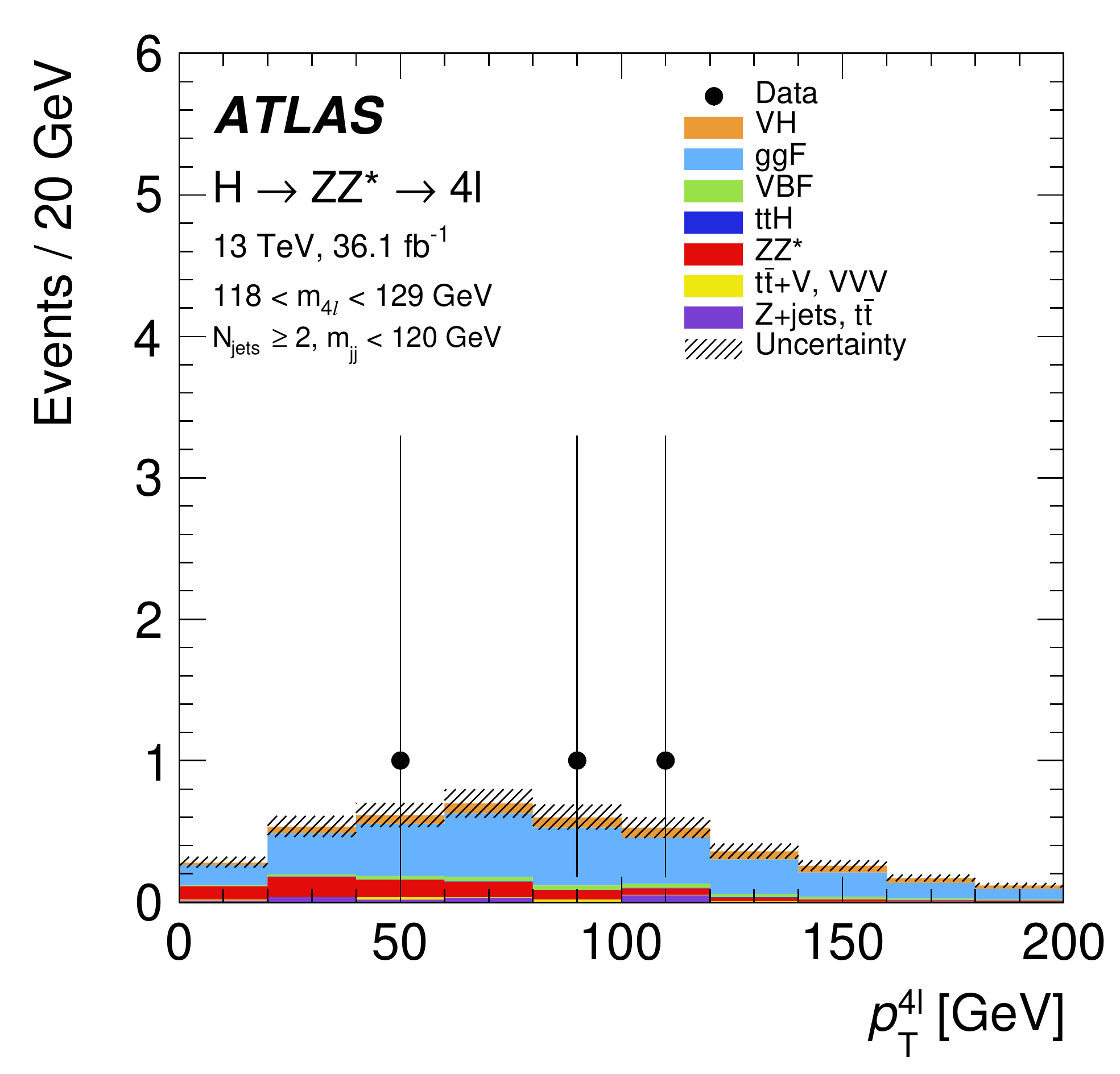}}
\end{center}
\vspace{-0.3cm}
\caption{The observed and expected distributions of (a) $N_{\mathrm{jet}}$ after the inclusive selection, (b) $p_{\mathrm{T}}^{4l}$ in the 1-jet categories, (c) $m_{jj}$ in the 2-jet categories, (d) $p_{\mathrm{T}}^{j1}$ in the \VBF-enriched categories and (e) $p_{\mathrm{T}}^{4l}$ in the \CatVHHad categories for an integrated luminosity of \Lum collected at $\sqrt{\mathrm{s}}=$~13~TeV assuming the SM Higgs boson signal with a mass $m_{H}$ = 125.09~\gev . }
\label{fig:categoryVars}
\end{figure}

\begin{figure}[!p]
\begin{center}
  
\subfloat[\label{fig:oBDTZZ}]{\includegraphics[width=0.4\linewidth]{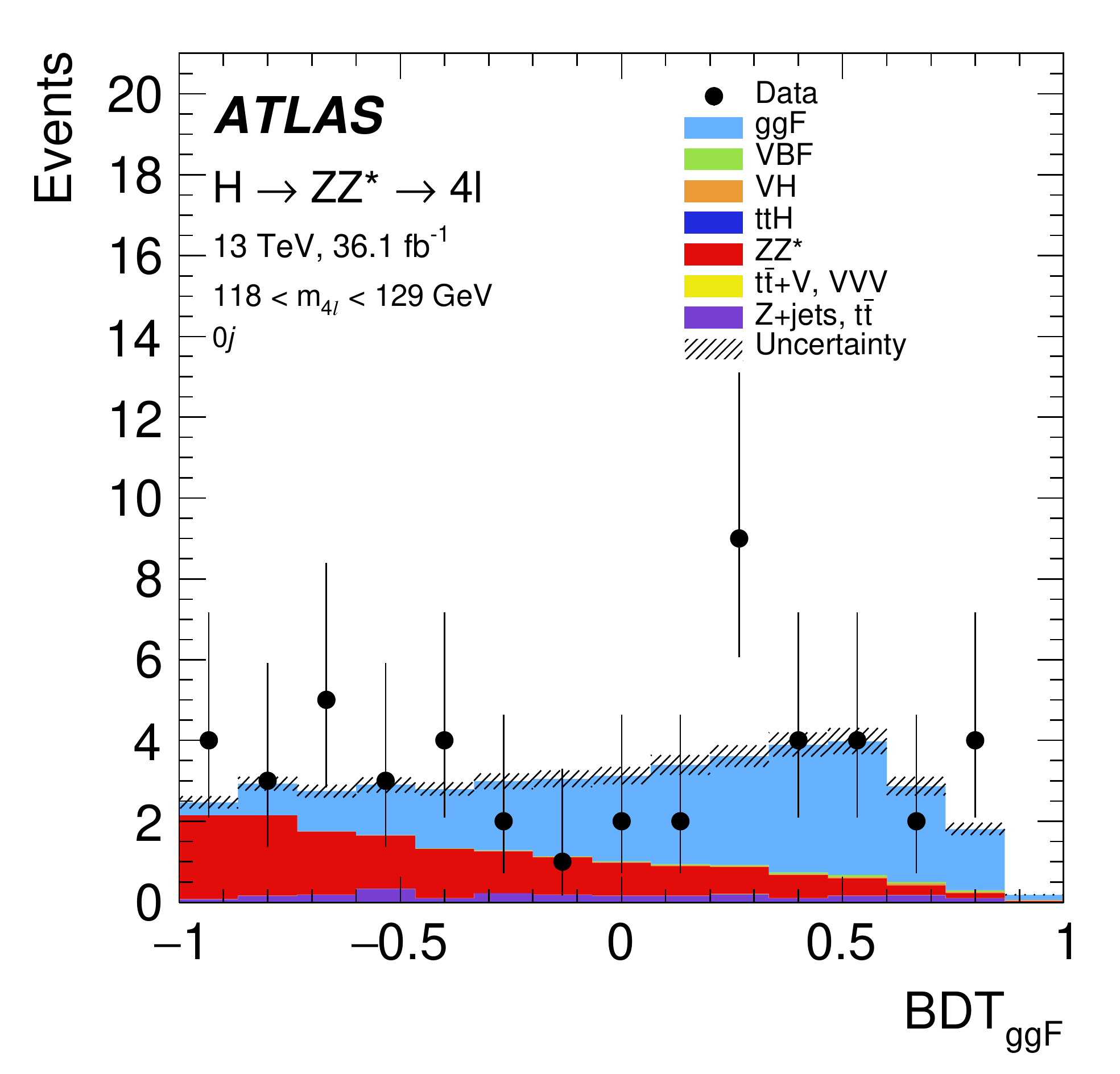}}
\subfloat[\label{fig:oBDT1jpth1}]{\includegraphics[width=0.4\linewidth]{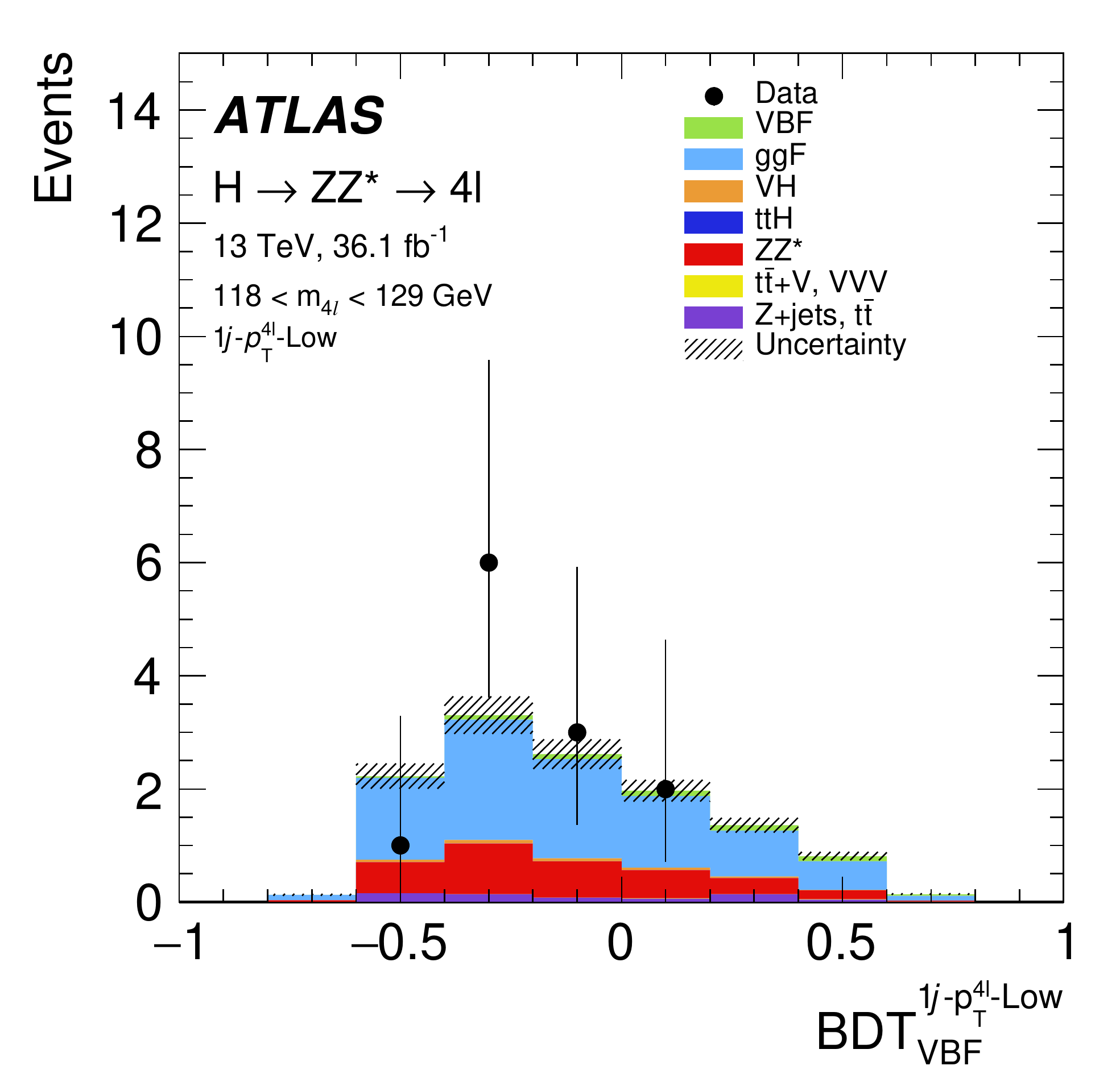}}\\
\subfloat[\label{fig:oBDTj1pth2}]{\includegraphics[width=0.4\linewidth]{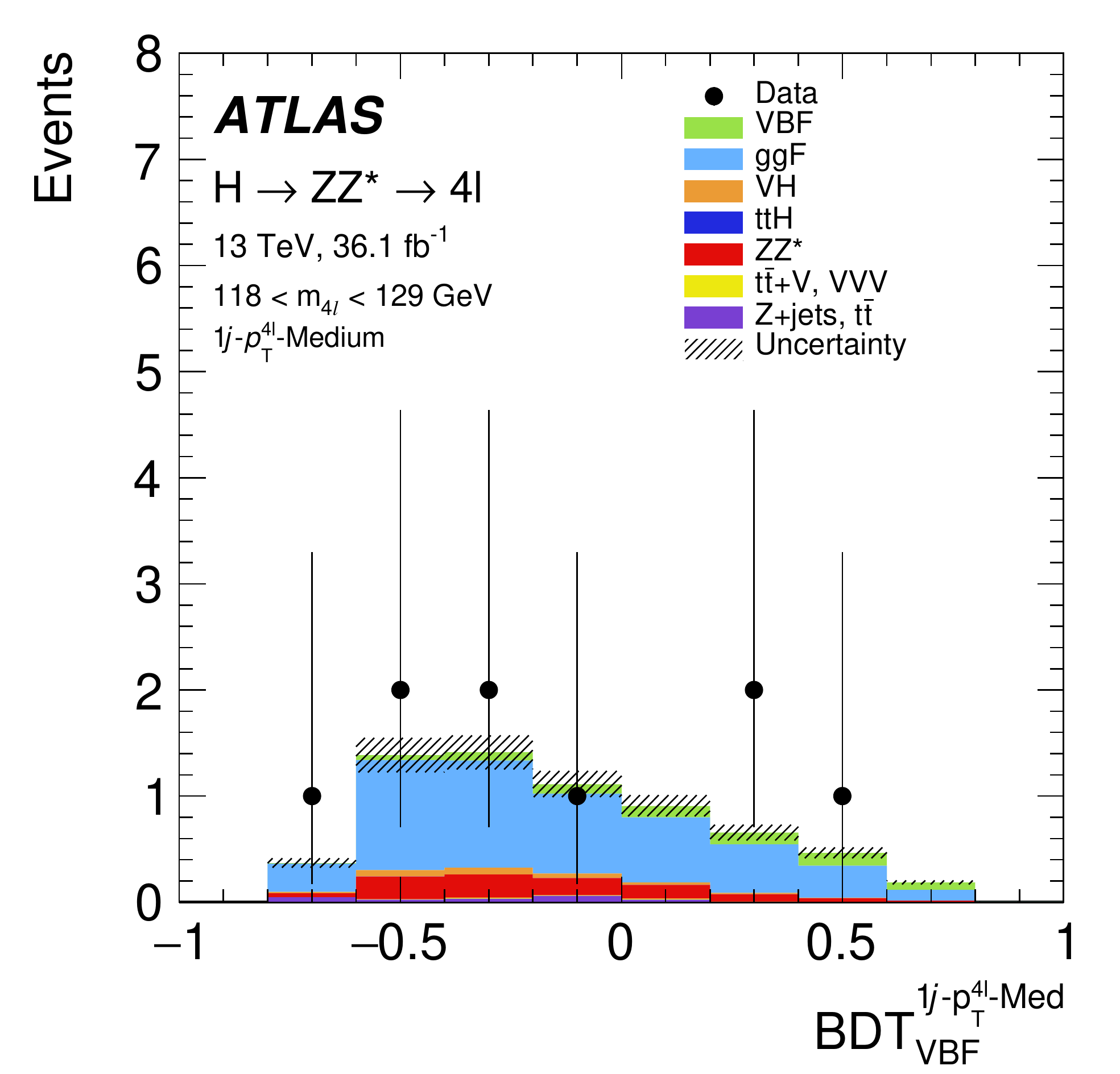}}
\subfloat[\label{fig:oBDTVBF}]{\includegraphics[width=0.4\linewidth]{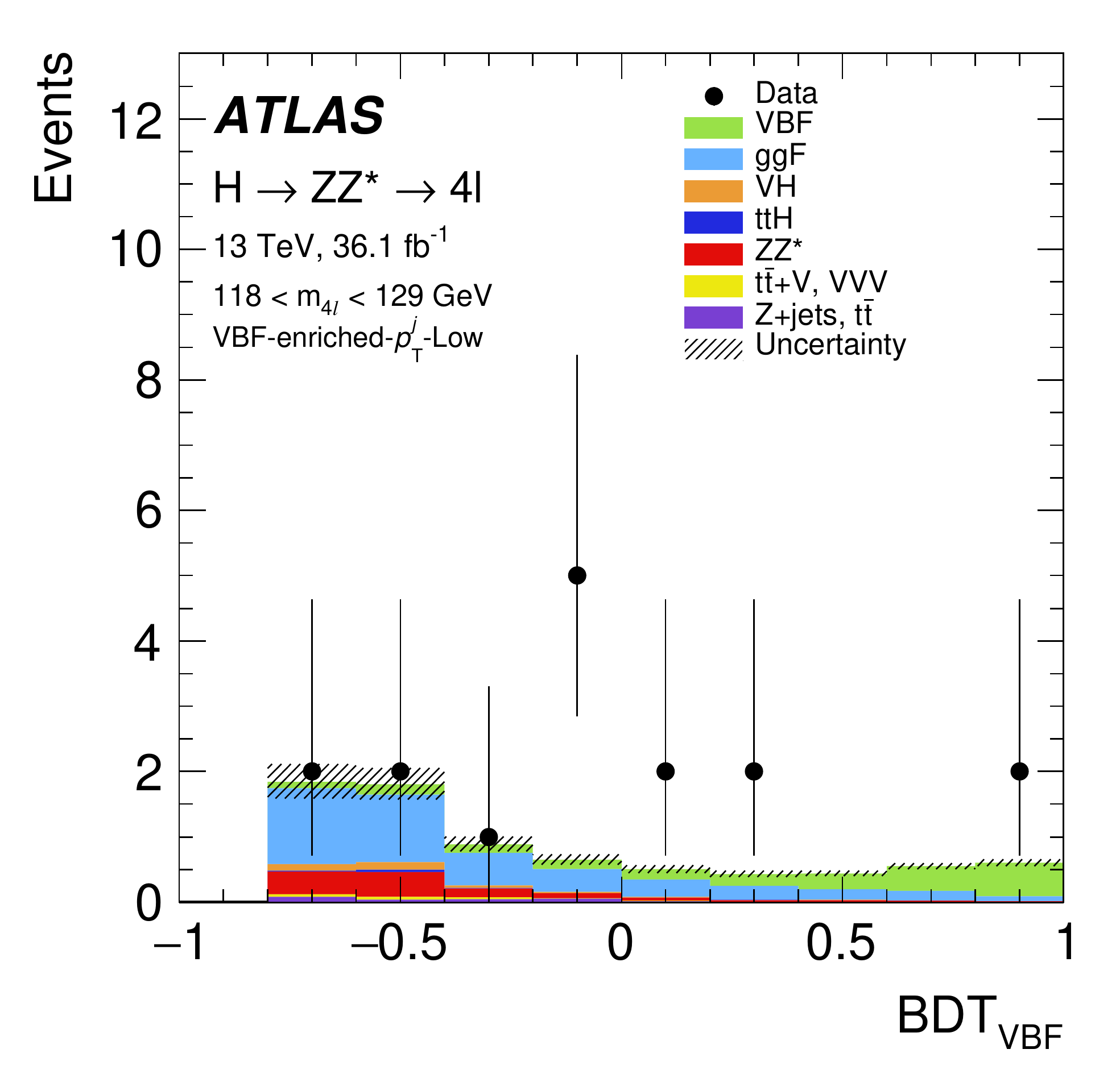}}\\
\subfloat[\label{fig:oBDTVH}]{\includegraphics[width=0.4\linewidth]{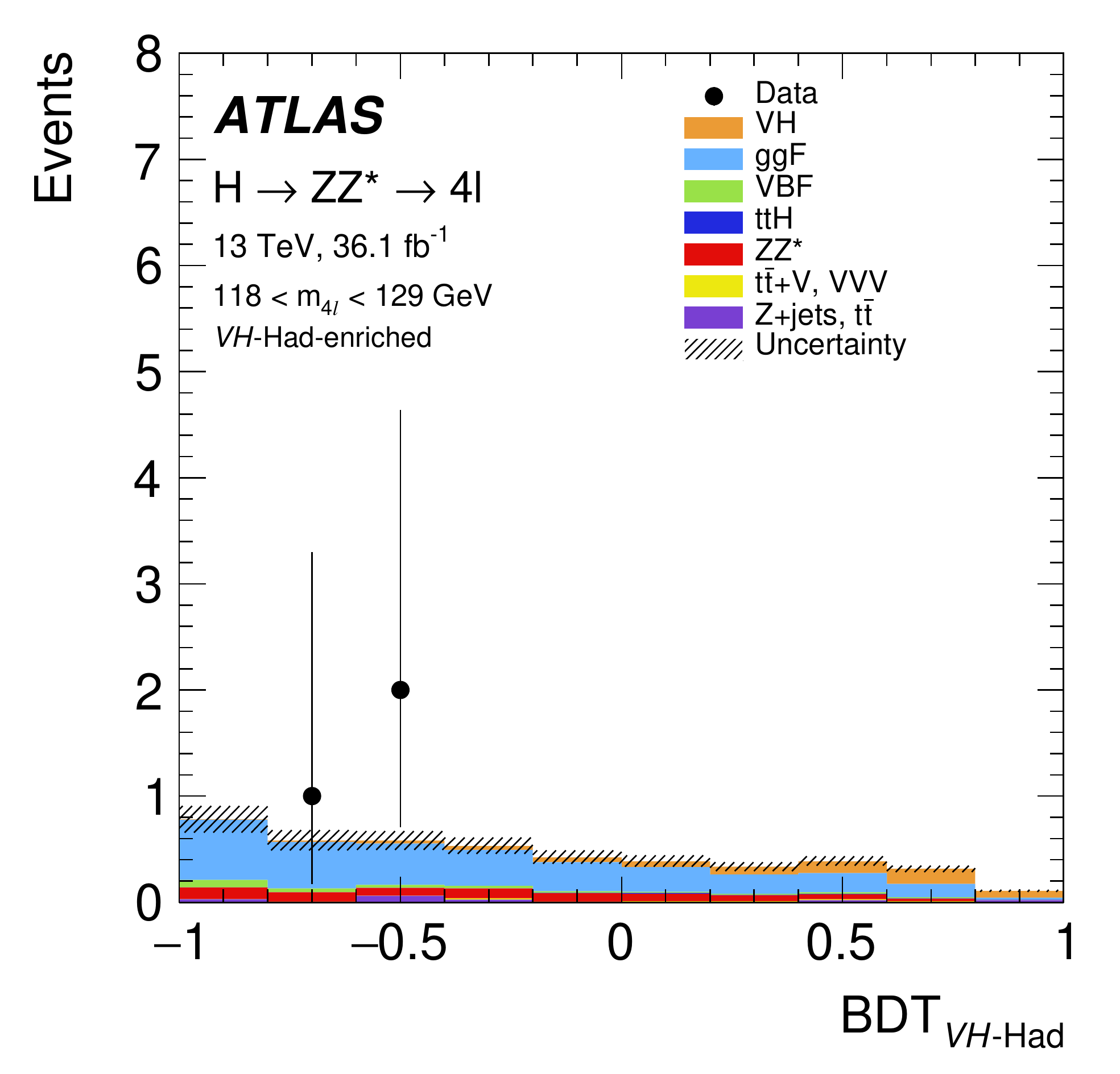}}

\end{center}
\vskip-0.2cm
\caption{The observed and expected \BDT output distributions in the (a) \CatZeroJ, (b) \CatOneJL, (c) \CatOneJM,  (d) \CatVBFLow and (e) \CatVHHad  categories for an integrated luminosity of \Lum and at $\sqrt{\mathrm{s}}=$~13~TeV assuming the SM Higgs boson signal with a mass $m_{H}$ = 125.09~\gev . } 
\label{fig:BDToutputs_observed}
\end{figure}

The expected and observed four-lepton invariant mass distribution of the selected Higgs boson candidates after the event selection with a constrained $Z$ boson mass is shown in Figure~\ref{fig:m4l}.
The corresponding expected and observed numbers of events are shown in Table~\ref{tab:event_yields_inclusive} separately for each of the four decay channels. The predicted event yields are in reasonable agreement with the data.
\begin{table*}[!htbp]
  \centering 
  \caption{The expected and observed numbers of signal and background events in the
    four-lepton decay channels for an integrated luminosity of \Lum and at $\sqrt{\mathrm{s}}=$~13~TeV, assuming the SM Higgs boson signal with a mass $m_{H}$~=~125.09~\gev . The second column shows
    the expected number of signal events for the full mass range while the subsequent columns correspond to the mass range 
    of $118 < m_{4\ell} < 129$~\gev.  In addition to the $ZZ^*$ background,  the contribution of other backgrounds is shown, comprising the data-driven estimate from Table~\ref{tab:redbkg_yields} and the simulation-based estimate of contributions from rare triboson and $t\bar{t}V$ processes. Statistical and systematic uncertainties are added in quadrature.
}
    \label{tab:event_yields_inclusive}
 
        \vspace{0.1cm}
  {\renewcommand{\arraystretch}{1.1}

         \resizebox{\linewidth}{!}{
 \begin{tabular}{c|c|rrr|rr}
   \hline\hline
   Decay & 
   \multicolumn{1}{c|}{Signal}             & 
   \multicolumn{1}{c}{Signal} & 
   \multicolumn{1}{c}{\zzstar} &  
   \multicolumn{1}{c|}{Other} &       
   \multicolumn{1}{c}{Total} & 
   \multicolumn{1}{c}{Observed} \\
   
   \multicolumn{1}{c|}{channel}  & 
   \multicolumn{1}{c|}{(full mass range)} &           
   & 
   \multicolumn{1}{c}{background}&  
   \multicolumn{1}{c|}{backgrounds}&   
   \multicolumn{1}{c}{expected}       &  
   \\
   \hline
   $4\mu$    & $  21.0 \pm   1.7 $ & $  19.7 \pm   1.6 $ & $   7.5 \pm   0.6\phantom{0} $ & $  1.00 \pm  0.21 $ & $  28.1 \pm   1.7 $ & 32 \\
   $2e2\mu$  & $  15.0 \pm   1.2 $ & $  13.5 \pm   1.0 $ & $   5.4 \pm   0.4\phantom{0} $ & $  0.78 \pm  0.17 $ & $  19.7 \pm   1.1 $ & 30 \\
   $2\mu2e$  & $  11.4 \pm   1.1 $ & $  10.4 \pm   1.0 $ & $  3.57 \pm  0.35 $ & $  1.09 \pm  0.19 $ & $  15.1 \pm   1.0 $ & 18 \\
   $4e$      & $  11.3 \pm   1.1 $ & $   \phantom{0}9.9 \pm   1.0 $ & $  3.35 \pm  0.32 $ & $  1.01 \pm  0.17 $ & $  14.3 \pm   1.0 $ & 15 \\
   \hline
   Total     & $    59 \pm     5 $ & $    54 \pm     4\phantom{.0} $ & $  19.7 \pm   1.5\phantom{0} $ & $   3.9 \pm   0.5\phantom{0} $ & $    77 \pm     4\phantom{.0} $ & 95 \\
   \hline\hline
 \end{tabular}
 }}
\end{table*}
The observed and expected distributions of the jet multiplicity, the dijet invariant mass, as well as the leading jet and the four-lepton transverse momenta, which are used for the categorization of reconstructed events, are shown in Figure~\ref{fig:categoryVars} for different stages of the event categorization. As shown in figures~\ref{fig:mjj} and ~\ref{fig:ptj1} there is an excess of events observed in the sample with $\geq 2$ jets (shown as a dijet invariant mass distribution) and also in the subset with $m_{jj} >$~120~GeV (shown as the jet \pt distribution) in comparison with the expectations. All other distributions are in good agreement with the data. 
The expected numbers of signal and background events in each reconstructed event category  (including the splitting of the \VH-enriched category for the tensor structure measurement) are shown in Table~\ref{tab:event_yields_all} together with the corresponding observed number of events.
\begin{table*}[!htbp]
    \centering
    \caption{The expected and observed numbers of signal and background events in the mass range \mbox{$118<m_{4\ell}<129$~\GeV\ } for an integrated luminosity of \Lum and at $\sqrt{\mathrm{s}}=$~13~TeV in each reconstructed event category (including the splitting of the \VH-enriched category for the tensor structure measurement), assuming the SM Higgs boson signal with a mass $m_{H}$ = 125.09~\gev . In addition to the $ZZ^*$ background,  the contribution of other backgrounds is shown, comprising the data-driven estimate from Table~\ref{tab:redbkg_yields} and the simulation-based estimate of contributions from rare triboson and $t\bar{t}V$ processes. Statistical and systematic uncertainties are added in quadrature.}
    \label{tab:event_yields_all}
    \vspace{0.1cm}
    
        \vspace{0.1cm}
  {\renewcommand{\arraystretch}{1.1}

     \resizebox{\linewidth}{!}{
        \begin{tabular}{l rrrrrrrrr}
        \hline\hline
           Reconstructed    & 
           \multicolumn{1}{c}{Signal}   & 
	    \multicolumn{1}{c}{\zzstar}          &
	   \multicolumn{1}{c}{Other}& 
	 \multicolumn{1}{c}{Total} &      
	   \multicolumn{1}{c}{Observed}\\

	   event category       &  
	           & 
	    \multicolumn{1}{c}{background}&
	    \multicolumn{1}{c}{backgrounds}& 
	     \multicolumn{1}{c}{expected}& 
	    \\

            \hline
	    
      \CatZeroJ  & $  26.8 \pm   2.5\phantom{00} $ & 
      $  13.7 \pm   1.0\phantom{00} $ & 
      $  2.23 \pm  0.31\phantom{00} $ & 
      $  42.7 \pm   2.7\phantom{00} $ & 
      49 \\
      
      \CatOneJL  & $   8.8 \pm   1.1\phantom{00} $ & 
      $   3.1 \pm   0.4\phantom{00} $ & 
      $  0.53 \pm  0.07\phantom{00} $ & 
      $  12.5 \pm   1.2\phantom{00} $ & 
      12 \\
      
      \CatOneJM  & $   5.4 \pm   0.7\phantom{00} $ & 
      $  0.88 \pm  0.12\phantom{0} $ & 
      $  0.38 \pm  0.05\phantom{00} $ & 
      $   6.7 \pm   0.7\phantom{00} $ & 
      9 \\
      
      \CatOneJH  & $  1.47 \pm  0.24\phantom{0} $ & 
      $ 0.139 \pm 0.022 $ & 
      $ 0.045 \pm 0.007\phantom{0} $ & 
      $  1.65 \pm  0.24\phantom{0} $ & 
      3 \\
     
     \CatVBFLow  & $   6.3 \pm   0.8\phantom{00} $ & 
     $  1.08 \pm  0.32\phantom{0} $ & 
     $  0.40 \pm  0.04\phantom{00} $ & 
     $   7.7 \pm   0.9\phantom{00} $ & 
     16 \\
    
    \CatVBFHigh  & $  0.58 \pm  0.10\phantom{0} $ & 
    $ 0.093 \pm 0.032 $ & 
    $ 0.054 \pm 0.006\phantom{0} $ & 
    $  0.72 \pm  0.10\phantom{0} $ & 3 \\
   
   \CatVHHadLow  & $   2.9 \pm   0.5\phantom{00} $ & 
   $  0.63 \pm  0.16\phantom{0} $ &
    $ 0.169 \pm 0.021\phantom{0} $ & 
    $   3.7 \pm   0.5\phantom{00} $ & 
    3 \\
  
  \CatVHHadHigh  & $  0.64 \pm  0.09\phantom{0} $ & 
  $ 0.029 \pm 0.008 $ & 
  $ 0.0182 \pm 0.0022 $ & 
  $  0.69 \pm  0.09\phantom{0} $ &
   0 \\
  
      \CatVHLep  & $ 0.318 \pm 0.019 $ & 
      $ 0.049 \pm 0.008 $ & 
      $ 0.0137 \pm 0.0019 $ & 
      $ 0.380 \pm 0.020 $ & 
      0 \\
      
        \CatttH  & $  0.39 \pm  0.04\phantom{0} $ & 
	$ 0.014 \pm 0.006 $ & 
	$  0.07 \pm  0.04\phantom{00} $ & 
	$  0.47 \pm  0.05\phantom{0} $ &
	 0 \\
	
	\hline
          Total  & $    54 \pm     4\phantom{.000} $ & 
	  $  19.7 \pm   1.5\phantom{00} $ & 
	  $   3.9 \pm   0.5\phantom{000} $ & 
	  $    77 \pm     4\phantom{.000} $ 
	  & 95 \\
          
	                        \hline\hline
            \noalign{\vspace{0.05cm}}
            \label{tab:yields_obs_sum} 
        \end{tabular}
	}}
    \end{table*}
The expected event yields are in reasonable agreement with the observed ones. The largest differences are again observed in the two \VBF-enriched categories.
The expected and observed distributions of the \BDT discriminants introduced in Section~\ref{subsec:discriminants} are shown in Figure~\ref{fig:BDToutputs_observed}, where a small excess is observed at larger values of the \VBF \BDT. All other distributions are in good agreement with the data.
Based on these results, the measurements of the Higgs boson production cross sections and of its tensor coupling structure are performed. The profile likelihood ratio~\cite{Cowan:2010js} is used for the interpretation of data with the effects of systematic uncertainties included as constrained nuisance parameters. If the same source of uncertainty affects two or more processes (e.g. the error in the integrated luminosity can affect the signal yield and the MC-based background estimates), the same nuisance parameter is assigned to each of these processes.

\subsection{Cross-section measurement by production modes} 
\label{subsec:results_SM}

In order to measure the Higgs boson production cross section times branching ratio for $H \rightarrow ZZ^*$ decay for each Stage-0 or reduced Stage-1 production bin, a fit to the data is performed using the likelihood function ${\cal L}( \vec{\sigma}, \vec{\theta})$ that depends on the Higgs boson production cross section $\vec{\sigma} = \{ \sigma_{1}, \sigma_{2}, \ldots , \sigma_{N} \} $ in each production bin and the nuisance parameters $\vec{\theta}$ accounting for the systematic uncertainties. The likelihood function is defined as a product of conditional probabilities $P$ over binned distributions of the discriminating observables in each reconstructed event category $j$,
\begin{equation*}
\begin{split}
& \mathcal{L}(\vec{\sigma}, \vec{\theta}) =  \displaystyle  \prod_{j}^{N_{\mathrm{categories}}} \prod_{i}^{N_{\mathrm{bins}}} 
 P\left(N_{i,j} ~| ~L\cdot\vec{{\sigma}} \cdot  \vec{A}_{i, j}(\vec{\theta}) ~+ B_{i,j}(\vec{\theta}) \right)  \times \prod_{m}^{N_{\mathrm{nuisance}}} \mathcal{C}_{m}(\vec{\theta}) \;,
\end{split}
\label{eq:likelihood_simplified}
\end{equation*}
with Poisson distributions $P$ corresponding to the observation of $N_{i,j}$ events in each bin $i$ of the\linebreak discriminating  observable
given the expectations for the background, $B_{i,j}(\vec{\theta})$, and for the signal,\linebreak $S_{i,j}(\vec{\theta}) = L \cdot\vec{{\sigma}} \cdot \vec{A}_{i,j}(\vec{\theta})$, where $L$ is the integrated luminosity and $\vec{A}_{i,j}(\vec{\theta})$ the signal acceptance in each production bin. The signal acceptance is defined as the number of simulated signal events satisfying the event selection criteria in a given reconstructed event category divided by the total number of events generated in the phase space defined by the production bin.
Constraints on the nuisance parameters corresponding to systematic uncertainties described in Section~\ref{sec:systematics} are
represented by the functions $\mathcal{C}_{m}(\vec{\theta})$. The cross sections are treated as independent parameters for each production bin and correlated among the different categories.
The test statistic used to compare the probabilities of different hypotheses
is the ratio of profile likelihoods~\cite{Cowan:2010js},
\begin{equation*}
q = -2 \ln \frac{\mathcal{L}(\vec{\sigma}, \hat{\hat{\vec{\theta}}}(\vec{\sigma})) } {\mathcal{L}(\hat{{\vec{\sigma}}}, \hat{{\vec{\theta}}} )} = -2 \ln \lambda(\vec{\sigma})\;,
\label{eq:lambda_simplified}
\end{equation*}
where $\vec{\sigma}$ represents only the cross section(s) considered as parameter(s) of interest in the given fit, 
while the likelihood is maximized with respect to all remaining cross sections and nuisance parameters. In the denominator the likelihood is maximized with respect to all other cross sections and nuisance parameters as well as the parameters of interest, which are fixed to hypothetical values in the numerator.
The parameter of interest $\sigma$ in each production bin is alternatively replaced by $\mu \cdot \sigma_{\mathrm{SM}}(\vec{\theta})$, allowing an interpretation in terms of the signal strength $\mu$ relative to the SM prediction $\sigma_{\mathrm{SM}}(\vec{\theta})$. In addition, the number of signal events is extracted from a simultaneous fit of the signal templates  in all reconstructed event categories, using a coarser \BDT binning with several bins merged into one and considering only the background systematic uncertainties in the fit.

The expected and observed numbers $N_{\mathrm{S}}$ of signal events are shown in Table~\ref{tab:NsInterpretation} together with the signal acceptances for the Stage-0 production bins.
  \begin{sidewaystable}
    
  \centering  

  \caption{The expected and observed numbers of signal events in reconstructed event categories for an integrated luminosity of \Lum at $\sqrt{\mathrm{s}}=$~13~TeV, together with signal acceptances for each Stage-0 production mode. Results are obtained in bins of \BDT discriminants using coarse binning with several bins merged into one. The symbol $-$ represents the cases in which the acceptance is $<0.01\%$.}
\label{tab:NsInterpretation}   
 
     \vspace{0.1cm}
  {\renewcommand{\arraystretch}{1.1}

    \begin{tabular}{l|cc|ccccc}
    \hline\hline
    
    Reconstructed& 
    \multicolumn{2}{c}{$N_{\mathrm{S}}$} & 
    \multicolumn{5}{|c}{Acceptance [\%]} \\
    
    event category& 
    Expected & 
    Observed & 
    \STXSggF & \STXSVBF & \STXSVH & \STXSttH & \bbH\\
    
    \hline

   \CatZeroJ, \BDT-Bin 0-4&   
   \multirow{1}{*}{ $5^{+4}_{-3}$ } & 
   \multirow{1}{*}{  $10^{+5}_{-4}$ } & 
   \multirow{1}{*}{  $2.40  \pm 0.23$ } & 
   \multirow{1}{*}{  $0.178 \pm 0.029$ } & 
   \multirow{1}{*}{ $0.20  \pm 0.05$ } & 
   \multicolumn{1}{c}{\multirow{1}{*}{ $-$ }} & 
   \multirow{1}{*}{ $\phantom{0}2.0   \pm 1.0\phantom{0}$ } \\

    \CatZeroJ, \BDT-Bin 5-9&
    \multirow{1}{*}{ $11\pm 4$ } & 
    \multirow{1}{*}{ $11\pm 4$ } & 
    \multirow{1}{*}{ $\phantom{0}5.4  \pm 0.5\phantom{0}$ } & 
    \multirow{1}{*}{ $\phantom{0}0.64 \pm 0.12\phantom{0}$ } & 
    \multirow{1}{*}{ $0.74  \pm 0.11$ } & 
    \multicolumn{1}{c}{\multirow{1}{*}{ $-$ }} & 
    \multirow{1}{*}{ $\phantom{0}6.3   \pm 3.2\phantom{0}$ } \\
    
    \CatZeroJ, \BDT-Bin 10-14& 
    \multirow{1}{*}{ $11^{+4}_{-3}$ } & 
    \multirow{1}{*}{ $12^{+4}_{-3}$ } & 
    \multirow{1}{*}{ $\phantom{0}5.2  \pm 0.6\phantom{0}$ } & 
    \multirow{1}{*}{ $\phantom{0}1.03 \pm 0.28\phantom{0}$ } & 
   \multirow{1}{*}{  $1.89  \pm 0.27$ } & 
   \multicolumn{1}{c}{ \multirow{1}{*}{ $-$ }} & 
    \multirow{1}{*}{ $\phantom{0}5.6   \pm 2.8\phantom{0}$ } \\ 
     
    \CatOneJL, \BDT-Bin 0-4& 
    \multirow{1}{*}{ $5.8^{+3.2}_{-2.6}$ } & 
    \multirow{1}{*}{ $7.6^{+3.5}_{-2.9}$ } &
    \multirow{1}{*}{ $\phantom{0}2.7 \pm 0.4\phantom{0} $ } & 
    \multirow{1}{*}{ $\phantom{0}1.25 \pm 0.10\phantom{0} $ } & 
    \multirow{1}{*}{ $1.72 \pm 0.19 $ } & 
    \multicolumn{1}{c}{\multirow{1}{*}{ $- $ }} & 
    \multirow{1}{*}{ $\phantom{0}3.1 \pm 1.6\phantom{0} $ }\\

    \CatOneJL, \BDT-Bin 5-9& 
   \multirow{1}{*}{ $3.1^{+2.4}_{-1.8}       $ } & 
   \multirow{1}{*}{ $0.8^{+1.8}               $ } &
   
   \multirow{1}{*}{ $1.33 \pm 0.20 $ } & 
   \multirow{1}{*}{ $\phantom{0}2.02 \pm 0.12\phantom{0} $ } & 
   \multirow{1}{*}{ $0.85 \pm 0.09 $ } & 
   \multicolumn{1}{c}{\multirow{1}{*}{ $- $ }} & 
   \multirow{1}{*}{ $\phantom{0}1.4 \pm 0.7\phantom{0} $ }\\

    \CatOneJM, \BDT-Bin 0-4&       
      \multirow{1}{*}{ $3.5^{+2.5}_{-1.8}       $ } & 
      \multirow{1}{*}{ $5.1^{+2.8}_{-2.1}       $ }&
      \multirow{1}{*}{ $1.53 \pm 0.25 $  }& 
      \multirow{1}{*}{ $\phantom{0}1.41 \pm 0.23\phantom{0} $  }& 
      \multirow{1}{*}{ $2.09 \pm 0.25 $  }& 
      \multicolumn{1}{c}{\multirow{1}{*}{ $-$  }}& 
      \multirow{1}{*}{ $\phantom{0}1.0 \pm 0.5\phantom{0} $ }\\

    \CatOneJM, \BDT-Bin 5-9& 
      \multirow{1}{*}{ $2.0^{+1.9}_{-1.2}       $  }& 
      \multirow{1}{*}{ $2.6^{+2.1}_{-1.4}       $ }&
      \multirow{1}{*}{ $0.74 \pm 0.11 $  }& 
      \multirow{1}{*}{ $\phantom{0}2.68 \pm 0.27\phantom{0} $  }& 
      \multirow{1}{*}{ $0.49 \pm 0.05 $  }& 
      \multicolumn{1}{c}{\multirow{1}{*}{ $- $ }}& 
      \multirow{1}{*}{ $0.27 \pm 0.14 $ }\\
     
    \multirow{1}{*}{\CatOneJH }&
     \multirow{1}{*}{ $1.5^{+1.7}_{-1.0}       $  }& 
     \multirow{1}{*}{ $2.8^{+2.1}_{-1.4}       $ }&
    \multirow{1}{*}{  $0.56 \pm 0.12 $  }& 
     \multirow{1}{*}{ $\phantom{0}1.74 \pm 0.24\phantom{0} $  }& 
     \multirow{1}{*}{ $1.06 \pm 0.10 $  }& 
     \multicolumn{1}{c}{\multirow{1}{*}{ $- $ }}& 
     \multirow{1}{*}{ $0.17 \pm 0.09 $ }\\

      \CatVBFLow, \BDT-Bin 0-4&
    \multirow{1}{*}{ $4.0^{+2.7}_{-2.0}       $  }& 
    \multirow{1}{*}{ $8.7^{+3.5}_{-2.8}       $ }&
     \multirow{1}{*}{ $1.53 \pm 0.30 $  }& 
     \multirow{1}{*}{ $\phantom{0}3.33 \pm 0.32\phantom{0} $  }& 
    \multirow{1}{*}{  $2.83 \pm 0.32 $  }& 
    \multirow{1}{*}{  $\phantom{00}2.8 \pm 0.4\phantom{00} $  }& 
     \multirow{1}{*}{ $\phantom{0}1.9 \pm 1.0\phantom{0} $ }\\
    
      \CatVBFLow, \BDT-Bin 5-9&
     \multirow{1}{*}{ $2.4^{+2.0}_{-1.3}       $ } & 
     \multirow{1}{*}{ $5.7^{+2.8}_{-2.1}       $ }&
     \multirow{1}{*}{ $0.43 \pm 0.11 $  }& 
     \multirow{1}{*}{ $\phantom{00}9.6 \pm 0.9\phantom{00} $  }& 
     \multirow{1}{*}{ $0.27 \pm 0.04 $  }& 
     \multirow{1}{*}{ $0.153 \pm 0.028 $  }& 
     \multirow{1}{*}{ $0.32 \pm 0.17 $ }\\
     
      \multirow{1}{*}{\CatVBFHigh }&
      \multirow{1}{*}{ $0.6^{+1.3}_{-0.6}       $ } & 
      \multirow{1}{*}{ $2.8^{+2.1}_{-1.4}       $ }&
      \multirow{1}{*}{ $0.16 \pm 0.05 $  }& 
      \multirow{1}{*}{ $\phantom{0}1.18 \pm 0.11\phantom{0} $  }& 
      \multirow{1}{*}{ $0.51 \pm 0.06 $  }& 
      \multirow{1}{*}{ $\phantom{0}0.71 \pm 0.09\phantom{0} $  }& 
      \multicolumn{1}{c}{\multirow{1}{*}{ $- $ }}\\

       \CatVHHad, \BDT-Bin 0-4&
       \multirow{1}{*}{ $2.4^{+2.1}_{-1.4}       $ }  & 
       \multirow{1}{*}{ $2.4^{+2.1}_{-1.4}       $ } &
       \multirow{1}{*}{ $1.00 \pm 0.23 $  }& 
       \multirow{1}{*}{ $\phantom{0}1.08 \pm 0.13\phantom{0} $  }& 
       \multirow{1}{*}{ $1.52 \pm 0.17 $  }& 
       \multirow{1}{*}{ $\phantom{0}0.33 \pm 0.07\phantom{0} $  }& 
       \multirow{1}{*}{ $\phantom{0}0.9 \pm  0.5\phantom{0} $ }\\

        \CatVHHad, \BDT-Bin 5-9&
	\multirow{1}{*}{ $1.3^{+1.6}_{-0.9}       $ } & 
	\multirow{1}{*}{ $0        $ }&
	\multirow{1}{*}{ $0.37 \pm 0.07 $  }& 
	\multirow{1}{*}{ $\phantom{0}0.29 \pm 0.06\phantom{0} $  }& 
	\multirow{1}{*}{ $\phantom{0}5.0 \pm 0.5\phantom{0} $  }& 
	\multirow{1}{*}{ $\phantom{0}1.06 \pm 0.19\phantom{0} $  }& 
	\multirow{1}{*}{ $0.29 \pm 0.15 $ }\\

       \multirow{1}{*}{\CatVHLep }&
       \multirow{1}{*}{ $0.3^{+1.0}               $  }& 
       \multirow{1}{*}{ $0        $ }&
       \multicolumn{1}{c}{\multirow{1}{*}{ $- $  }}& 
       \multicolumn{1}{c}{\multirow{1}{*}{ $- $  }}& 
       \multirow{1}{*}{ $2.94 \pm 0.24 $  }& 
       \multirow{1}{*}{ $\phantom{0}1.70 \pm 0.19\phantom{0} $  }& 
       \multicolumn{1}{c}{\multirow{1}{*}{ $- $ }}\\

        \multirow{1}{*}{\CatttH }&
 	\multirow{1}{*}{ $0.4^{+1.1}               $  }& 
	\multirow{1}{*}{ $0        $ }&
 	\multicolumn{1}{c}{\multirow{1}{*}{ $- $  }}& 
	\multicolumn{1}{c}{\multirow{1}{*}{ $- $  }}& 
	\multirow{1}{*}{ $0.19 \pm 0.04 $  }& 
	\multirow{1}{*}{ $\phantom{0}13.5 \pm 2.9\phantom{00} $  }& 
	\multirow{1}{*}{ $0.31 \pm 0.17 $ }\\
 
 \hline
        \multicolumn{2}{l}{Combined acceptance }&
	\multirow{1}{*}{ ~ }&
	\multirow{1}{*}{ $23.5 \pm 1.9\phantom{0} $  }& 
	\multirow{1}{*}{ $26.4 \pm 1.4\phantom{00} $  }& 
	\multirow{1}{*}{ $22.3 \pm 1.9\phantom{0} $  }& 
	\multirow{1}{*}{ $20.3 \pm 2.0\phantom{00} $  }& 
	\multirow{1}{*}{ $24 \pm 12\phantom{.0} $ }\\

     \hline\hline
     
     \end{tabular}
  }
  \end{sidewaystable}

Assuming that the relative signal fractions in each production bin are given by the predictions for the SM Higgs boson,
the inclusive production cross section of 
\begin{equation*}
  \sBR \equiv \sBRZZ~=~{\color{black}{1.73^{+0.24}_{-0.23}(\mathrm{stat.})^{+0.10}_{-0.08}(\mathrm{exp.})\pm 0.04(\mathrm{th.})~\mathrm{pb}~=~1.73^{+0.26}_{-0.24}~ \mathrm{pb}}}
\end{equation*}
is measured in the rapidity range $|y_H|<$~2.5, compared  to the SM prediction of  \linebreak
\sBRsm~$\equiv~$\sBRZZsm~$= 1.34 \pm 0.09$~pb.
The data are also interpreted in terms of the global signal strength, yielding
\begin{equation*}
 \mu = 1.28^{+0.18}_{-0.17}(\mathrm{stat.})^{+0.08}_{-0.06}(\mathrm{exp.})^{+0.08}_{-0.06}(\mathrm{th.}) = 1.28^{+0.21}_{-0.19}.
\end{equation*}

\begin{figure}[!htbp]
\begin{center}
  \subfloat[]{\includegraphics[width=0.45\linewidth]{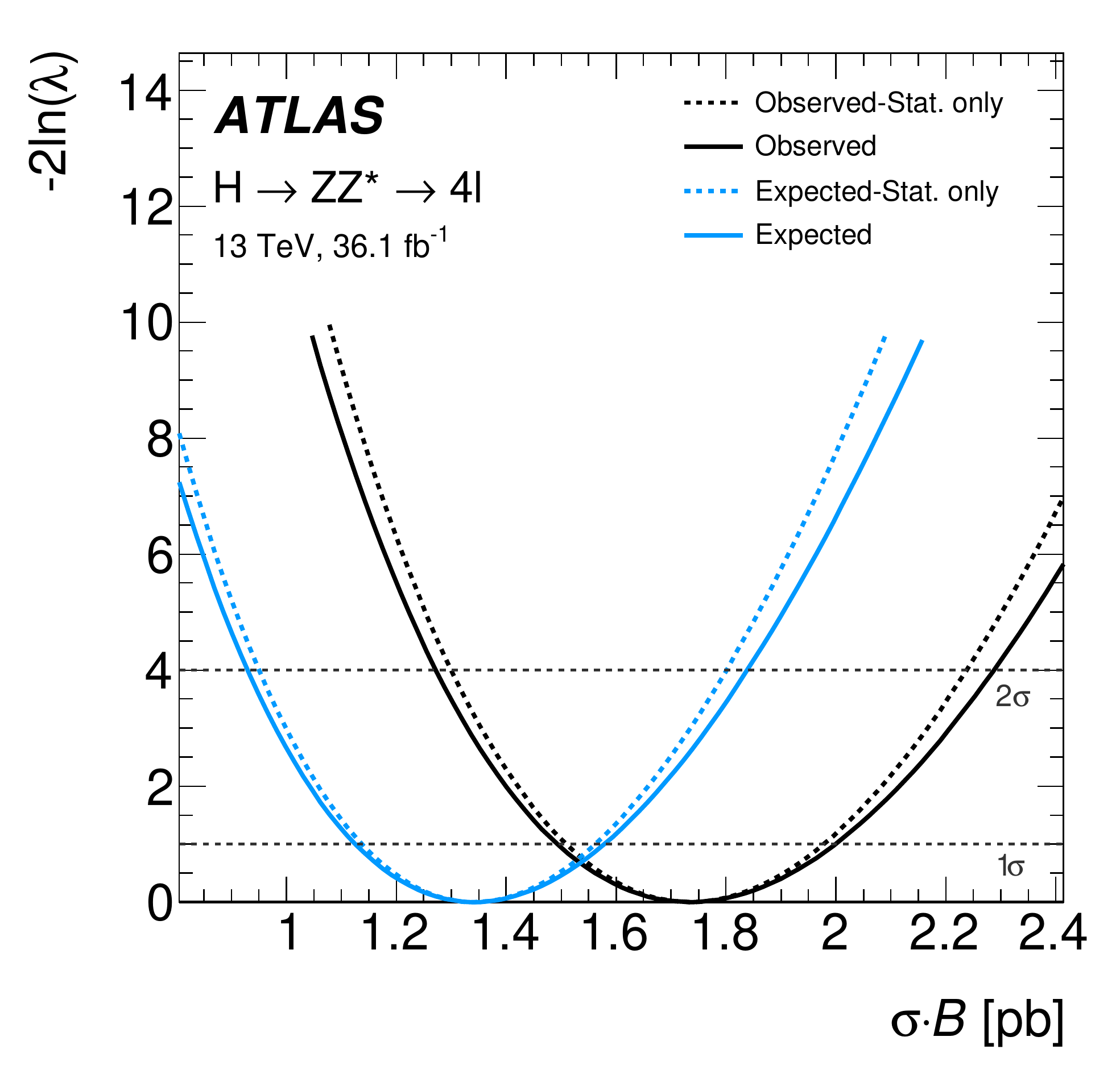}\label{fig:totXS}} 
  \subfloat[]{\includegraphics[width=0.45\linewidth]{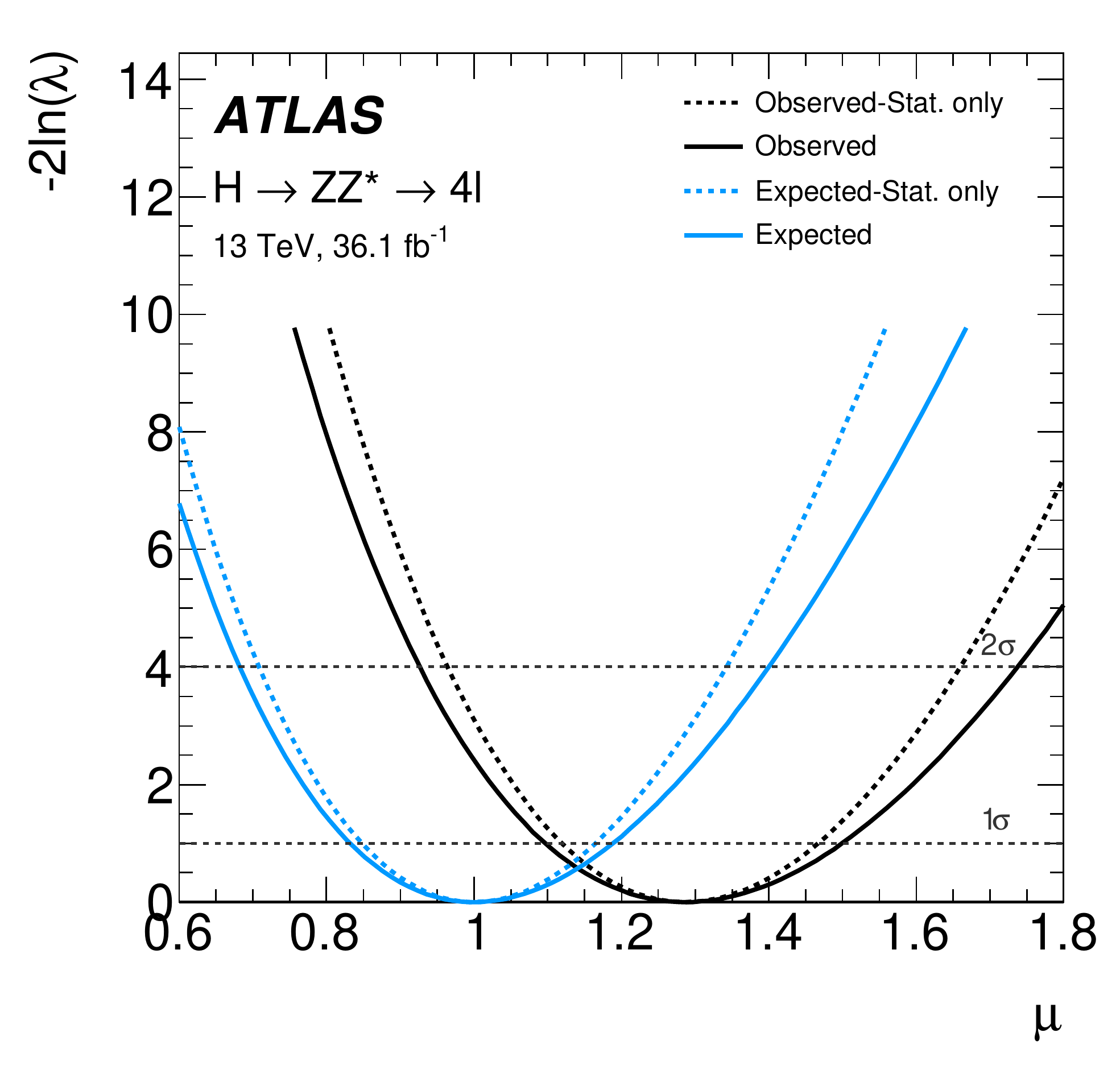}\label{fig:totMU}}
\end{center}
\vskip-0.4cm
\caption{The profile likelihood as a function of (a) \sBRZZ and (b) the inclusive signal strength $\mu$;
  the scans are shown both with (solid line) and without (dashed line) systematic uncertainties.}
  \label{fig:incl_results}
\end{figure}

 The measured cross section and signal strength agree with the SM prediction at the level of 1.7$\sigma$ and 1.6$\sigma$, respectively.
The corresponding likelihood scans are shown in Figure~\ref{fig:incl_results}. The dominant systematic uncertainty of the cross-section measurement is the experimental uncertainty in the integrated luminosity and lepton efficiency measurements. The signal strength measurement is also equally affected by the theoretical uncertainty of the \ggF signal yield due to QCD scale variations.
This theory uncertainty in the predicted signal yield cancels out when expressing the results in terms of the ratio of the observed to expected cross section times the branching ratio \sBRratio~$= 1.29^{+0.20}_{-0.18}$, with no uncertainty assigned to the denominator.

\begin{figure}[!htbp]
   \begin{center}
   
   \begin{minipage}[c]{0.57\linewidth}
   \subfloat[]{\hskip0.6cm\includegraphics[width=\linewidth]{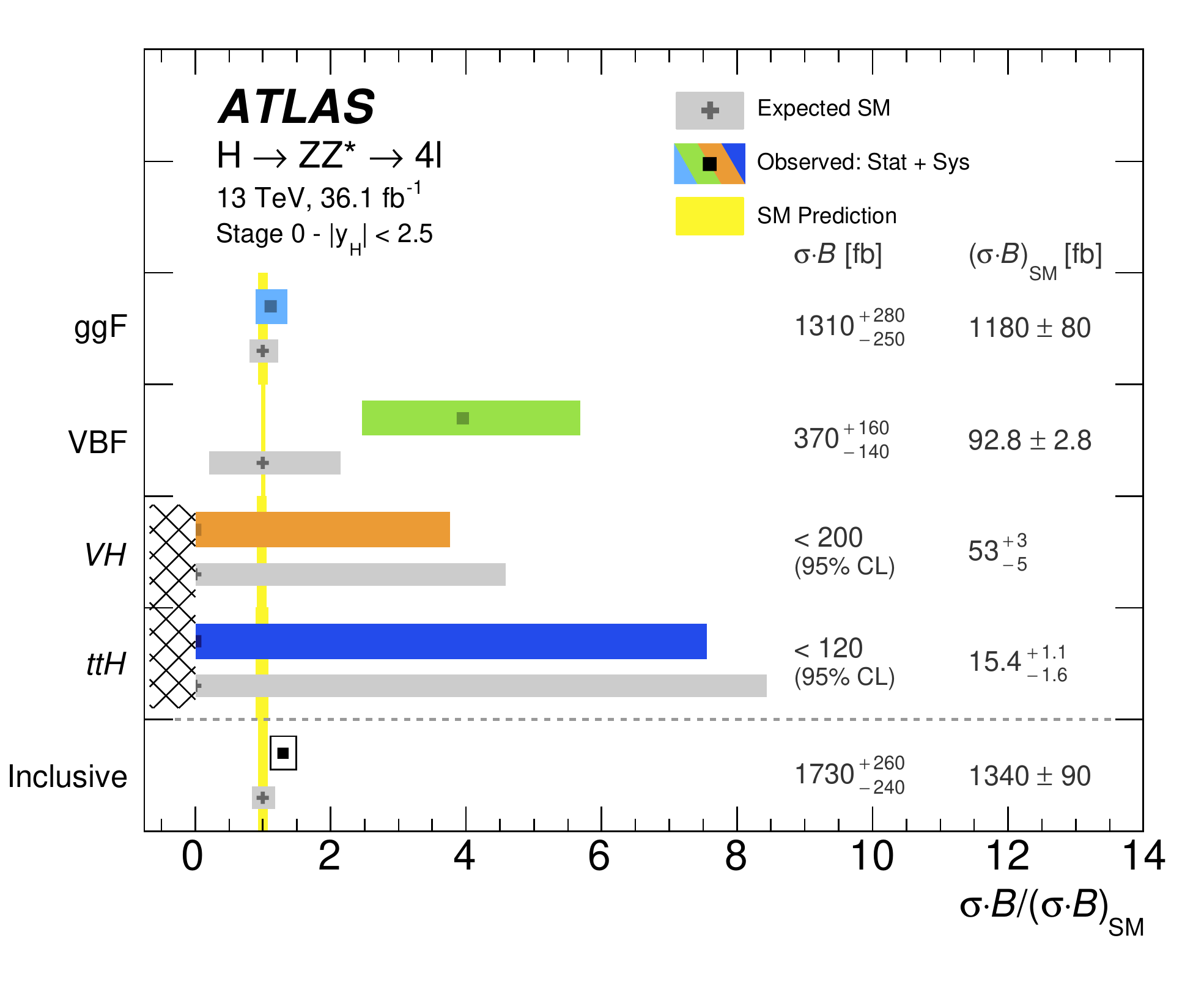}\label{fig:st0}}
   \end{minipage}
   
   \begin{minipage}[c]{0.66\linewidth}
   \subfloat[]{\includegraphics[width=\linewidth]{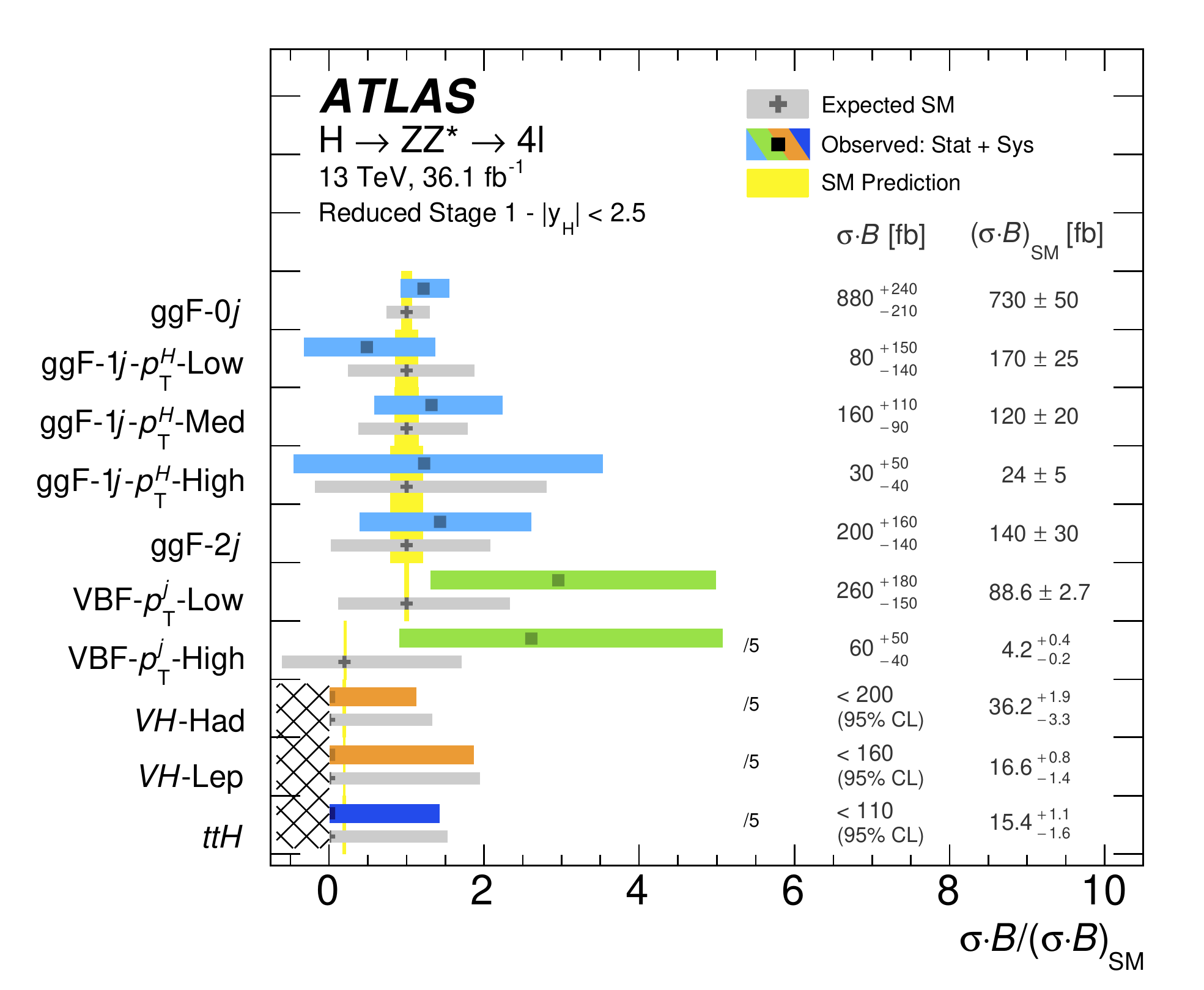}\label{fig:st1}}
   \end{minipage}
   
   \end{center}
   \caption{ The observed and expected SM values of the cross-section ratios  \sBR normalized by the SM expectation \sBRsm for the inclusive production and in the (a) Stage-0 and (b) reduced Stage-1 production bins for an integrated luminosity of \Lum at $\sqrt{\mathrm{s}}=13$~\TeV. Different colors for the observed results indicate different Higgs boson production modes. The hatched area indicates that the \VH and \ttH parameters of interest are constrained to positive values. For visualization purposes, the \STXSVBFHigh value and the limits for the three reduced Stage-1 production bins \STXSVHHad, \STXSVHLep and \STXSttH are divided by a factor of five when shown normalized to \sBRsm. The yellow vertical band represents the theory uncertainty in the signal prediction, while the horizontal grey bands represent the expected measurement uncertainty.}
   \label{fig:Stage01_results}
   \end{figure}

The SM expected cross section, the observed values of \sBRZZ and their ratio for the inclusive production and in each Stage-0 and reduced Stage-1 production bin are shown in Table~\ref{tab:Stage01}. 
The corresponding values are summarized in Figure~\ref{fig:Stage01_results}. The $\bbH$ production process is treated as a part of the \ggF production bins.
In the ratio calculation uncertainties on the SM expectation are not taken into account.
\begin{table*}[!htbp]
  \def\arraystretch{1.8}
  \centering
  \caption{The SM expected cross section \sBRsm, the observed values of \sBRZZ, and their ratio  \sBRratio 
  for the inclusive production and in each Stage-0 and reduced Stage-1 production bin for an integrated luminosity of \Lum and at $\sqrt{\mathrm{s}}=13$~\TeV. The \bbH contribution is considered as a part of the  \ggF production bins.
    The upper limits correspond to the 95\%~CL obtained with pseudo-experiments using the  CL$_{\textrm s}$  method. The uncertainties are given as (stat.)+(exp.)+(th.) for Stage 0 and as (stat.)+(syst.) for reduced Stage 1.  
  }
  \label{tab:Stage01}
  \label{tab:XS_cat_result}

     \vspace{0.1cm}
  {\renewcommand{\arraystretch}{1.5}

  \begin{tabular}{l | l c | c }
    \hline\hline
   Production bin &  \multicolumn{2}{c|}{ Cross section (\sBR) [pb]} &  \sBRratio 
   \\
     &  \multicolumn{1}{c}{SM expected} & \multicolumn{1}{c|}{Observed} &  \multicolumn{1}{c}{Observed}   \\
    \hline
    \multicolumn{4}{l}{Inclusive production, $|y_{H}|<2.5$}\\
    \hline
                        & 
    $1.34 \pm 0.09\phantom{00} $ &
    $1.73^{+0.24}_{-0.23}$$^{+0.10}_{-0.08}$$\pm 0.04$ &
    $1.29^{+0.18}_{-0.17}$$^{+0.07}_{-0.06}$$\pm 0.03$ 
    \\

    \hline
    \multicolumn{4}{l}{Stage-0 production bins, $|y_{H}|<2.5$}\\
    \hline
    \STXSggF                    & 
    $1.18 \pm 0.08\phantom{00} $ &
    $1.31^{+0.26}_{-0.24}$$^{+0.09}_{-0.07}$$\pm 0.05$ &
    $1.11^{+0.22}_{-0.20}$$^{+0.07}_{-0.06}$$\pm 0.04$ 
    \\
      
    \STXSVBF                    & 
    $0.0928 \pm 0.0028$ & 
    $0.37^{+0.15}_{-0.13}$$\pm 0.03$$\pm 0.03$ &
    $4.0^{+1.7}_{-1.4}$$\pm 0.3$$\pm 0.3$ 
    \\
    
    \STXSVH                     & 
    \multicolumn{1}{l}{$0.053^{+0.003}_{-0.005}$} & 
    \multicolumn{1}{c|}{$<0.20                  $}                                  &
    \multicolumn{1}{c}{$<3.7                  $} 
    \\
    
    \STXSttH                    & 
    \multicolumn{1}{l}{$0.0154^{+0.0011}_{-0.0016}$} & 
    \multicolumn{1}{c|}{$<0.12                  $ }                                 &
    \multicolumn{1}{c}{$<7.5                 $} 
    \\
    
    \hline
    \multicolumn{4}{l}{Reduced Stage-1 production bins, $|y_{H}|<2.5$}\\
    \hline
    
    \STXSggFZeroJ               & 
    $0.73 \pm 0.05\phantom{00}$ &
    $0.88^{+0.22}_{-0.20}$$^{+0.09}_{-0.07}$ &
    $1.22^{+0.30}_{-0.27}$$^{+0.13}_{-0.09}$   
    \\
    
    \STXSggFOneJL               & 
    $0.174 \pm 0.025\phantom{0}$ & 
    $0.08^{+0.15}_{-0.12}$$^{+0.04}_{-0.06}$ &
    $0.5^{+0.8}_{-0.7}$$^{+0.3}_{-0.4}$   
    \\
    
    \STXSggFOneJM               & 
    $0.120 \pm 0.018\phantom{0}$ & 
    $0.16^{+0.11}_{-0.09}$$^{+0.03}_{-0.01}$ &
    $1.3^{+0.9}_{-0.7}$$\pm 0.2$   
    \\
    
    \STXSggFOneJH               & 
    $0.024 \pm 0.005\phantom{0}$ & 
    $0.03^{+0.05}_{-0.04}$$\pm 0.01$ &
    $1.2^{+2.3}_{-1.7}$$\pm 0.3$   
    \\
    
    \STXSggFTwoJ                & 
    $0.137 \pm 0.029\phantom{0}$ & 
    $0.20^{+0.16}_{-0.14}$$\pm 0.03$ &
    $1.4^{+1.2}_{-1.0}$$\pm 0.2$   
    \\
    
    \STXSVBFLow                 & 
$0.0886 \pm 0.0027$ & 
$0.26^{+0.18}_{-0.14}$$^{+0.03}_{-0.02}$ &
$3.0^{+2.0}_{-1.6}$$^{+0.4}_{-0.2}$  
\\
    
    \STXSVBFHigh                & 
    \multicolumn{1}{l}{$0.0042^{+0.0004}_{-0.0002} $} & 
    $0.06^{+0.05}_{-0.04}$$\pm 0.01$ &
    $13^{+12}_{-8} $$\pm 1$     
    \\
    
    \STXSVHHad                  & 
    \multicolumn{1}{l}{$0.0362^{+0.0019}_{-0.0033}$} & 
    \multicolumn{1}{c|}{$<0.20                  $}                &
    \multicolumn{1}{c}{$<5.6                  $} 
    \\
    
    \STXSVHLep                  & 
    \multicolumn{1}{l}{$0.0166^{+0.0008}_{-0.0014}$} & 
    \multicolumn{1}{c|}{$<0.16                  $}                &
    \multicolumn{1}{c}{$<9.3                  $} 
    \\
    
    \STXSttH                    & 
    \multicolumn{1}{l}{$0.0154^{+0.0011}_{-0.0016}$} & 
    \multicolumn{1}{c|}{$<0.11$}                &
    \multicolumn{1}{c}{$<7.1$} 
    \\
    
    \hline\hline
  \end{tabular}
  }
  \end{table*}

All measured Stage-0 and reduced Stage-1 \ggF measurements agree with the predictions for the SM Higgs boson within 1$\sigma$. Somewhat worse agreement is obtained for the \VBF bins due to the observed excess of events in the two \VBF-enriched reconstructed event categories. The largest deviation of 2.2$\sigma$ is observed for the Stage-0 \VBF production bin due to an observed excess of events characterized by the presence of at least two jets and a dijet invariant mass above 120~GeV.  Due to the limited number of events in the \VH- and \CatttH categories, only upper limits are set on the cross sections and signal strengths for these production modes. The limits are based on the CL$_{\textrm s}$ prescription~\cite{Read:2002} and derived using pseudo-experiments. The \VH and \ttH parameters of interest are constrained to positive values to avoid the fit's prediction of negative total event yields in the \CatVHLep and \CatttH categories and provide a stable fit configuration. It was found that the impact of this constraint on the final fit results is negligible. 

The dominant contribution to the measurement uncertainty in the Stage-0 \ggF production bin originates from the same sources as in the inclusive measurement. In the remaining three Stage-0 production bins, similar sources of uncertainty are relevant for both the cross section and the signal strength measurements: the jet energy scale or resolution uncertainties in all production bins and, additionally, jet bin migrations for the \VBF and \VH processes. For the reduced Stage-1 categories the dominant cross-section uncertainties are the integrated luminosity and lepton efficiency measurements for the \STXSggFZeroJ and \STXSVHLep bins and the jet energy scale or resolution for all other categories. The \STXSVBFLow bin is additionally affected by  parton shower uncertainties, while the effects of the finite top quark mass have a dominant impact on the \STXSVBFHigh bin, together with migrations between transverse momentum bins. The signal strength measurement of the reduced Stage-1 \ggF processes is also strongly affected by the theory uncertainties from event migrations between different jet multiplicity and Higgs boson transverse momentum bins, while parton shower and QCD scale uncertainties affect the remaining reduced Stage-1 production bins.

\begin{figure}[!htbp]
\begin{center}
  \subfloat[]{\includegraphics[width=0.5\linewidth]{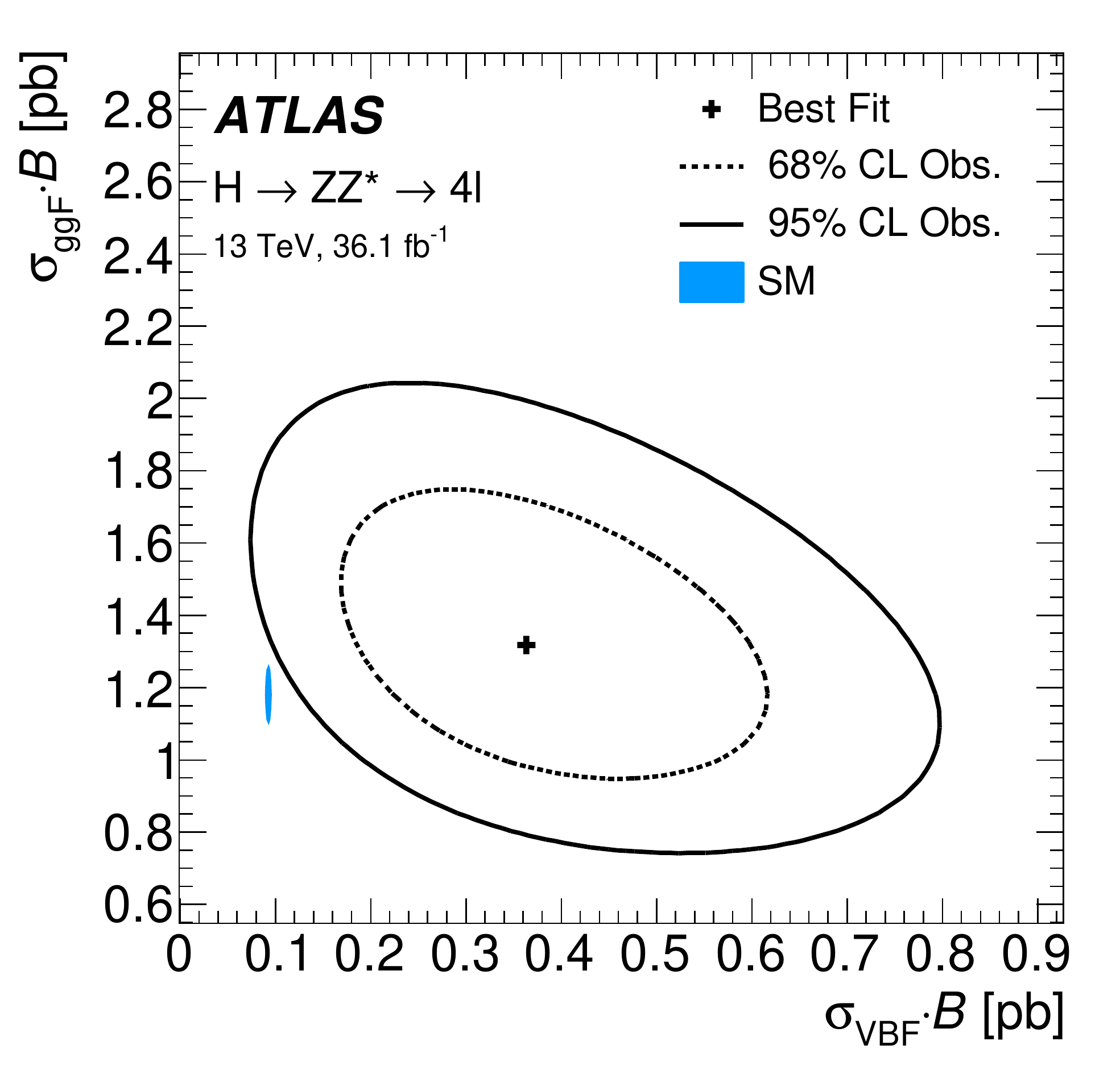}\label{fig:sigGGFVBF}}
  \subfloat[]{\includegraphics[width=0.5\linewidth]{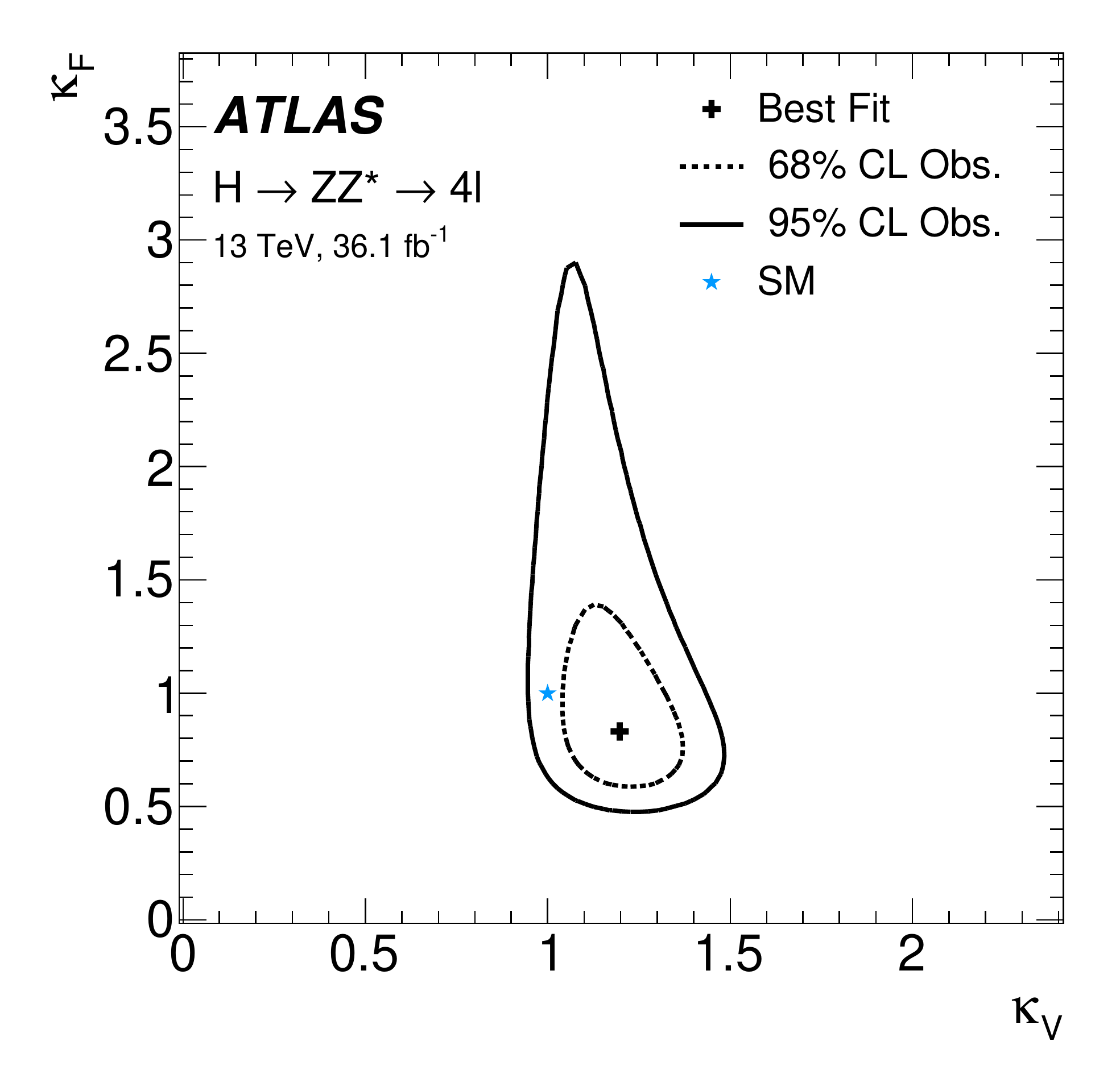}\label{fig:kvkf}}

\end{center}
\caption{ (a) Likelihood contours at 68\% CL (dashed line) and 95\% CL (solid line) in the ($\sigma_{\textrm{ggF}}\cdot B$ , $\sigma_{\textrm{VBF}}\cdot B$) plane and (b) likelihood contours in the $\kappa_{V}$--$\kappa_{F}$  plane. The best fits to
  the data (solid cross) and the SM predictions are also indicated.  In (a), the SM prediction is shown together with its theory uncertainty (filled blue elipse), while in (b) only the central value of the SM prediction (solid blue star) is shown. }
\end{figure}

Figure~\ref{fig:sigGGFVBF} shows the likelihood contours in the ($\sigma_{\textrm{{ggF}}}\cdot B$, $\sigma_{\textrm{{VBF}}}\cdot B$ ) plane.
The \VH and \ttH cross section parameters are left free in the fit, i.e. they are not treated as parameters of interest.
The compatibility with respect to the Standard Model expectation is
at the level of 2.3$\sigma$, due to the discrepancies observed in the \VBF-related distributions in Figures~\ref{fig:categoryVars} and~\ref{fig:BDToutputs_observed}.
The cross-section results by production mode (Stage 0) can also be interpreted in the $\kappa$ framework~\cite{Heinemeyer:2013tqa,YR4} in which coupling modifiers, $\kappa_{i}$, are introduced to parameterize possible deviations from the SM predictions of the Higgs boson couplings to SM bosons and fermions. One interesting benchmark allows two different Higgs boson coupling strength modifiers to fermions and bosons, reflecting the different structure of the interactions of the SM Higgs sector with gauge bosons and fermions. The universal coupling-strength scale factors $\kappa_{F}$ for all fermions and $\kappa_{V}$ for all vector bosons are defined as $\kappa_{V}= \kappa_{W}=\kappa_{Z}$ and  $\kappa_{F}=\kappa_{t}=\kappa_{b}=\kappa_{c}=\kappa_{\tau}=\kappa_{\mu}=\kappa_{g}$. It is assumed that there are no undetected or invisible Higgs boson decays. The  observed likelihood contours in the $\kappa_{V}-\kappa_{F}$ plane are shown in Figure~\ref{fig:kvkf} (only the quadrant $\kappa_{F}>0$ and $\kappa_{V}>0$ is shown since this channel is not sensitive to the relative sign of the two coupling modifiers). The compatibility with the Standard Model expectation is at the level of 1.4$\sigma$. Better agreement is observed here compared to the likelihood contours for the cross sections, since the lower observed yield in the \VH categories compared with SM expectations compensates for the observed excess in the \VBF categories.

\subsection{Tensor structure of Higgs boson couplings to vector bosons} 
\label{subsec:results_BSM}

In order to probe the tensor structure of the Higgs boson couplings to vector bosons, a likelihood function is constructed as a product of conditional probabilities over  the event yield  $N_j$ in each reconstructed event category $j$,
\begin{equation*}
\begin{split}
& \mathcal{L}(\vec{\kappa}, \vec{\theta}) =  \displaystyle  \prod_{j}^{N_{\mathrm{categories}}}
 P\big(N_{j} ~| S^{(\vec{\kappa})}_{j}(\vec{\theta}) ~+ B_{j}(\vec{\theta}) \big)  \times  \prod_{m}^{N_{\mathrm{nuisance}}} \mathcal{C}_{m}(\vec{\theta}) \;,
\end{split}
\label{eq:likelihood}
\end{equation*}
with the set of coupling parameters $\vec{\kappa}$ representing the parameters of interest for a specific hypothesis test. 
The expected number of signal events $S_{j}^{(\vec{\kappa})}(\vec{\theta})$ is parameterized in terms of the SM and BSM couplings using the signal modelling described in Section~\ref{sec:signal_bkg}, while the expected background event yields $B_{j}(\vec{\theta})$ are given by the background estimates detailed in Section~\ref{sec:background}. As in the case of the cross-section measurements, the test statistic is based on a profile likelihood ratio,
\begin{equation*}
q = -2 \ln \frac{\mathcal{L}( \vec{\kappa}, \hat{\hat{\vec{\theta}}}(\vec{\kappa})) } {\mathcal{L}(\hat{{\vec{\kappa}}}, \hat{{\vec{\theta}}}(\hat{\vec{\kappa}}) )} = -2 \ln \lambda(\vec{\kappa})\;,
\label{eq:lambda}
\end{equation*}
with the conditional and the unconditional maximum-likelihood estimators in the numerator and the denominator, respectively.
The coupling parameter \kAgg is measured assuming that all other BSM couplings are equal to zero. The coupling parameters \kHVV and \kAVV are probed both simultaneously 
and one at a time assuming that all other BSM couplings vanish. If not stated otherwise, the SM couplings \kSM and \kHgg described in Section~\ref{subsec:strategy_BSM} are fixed to the SM value of one. The BSM changes in the Higgs sector are assumed not to affect the SM background processes.

\begin{figure}[!htbp]
  \centering
  \subfloat[\label{fig:kAgg}]{\includegraphics[width=0.5\linewidth]{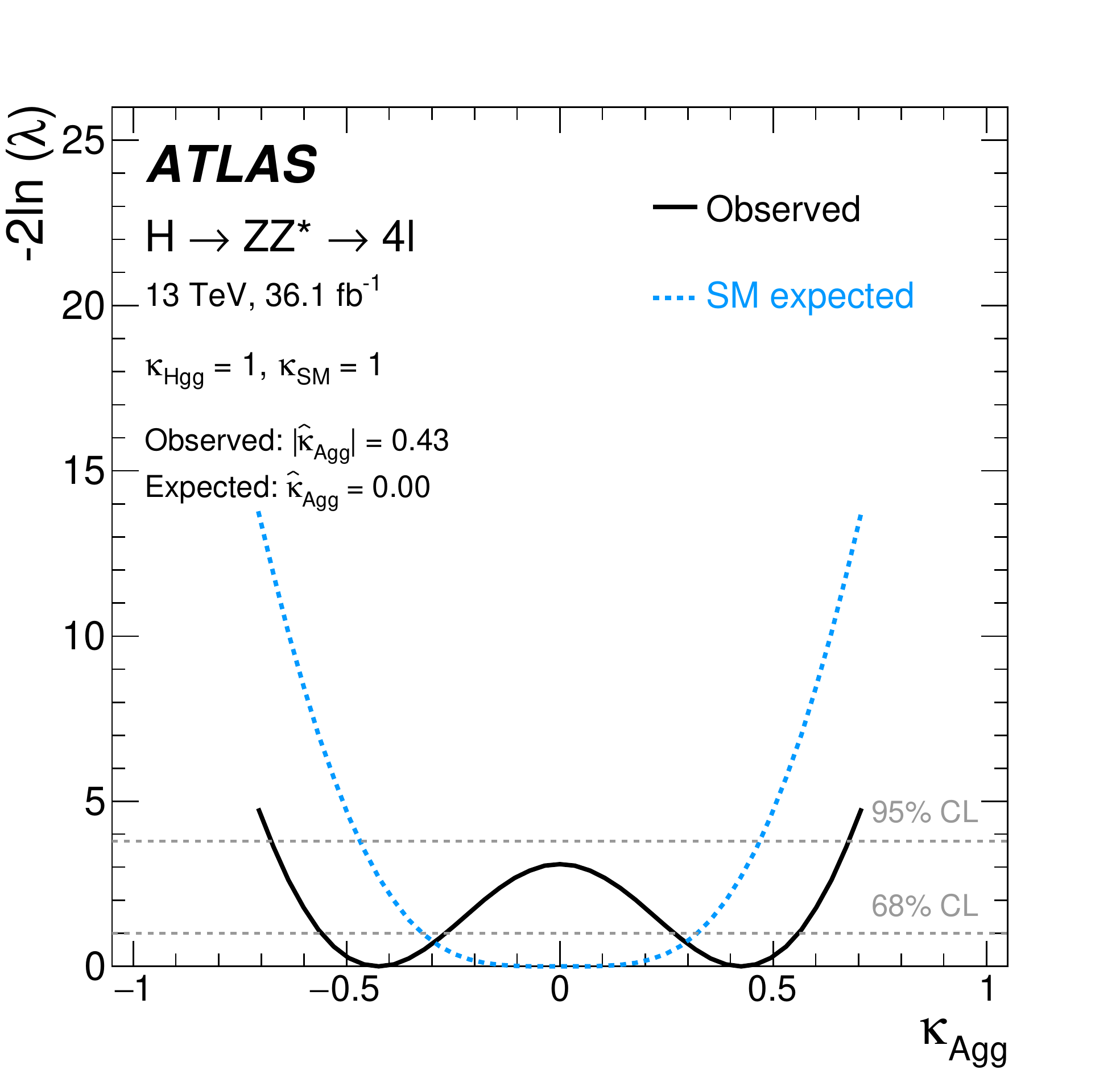}}\\
  \subfloat[\label{fig:kHVV}]{\includegraphics[width=0.5\linewidth]{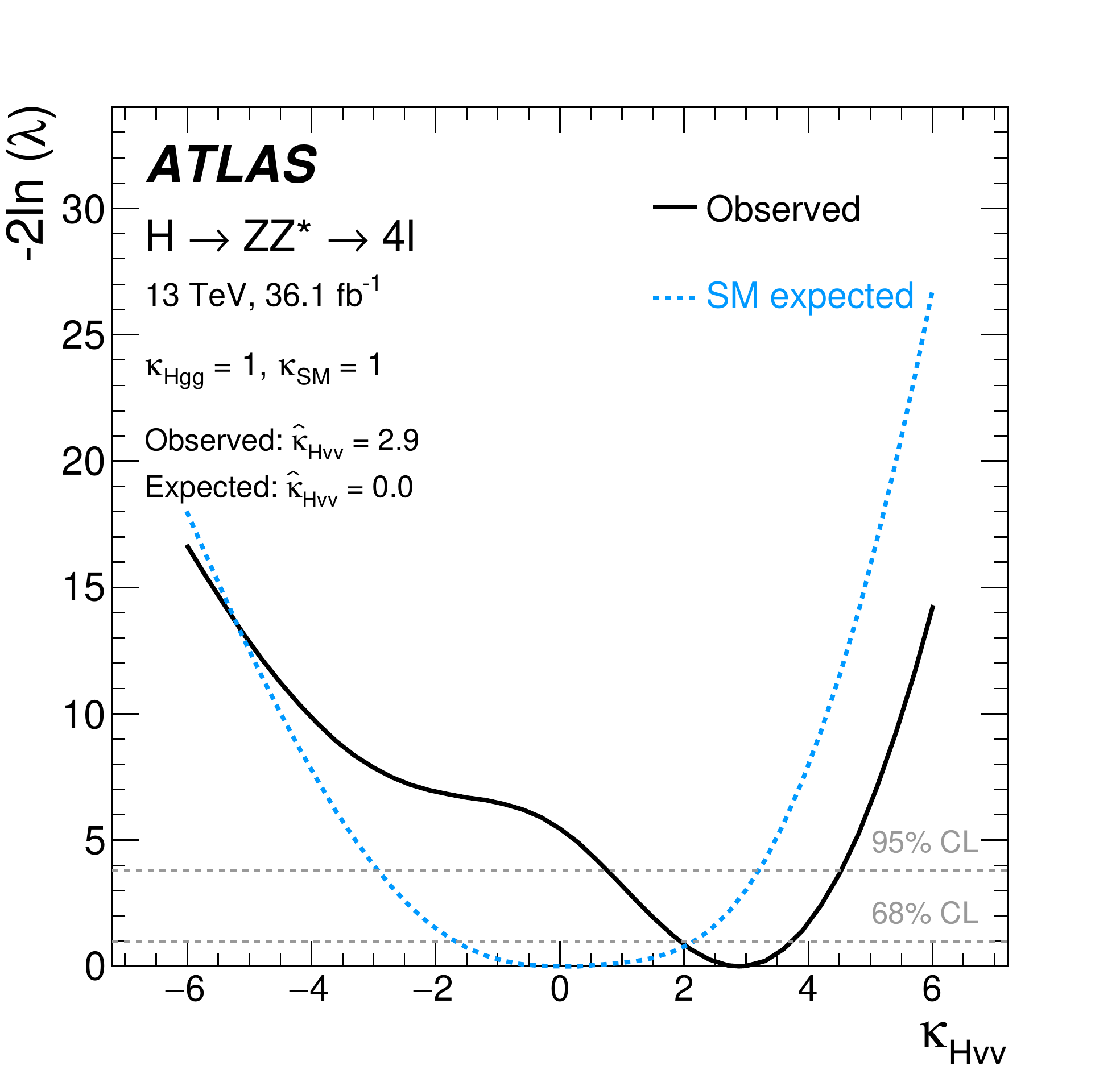}}
  \subfloat[\label{fig:kAVV}]{\includegraphics[width=0.5\linewidth]{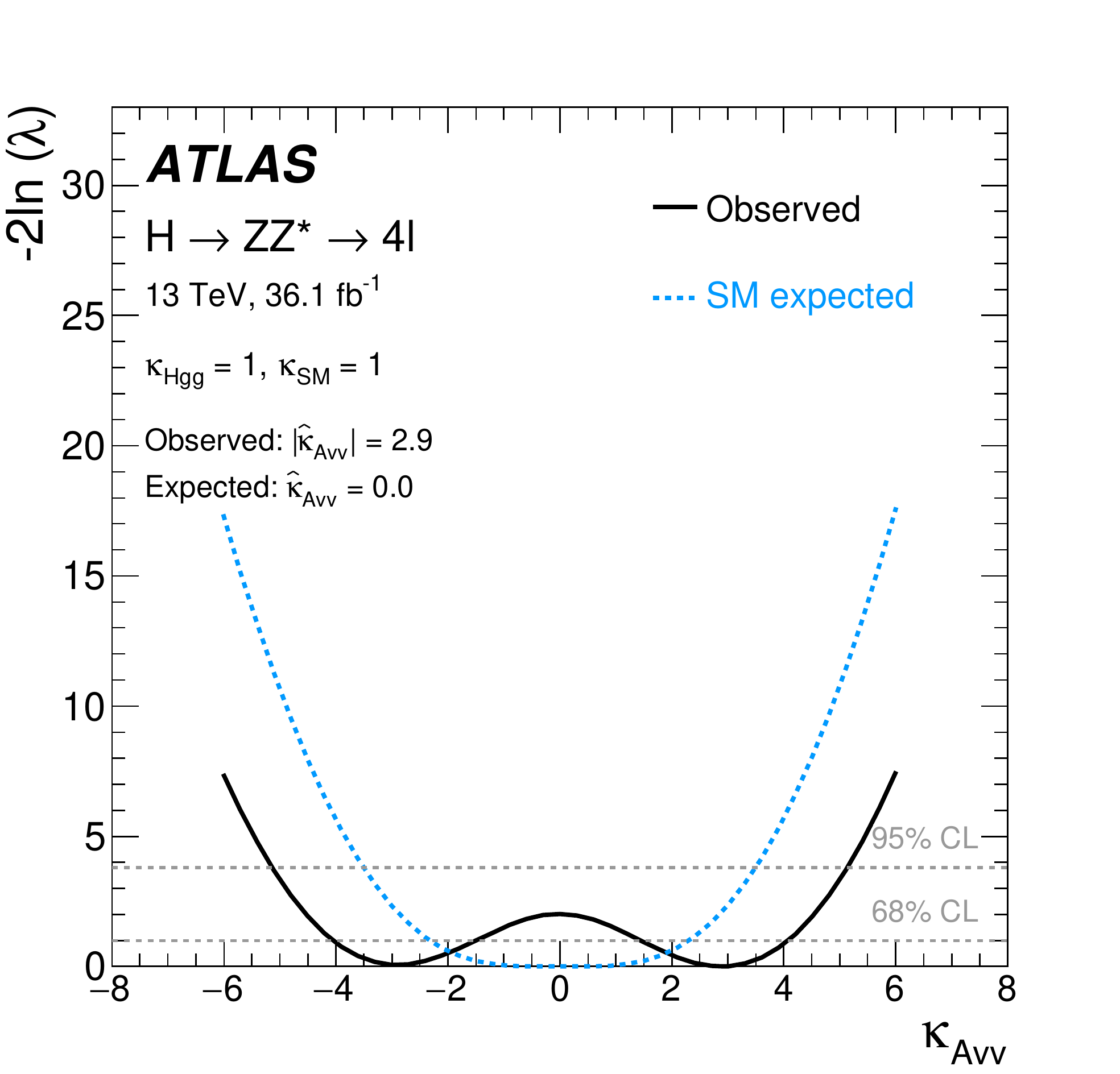}}
   \caption{Observed (solid black line) and SM expected (dashed blue line) negative log-likelihood scans for (a) \kAgg,  (b) \kHVV and (c) \kAVV coupling parameters using \Lum of data at $\sqrt{\mathrm{s}}=$~13~TeV. The horizontal lines indicate the value of the profile likelihood ratio corresponding to the 68\%  and 95\%  CL intervals for the parameter of interest, assuming the asymptotic $\chi^2$ distribution of the test statistic. 
    \label{fig:kBSM_1D_paperbody} }
\end{figure}

Figure~\ref{fig:kBSM_1D_paperbody} shows the observed negative log-likelihood as function of one BSM
coupling at a time, together with the expectation for the SM Higgs boson. 
 The corresponding exclusion limits at a 95\% confidence level (CL), the best-fit values and the size of the deviation from the SM are summarized in Table~\ref{tab:fit_eft}.  
\begin{table*}[t]
  \centering
  \caption{Expected and observed confidence intervals at 95\% CL on the \kAgg,  \kHVV and \kAVV coupling parameters, their best-fit values and corresponding compatibility with the SM expectation, as obtained from the negative log-likelihood scans performed with \Lum of data at $\sqrt{\mathrm{s}}=$~13~TeV. The coupling \kHgg is fixed to the SM value of one in the fit, while the coupling \kSM is either fixed to the SM value of one or left as a free parameter of the fit.}
  \label{tab:fit_eft}

     \vspace{0.1cm}
  {\renewcommand{\arraystretch}{1.1}
 
{
  
\begin{tabular}{cc|cc|rc|c}
    \hline\hline
    BSM coupling&  Fit&  Expected&  Observed&  \multicolumn{1}{c}{Best-fit}&  Best-fit& Deviation\\

    $\kappa_{\mathrm{BSM}}$&  configuration&  conf. inter.&  conf. inter.&  \multicolumn{1}{c}{$\hat{\kappa}_{\mathrm{BSM}}$}&  $\hat{\kappa}_{\mathrm{SM}}$& from SM\\
   \hline
   \hline
   \kAgg& (\kHgg~=~1, \kSM~=~1)& 
   [$-0.47$, 0.47]& [$-0.68$, 0.68]&
   $\pm$0.43& -&
   1.8$\sigma$\\

   \hline
   \kHVV& (\kHgg~=~1, \kSM~=~1)& 
   [$-2.9$, 3.2]& [0.8, 4.5]&
   2.9\phantom{0}& -&
   2.3$\sigma$\\

   \kHVV& (\kHgg~=~1, \kSM~free)& 
   [$-3.1$, 4.0]& [$-0.6$, 4.2]&
   2.2\phantom{0}&  1.2&
   1.7$\sigma$\\

   \hline
   \kAVV& (\kHgg~=~1, \kSM~=~1)& 
   [$-3.5$, 3.5]& [$-5.2$, 5.2]&
    $\pm$2.9\phantom{0}& -&
   1.4$\sigma$\\

   \kAVV& (\kHgg~=~1, \kSM~free)& 
   [$-4.0$, 4.0]& [$-4.4$, 4.4]&
    $\pm$1.5\phantom{0}& 1.2&
   0.5$\sigma$\\

   \hline\hline
  \end{tabular}
  }}
 \end{table*}
The event yields measured in the introduced reconstructed event categories do not provide any sensitivity to the sign of the \kAgg and \kAVV coupling parameters. On the other hand, event yields are expected to be larger for positive \kHVV values compared to the negative ones due to large interference effects with the CP-even SM coupling interactions.
Due to the larger number of events observed compared with expectation in the reconstructed
\VBF-enriched event categories, the best-fit values for the coupling parameters \kAgg, \kHVV and \kAVV differ from zero and deviate from the SM expectation at the level of 1.8$\sigma$, 2.3$\sigma$ and 1.4$\sigma$, respectively.  
If the coupling parameter \kSM of the SM interaction is left free in the fit, the expected limits on the BSM $HVV$ and $AVV$ couplings decrease by up to 10\%. The observed deviation from the SM hypothesis decreases to below 2$\sigma$ (1$\sigma$) for the BSM $HVV$ ($AVV$) coupling, since the observed excess of events is at least partially absorbed by a 20\% increase of the SM coupling parameter \kSM. The best-fit \kHVV and \kAVV values decrease correspondingly. Due to the mentioned interference effects for CP-even couplings, the expected yields decrease more steeply with decreasing \kHVV, so that the increasing \kSM value cannot fully compensate for the observed excess. The best-fit \kHVV value therefore decreases less than the best-fit \kAVV value compared to the fit configuration with \kSM~=~1.

\begin{figure}[!htbp]
  \centering
  \subfloat[\label{fig:2DkSMfixed}]{\includegraphics[width=0.5\linewidth]{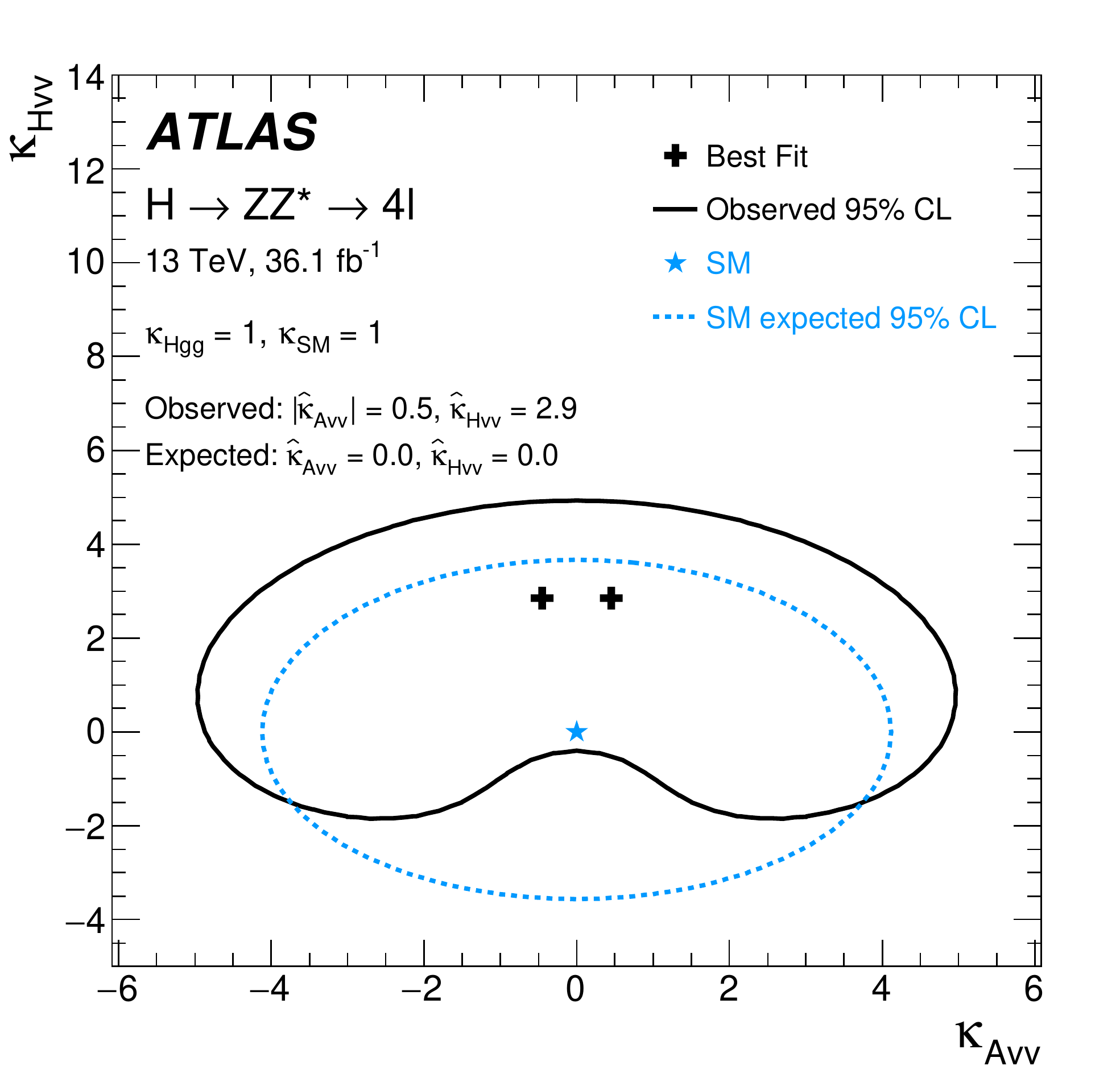}}
  \subfloat[\label{fig:2DkSMfloating}]{\includegraphics[width=0.5\linewidth]{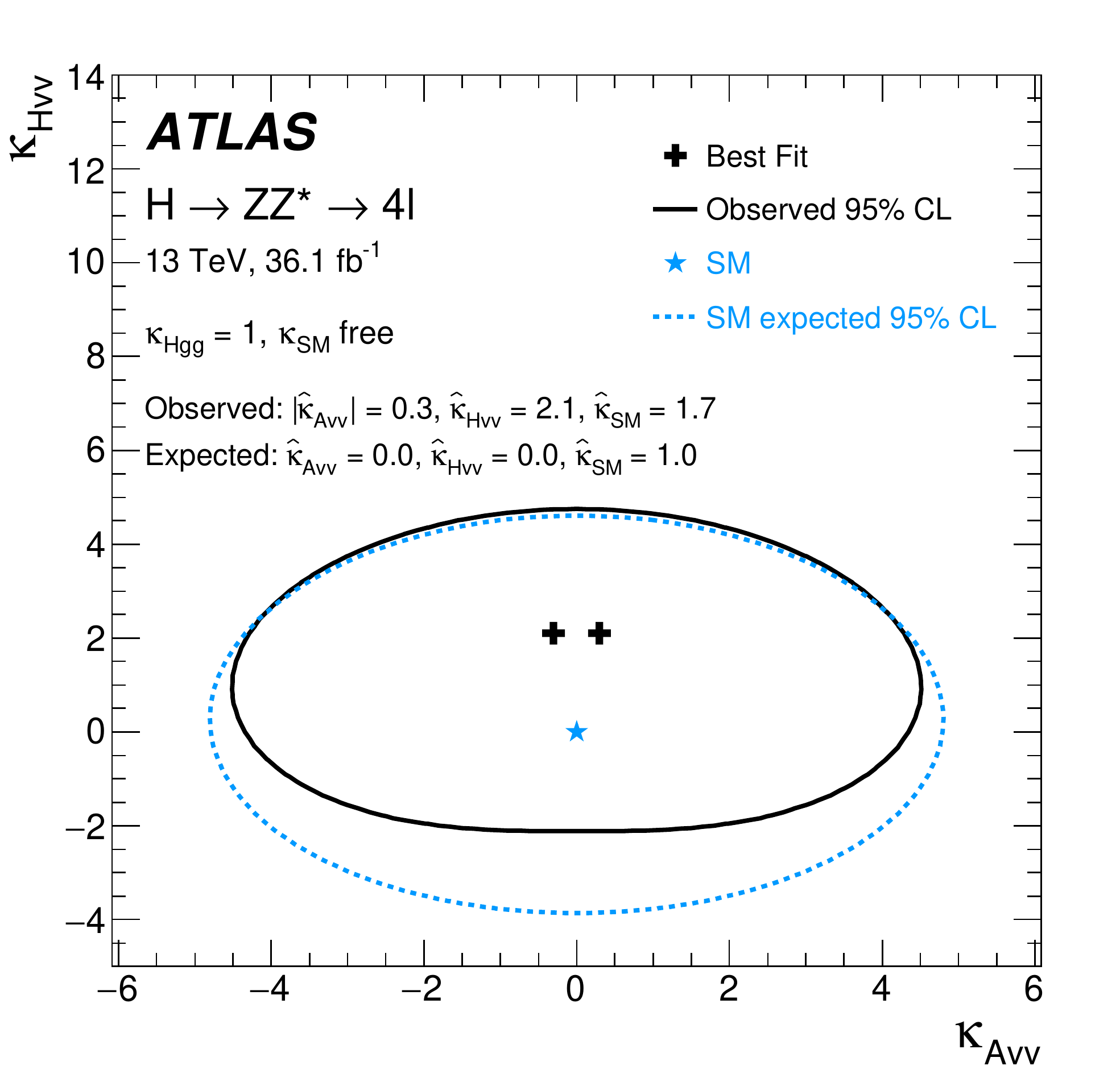}}
   \caption{Observed (black) and SM expected (blue) contours of the two-dimensional negative log-likelihood at  95\% CL for the \kHVV and \kAVV coupling parameters with \Lum of data at $\sqrt{\mathrm{s}}=$~13~TeV. The coupling \kHgg is fixed to the SM value of one in the fit. The coupling \kSM is (a) fixed to the SM value of one or (b) left as a free parameter of the fit (b). 
    \label{fig:kBSM_2D_paperbody} }
\end{figure}

The CP-even and CP-odd BSM couplings to heavy vector bosons are also probed simultaneously in a two-dimensional contour analysis of the negative log-likelihood. The results are shown in Figure~\ref{fig:kBSM_2D_paperbody} and summarized in Table~\ref{tab:kBSM_2D_paperbody}.
\begin{table*}[hbt!]
  \centering
  \caption{The best-fit coupling values and corresponding deviation from the SM expectation, as obtained from the two-dimensional \kHVV~--~\kAVV negative log-likelihood scans performed with \Lum of data at $\sqrt{\mathrm{s}}=$~13~TeV.}
  \label{tab:kBSM_2D_paperbody}

     \vspace{0.1cm}
  {\renewcommand{\arraystretch}{1.1}
 
{
  
\begin{tabular}{c|ccc|c}
\hline\hline
   Fit configuration&   Best-fit $\hat{\kappa}_{HVV}$&  Best-fit $\hat{\kappa}_{AVV}$&  Best-fit $\hat{\kappa}_{\mathrm{SM}}$& Deviation from SM\\
    \hline
    
    \kHgg~=~1, \kSM~=~1&
    2.9&  $\pm$0.5& -& 1.9$\sigma$\\
    
    \kHgg~=~1, \kSM~free&
    2.1&  $\pm$0.3& 1.7& 1.2$\sigma$\\

   \hline\hline
  \end{tabular}
  }}
 \end{table*}

The best-fit value $\hat{\kappa}_{HVV}$ obtained from the two-dimensional scan is similar to the one obtained in the one-dimensional scan. The value of $\hat{\kappa}_{AVV}$ from the two-dimensional scan is closer to the SM expectation than the corresponding value from the one-dimensional scan.
The obtained result is compatible with the SM prediction within 2$\sigma$. 

The coupling parameter \kAgg is also probed directly by the cross sections measured in the reduced Stage-1 production bins. The largest sensitivity to this coupling is obtained from the \STXSggFZeroJ production bin. Here one can neglect the impact of the BSM gluon coupling on the \BDTZeroJ observable that is based solely on the Higgs boson decay topology. The cross-section dependence on the BSM coupling is parameterized using simulated\progname{MadGraph5\_aMC@NLO} samples and fitted to the measured values. The fit results agree with those presented in Table~\ref{tab:fit_eft}.

\section{Summary}
\label{sec:summary}

The coupling properties of the Higgs boson are studied in the four-lepton decay channel using \Lum of LHC $pp$ collision data at $\sqrt{\mathrm{s}}=$13~TeV collected by the ATLAS experiment.
The Higgs boson candidate events are categorized into several topologies, providing sensitivity to different production modes in various regions of phase space. Additional \BDT discriminants are used to further improve the sensitivity in reconstructed event categories with a sufficiently large number of events. 

The cross sections times branching ratio for $H \rightarrow ZZ^*$ decay measured in dedicated production bins are in agreement with the SM predictions. The largest deviation of 2.2$\sigma$ is observed for the \VBF production due to an observed excess of events characterized by the presence of at least two jets and a dijet invariant mass above 120~GeV.  The inclusive cross section in the Higgs boson rapidity range of $|y_H|<$~2.5 is measured to be \sBRZZ~$=1.73^{+0.26}_{-0.24}$~pb compared to the SM prediction of $1.34\pm0.09$~pb. Results are also interpreted within the $\kappa$ framework with coupling modifiers $\kappa_V$ and $\kappa_F$, showing compatibility with the SM.
Based on event yields observed in each reconstructed event category, constraints are placed on possible BSM interactions
of the Higgs boson within the framework of an effective Lagrangian extension of the SM.
The data are shown to be consistent with the SM hypothesis, with the largest deviations of about 2$\sigma$ due to the excess of observed events in the VBF categories. Exclusion limits are set on the CP-even and CP-odd BSM couplings of the Higgs boson to vector bosons  and on the CP-odd BSM Higgs boson coupling to gluons.


\section*{Acknowledgements}


We thank CERN for the very successful operation of the LHC, as well as the
support staff from our institutions without whom ATLAS could not be
operated efficiently.

We acknowledge the support of ANPCyT, Argentina; YerPhI, Armenia; ARC, Australia; BMWFW and FWF, Austria; ANAS, Azerbaijan; SSTC, Belarus; CNPq and FAPESP, Brazil; NSERC, NRC and CFI, Canada; CERN; CONICYT, Chile; CAS, MOST and NSFC, China; COLCIENCIAS, Colombia; MSMT CR, MPO CR and VSC CR, Czech Republic; DNRF and DNSRC, Denmark; IN2P3-CNRS, CEA-DRF/IRFU, France; SRNSFG, Georgia; BMBF, HGF, and MPG, Germany; GSRT, Greece; RGC, Hong Kong SAR, China; ISF, I-CORE and Benoziyo Center, Israel; INFN, Italy; MEXT and JSPS, Japan; CNRST, Morocco; NWO, Netherlands; RCN, Norway; MNiSW and NCN, Poland; FCT, Portugal; MNE/IFA, Romania; MES of Russia and NRC KI, Russian Federation; JINR; MESTD, Serbia; MSSR, Slovakia; ARRS and MIZ\v{S}, Slovenia; DST/NRF, South Africa; MINECO, Spain; SRC and Wallenberg Foundation, Sweden; SERI, SNSF and Cantons of Bern and Geneva, Switzerland; MOST, Taiwan; TAEK, Turkey; STFC, United Kingdom; DOE and NSF, United States of America. In addition, individual groups and members have received support from BCKDF, the Canada Council, CANARIE, CRC, Compute Canada, FQRNT, and the Ontario Innovation Trust, Canada; EPLANET, ERC, ERDF, FP7, Horizon 2020 and Marie Sk{\l}odowska-Curie Actions, European Union; Investissements d'Avenir Labex and Idex, ANR, R{\'e}gion Auvergne and Fondation Partager le Savoir, France; DFG and AvH Foundation, Germany; Herakleitos, Thales and Aristeia programmes co-financed by EU-ESF and the Greek NSRF; BSF, GIF and Minerva, Israel; BRF, Norway; CERCA Programme Generalitat de Catalunya, Generalitat Valenciana, Spain; the Royal Society and Leverhulme Trust, United Kingdom.

The crucial computing support from all WLCG partners is acknowledged gratefully, in particular from CERN, the ATLAS Tier-1 facilities at TRIUMF (Canada), NDGF (Denmark, Norway, Sweden), CC-IN2P3 (France), KIT/GridKA (Germany), INFN-CNAF (Italy), NL-T1 (Netherlands), PIC (Spain), ASGC (Taiwan), RAL (UK) and BNL (USA), the Tier-2 facilities worldwide and large non-WLCG resource providers. Major contributors of computing resources are listed in Ref.~\cite{ATL-GEN-PUB-2016-002}.

\clearpage

\printbibliography

%

\clearpage
\begin{flushleft}
{\Large The ATLAS Collaboration}

\bigskip

M.~Aaboud$^\textrm{\scriptsize 137d}$,
G.~Aad$^\textrm{\scriptsize 88}$,
B.~Abbott$^\textrm{\scriptsize 115}$,
O.~Abdinov$^\textrm{\scriptsize 12}$$^{,*}$,
B.~Abeloos$^\textrm{\scriptsize 119}$,
S.H.~Abidi$^\textrm{\scriptsize 161}$,
O.S.~AbouZeid$^\textrm{\scriptsize 139}$,
N.L.~Abraham$^\textrm{\scriptsize 151}$,
H.~Abramowicz$^\textrm{\scriptsize 155}$,
H.~Abreu$^\textrm{\scriptsize 154}$
% ...
\bigskip
\\
% ...
$^{12}$ Institute of Physics, Azerbaijan Academy of Sciences, Baku, Azerbaijan\\
% ...
$^{88}$ CPPM, Aix-Marseille Universit{\'e} and CNRS/IN2P3, Marseille, France\\
% ...
$^{115}$ Homer L. Dodge Department of Physics and Astronomy, University of Oklahoma, Norman OK, United States of America\\
% ...
$^{119}$ LAL, Univ. Paris-Sud, CNRS/IN2P3, Universit{\'e} Paris-Saclay, Orsay, France\\
% ...
$^{137}$ $^{(a)}$ Facult{\'e} des Sciences Ain Chock, R{\'e}seau Universitaire de Physique des Hautes Energies - Universit{\'e} Hassan II, Casablanca; $^{(b)}$ Centre National de l'Energie des Sciences Techniques Nucleaires, Rabat; $^{(c)}$ Facult{\'e} des Sciences Semlalia, Universit{\'e} Cadi Ayyad, LPHEA-Marrakech; $^{(d)}$ Facult{\'e} des Sciences, Universit{\'e} Mohamed Premier and LPTPM, Oujda; $^{(e)}$ Facult{\'e} des sciences, Universit{\'e} Mohammed V, Rabat, Morocco\\
% ...
$^{139}$ Santa Cruz Institute for Particle Physics, University of California Santa Cruz, Santa Cruz CA, United States of America\\
% ...
$^{151}$ Department of Physics and Astronomy, University of Sussex, Brighton, United Kingdom\\
% ...
$^{154}$ Department of Physics, Technion: Israel Institute of Technology, Haifa, Israel\\
$^{155}$ Raymond and Beverly Sackler School of Physics and Astronomy, Tel Aviv University, Tel Aviv, Israel\\
% ...
$^{161}$ Department of Physics, University of Toronto, Toronto ON, Canada\\
% ...
$^{a}$ Also at Department of Physics, King's College London, London, United Kingdom\\
$^{b}$ Also at Institute of Physics, Azerbaijan Academy of Sciences, Baku, Azerbaijan\\
$^{c}$ Also at Novosibirsk State University, Novosibirsk, Russia\\
$^{d}$ Also at TRIUMF, Vancouver BC, Canada\\
$^{e}$ Also at Department of Physics {\&} Astronomy, University of Louisville, Louisville, KY, United States of America\\
% ...
$^{*}$ Deceased
\end{flushleft}







\end{document}